\documentclass[11pt,a4paper]{article}
\pdfoutput=1
\usepackage{dsfont}
\usepackage{slashed}
\usepackage{setspace}
\pdfoutput=1
\usepackage[utf8]{inputenc} 
\usepackage{a4wide}
\usepackage{cite}
\usepackage{xcolor}
\usepackage{amssymb}
\usepackage{amsfonts}
\usepackage{graphicx}
\usepackage{subcaption}
\usepackage{hyperref}
\usepackage{mathbbol}
\usepackage{amssymb}
\usepackage{booktabs}
\usepackage{colortbl}
\usepackage{multirow}
\usepackage{appendix}
\DeclareSymbolFontAlphabet{\amsmathbb}{AMSb}
\usepackage{amsfonts}
\usepackage{amssymb}
\usepackage[centertags]{amsmath}
 
 \def\1{\mathbf{1}}
 \def\3{\mathbf{3}}
 \def\2{\mathbf{2}}

\def\gtap{\ \raisebox{-.4ex}{\rlap{$\sim$}} \raisebox{.4ex}{$>$}\ }

\allowdisplaybreaks[1]

\def\3{\mathbf{3}}

\usepackage{a4wide}
\usepackage{colordvi}

\newcommand{\bec}{\begin{cases}}
\newcommand{\eec}{\end{cases}}
\newcommand{\beq}{\begin{equation*}}
\newcommand{\eeq}{\end{equation*}}
\newcommand{\be}{\begin{equation}}
\newcommand{\ee}{\end{equation}}
\newcommand{\ba}{\begin{eqnarray}}
\newcommand{\ea}{\end{eqnarray}}

\begin{document}

\begin{titlepage}

${}$\vskip 3cm
\vspace*{-15mm}
\begin{flushright}
SISSA 14/2021/FISI\\
IPPP/20/118\\
IPMU21-0044
\end{flushright}
\vspace*{0.7cm}

\vskip 9mm
\begin{center}
{\bf\Large 
Aspects of High Scale Leptogenesis with Low-Energy Leptonic\\
[4mm]
 CP Violation
} \\
[5mm]
A. Granelli$^{~a}$, K. Moffat$^{~b}$ and 
S.~T.~Petcov$^{~a,c,}$\footnote{Also at:
Institute of Nuclear Research and Nuclear Energy,
Bulgarian Academy of Sciences, 1784 Sofia, Bulgaria.} \\
\vspace{8mm}
$^{a}$\,{\it SISSA/INFN, Via Bonomea 265, 34136 Trieste, Italy.} \\
\vspace{2mm}
$^{b}$\,{\it Institute for Particle Physics Phenomenology, Department of
Physics, Durham University, South Road, Durham DH1 3LE, United Kingdom.}\\
\vspace{2mm}
$^{c}$\,{\it Kavli IPMU (WPI), University of Tokyo, 5-1-5 Kashiwanoha, 277-8583 Kashiwa, Japan.}
\end{center}
\vspace{2mm}

\begin{abstract}
Using the density matrix equations (DME) for high scale leptogenesis 
based on the type I seesaw mechanism,
in which the CP violation (CPV) is provided by the low-energy Dirac 
or/and Majorana phases of the neutrino mixing (PMNS) matrix, 
we investigate the 1-to-2 and the 2-to-3 flavour regime transitions, 
where the 1, 2 and 3 leptogenesis flavour regimes in the generation 
of the baryon asymmetry of the Universe $\eta_B$ are  
described by the Boltzmann equations. Concentrating on the 
1-to-2 flavour transition we determine the general conditions 
under which $\eta_B$ goes through zero 
and changes sign in the transition. Analysing in detail the behaviour 
of $\eta_B$ in the transition in the case of two heavy Majorana 
neutrinos $N_{1,2}$ with hierarchical masses, $M_1 \ll M_2$, we find, 
in particular, that i) the Boltzmann equations  in many  cases 
fail to describe correctly the generation of $\eta_B$ in 
the 1, 2 and 3 flavour regimes,  
ii) the 2-flavour regime can persist above (below) 
$\sim 10^{12}$ GeV ($\sim 10^9$ GeV), 
iii) the flavour effects in leptogenesis persist beyond 
the typically considered maximal for these effects 
leptogenesis scale of $10^{12}$ GeV.  
We further determine the minimal scale $M_{1\text{min}}$ 
at which we can have successful leptogenesis 
when the CPV is provided only by the Dirac 
or Majorana 
phases of the PMNS matrix as well as the ranges of scales 
and values of the phases for having successful leptogenesis.
We show, in particular, that when the CPV is due to the Dirac 
phase $\delta$, there is a direct relation 
between the sign of $\sin \delta$ 
and the sign of $\eta_B$ in the regions of 
viable leptogenesis 
in the case of normal hierarchical 
light neutrino mass spectrum;
for the inverted hierarchical spectrum 
the same result holds for $M_1 \lesssim 10^{13}$ GeV.
The considered different scenarios of leptogenesis
are testable and falsifiable in low-energy neutrino experiments. 

\end{abstract}

\end{titlepage}
\setcounter{footnote}{0}
\setcounter{page}{2}

\section{Introduction}
\label{sec:intro}

\indent
  In spite of the fact that the leptogenesis idea of the origin 
of the matter-antimatter, or baryon, asymmetry of the Universe  
is 35 years old  \cite{Fukugita:1986hr,Kuzmin:1985mm},
the leptogenesis scenario of the asymmetry generation continues 
to be actively investigated 
(see, e.g., the recent review article \cite{bodeker2021baryogenesis}
which includes also an extended list of references).
A very attractive version of leptogenesis is that based on 
type I seesaw mechanism of neutrino mass generation
\cite{Minkowski:1977sc,Yanagida:1979as,GellMann:1980vs,Glashow:1979nm,
Mohapatra:1979ia}, which also corresponds to the original scenario proposed 
in \cite{Fukugita:1986hr}. The type I seesaw mechanism provides a natural 
explanation of the smallness of neutrino masses and via leptogenesis 
establishes a link between the existence and smallness of neutrino masses 
and the existence of the baryon asymmetry. 
A basic ingredient of the seesaw scenario
are the singlet RH neutrinos 
$\nu_{lR}$ (singlet RH neutrino fields $\nu_{lR}(x)$), 
by which the Standard Model (SM) can be extended without 
modifying its fundamental properties.  
Such an extension with two RH neutrinos 
is the minimal set-up in which 
leptogenesis can take place, satisfying the three Sakharov 
conditions \cite{Sakharov:1967dj}
for a dynamical generation of the baryon asymmetry. 
In leptogenesis the requirement of lepton charge non-conservation 
is satisfied, as is well known, due to a Majorana mass 
term of the RH neutrinos $\nu_{lR}$ and the Yukawa coupling 
${\cal L}_{\rm Y}(x)$ of $\nu_{lR}$ 
with the Standard Model lepton 
and Higgs doublets, $\psi_{lL}(x)$ and $\Phi(x)$, 
while the requisite C- and CP-symmetry violations 
are ensured by the $\nu_{lR}$ Majorana mass term 
and/or the Yukawa coupling ${\cal L}_{\rm Y}(x)$.
Both the $\nu_{lR}$ Majorana mass term and ${\cal L}_{\rm Y}(x)$
respect the $SU(2)_L\times U(1)_{Y_W}$ symmetry of the SM.
In the diagonal mass basis of the RH neutrinos 
$\nu_{lR}$ and the charged leptons $l^\pm$, $l=e,\mu,\tau$, 
${\cal L}_{\rm Y}(x)$ and the Majorana mass term 
are given by:
\be
\label{YnuIntro}
{\cal L}_{\rm Y,M}(x) =
-\,\left (Y_{li} \overline{\psi_{l L}}(x)\,i\tau_2\,\Phi^*(x)\,N_{iR}(x)
+ \hbox{h.c.} \right )
-\,\frac{1}{2}\,M_{i}\,\overline{N_i}(x)\, N_i(x)\,,
\ee
%
\noindent
where $Y_{li}$ is the matrix of neutrino Yukawa couplings 
(in the chosen basis) and $N_i$ ($N_i(x)$) is the heavy Majorana neutrino
(field) possessing a mass $M_i > 0$. 

 In the present article we revisit the 
high (GUT) scale flavoured leptogenesis scenario, 
which is realised for masses of the heavy Majorana neutrinos 
in the range $M_i \sim (10^9 - 10^{14})$ GeV, i.e., 
for $M_i$ by a few to several orders of magnitude smaller than 
the unification scale of electroweak and
strong interactions, $M_{GUT}\cong 2\times 10^{16}$ GeV. 
The values of the masses $M_i$, i.e., their scale and spectrum, 
set the scale of leptogenesis. 
In this scenario, the out-of-equilibrium decays 
$N_j \rightarrow l^{+} + \Phi^{(-)}$ and 
$N_j \rightarrow l^{-} + \Phi^{(+)}$
of the heavy Majorana neutrinos  $N_j$, 
caused by the CP non-conserving 
neutrino Yukawa couplings in Eq. \eqref{YnuIntro}, 
proceed with different rates,
producing CP violating (CPV) 
asymmetries in the flavour lepton charges $L_l$, $l=e,\mu,\tau$,
and in the integral lepton charge $L = L_e + L_\mu + L_\tau$,   
of the Universe. The lepton asymmetries thus generated 
are converted into a baryon asymmetry 
by $(B-L)$-conserving, but $(B+L)$-violating,
sphaleron processes which exist in the SM and are effective at 
temperatures $T \sim (132 -  10^{12})$ GeV.
  
For heavy Majorana neutrinos $N_{1,2,3}$ with 
hierarchical mass spectrum, $M_1 \ll M_2 \ll M_3$, 
there exists, as is well known, a lower bound on the mass 
of the lightest $N_1$ for which the matter-antimatter asymmetry 
can be generated in leptogenesis: $M_1 \gtap 10^{9}$ GeV \cite{Davidson:2002qv}.
Furthermore, when flavour effects in leptogenesis 
\cite{Nardi:2006fx,Abada:2006fw,Abada:2006ea} 
(see also \cite{Barbieri:1999ma,Nielsen:2002pc,Endoh:2003mz}) 
are taken into account, leptogenesis was shown to be 
possible at scales compatible with the quoted lower bound 
with the requisite CP violation provided exclusively by 
the Dirac and/or  Majorana phases in the Pontecorvo, 
Maki, Nakagawa, Sakata (PMNS) neutrino (lepton) mixing matrix 
$U_{\text{PMNS}}$  \cite{Pascoli:2006ie,Pascoli:2006ci,Blanchet:2006be,Branco:2006ce,Anisimov:2007mw,Molinaro:2008rg,Molinaro:2008cw,Dolan:2018qpy}. 
Rather detailed studies performed in 2018 
in \cite{Moffat:2018wke,Moffat:2018smo} have shown 
that in the case of spectrum of masses of the 
heavy Majorana neutrinos with mild hierarchy, 
$M_2 \sim 3M_1$, $M_3\sim 3M_2$, 
the ``flavoured'' leptogenesis can successfully 
generate the observed baryon asymmetry at scales as 
low as $\sim 10^{6}$ GeV, i.e., 
for $M_1 \gtap 10^{6}$ GeV, 
and that in this case as well 
the Dirac or Majorana CPV phases present in the PMNS matrix 
can be the unique source of the required CP violation. 
Moreover, in \cite{Brivio:2019hrj} 
the same conclusion concerning 
the Dirac or Majorana phases being the 
source of CP violation in leptogenesis, 
was shown to hold in the case of the so-called ``Neutrino Option" 
seesaw scenario \cite{Brivio:2017dfq} 
in which the term quadratic in the Higgs field 
in the Higgs potential, responsible for the breaking of the 
SM electroweak symmetry, 
is generated radiatively at one loop by the neutrino 
Yukawa coupling in Eq. (\ref{Ynu}).  

 The studies performed in 
\cite{Moffat:2018wke,Moffat:2018smo}
were based on the density matrix equations (DME)
for leptogenesis \cite{Simone_2007,Blanchet_2007,Blanchet_2013}, in which, 
in particular, the flavour decoherence effects associated with the 
charged lepton Yukawa couplings are accounted for in the various regimes 
they go through continuously (from negligible to non-negligible but 
non-thermalised to fully thermalised)
in the expanding and cooling Universe. 
In contrast, the Boltzmann equations (BE) 
describe flavour effects in leptogenesis only for 
either negligible or fully thermalised charged lepton Yukawas.
As a consequence, the DME approach to leptogenesis opened up 
the possibility to investigate quantitatively the 
behaviour of the baryon asymmetry $\eta_B$ 
in transitions between the different flavour regimes 
in leptogenesis. The analysis done in \cite{Moffat:2018smo} 
revealed that in the case of heavy Majorana neutrinos with 
hierarchical mass spectrum, $M_1 \ll M_2 \ll M_3$,
the baryon asymmetry $\eta_B$ can go through zero,  
changing sign at certain scale $M_1$ 
in the transition between the 
unflavoured (or single flavoured) and the two flavoured 
leptogenesis regimes, associated with the $\tau$-Yukawa coupling.
Moreover, when the CP violation is provided in leptogenesis 
by the low-energy CPV 
phases present in the PMNS matrix, the width of the transition 
was shown to become extremely large and to lead 
to the existence of a ``plateau'' in the baryon asymmetry 
dependence on the scale $M_1$. 

 In the present article we continue to investigate 
the transitions between the different flavour regimes 
in high scale leptogenesis based on type I 
seesaw mechanism with hierarchical heavy Majorana neutrinos, 
$M_1 \ll M_2 \ll M_3$, began in  \cite{Moffat:2018smo}. 
We consider the case in which the CP violation (CPV) in leptogenesis 
is provided by the low-energy Dirac 
or/and Majorana phases of the PMNS neutrino mixing matrix.  
Using the density matrix equations (DME) 
to describe high scale leptogenesis,
we investigate in detail the 1-to-2 and the 2-to-3 flavour 
regime transitions, where the 1, 2 and 3 leptogenesis flavour regimes 
in the generation of the baryon asymmetry $\eta_B$ are  
described by the Boltzmann equations. Concentrating on the 
1-to-2 flavour transition we determine the general conditions 
under which $\eta_B$ goes through zero and changes 
sign in the transition and $|\eta_B|$ 
reaches a plateau as $M_1$ increases. 
We analyse further in detail 
the behaviour of $\eta_B$ in the transition under 
these conditions in the case of two heavy Majorana 
neutrinos $N_{1,2}$ with hierarchical masses, $M_1 \ll M_2$, 
and identify, in particular, cases in which the 
baryon asymmetry exhibits a ``non-standard'' behaviour 
in the transition. We determine the minimal scale 
$M_{1\text{min}}$ as well as the 
corresponding ranges of $M_1$ and of the 
Dirac and Majorana CPV phases
for which we can have successful leptogenesis 
when the requisite CP violation in leptogenesis 
is provided only either by the Dirac or by the Majorana phases 
of the PMNS matrix.

 The paper is organised as follows. In Section \ref{Notation} 
we summarise the existing data on the neutrino masses and mixing 
 that we use in our analysis, the basics of the type I seesaw 
scenario, and introduce the 1, 2 and 3 flavour 
Boltzmann equations as well as the  density matrix equations 
that are employed in our study, elucidating the role 
of the charged lepton Yukawa couplings.
In Section \ref{sec:sign} we investigate 
the baryon asymmetry ($\eta_B$) sign change 
in the transition between the one and two flavour regimes  
in the case of three heavy Majorana  neutrinos $N_{1,2,3}$ 
with hierarchical masses, $M_1 \ll M_2\ll M_3$, 
$M_1 \gtrsim 10^{9}$ GeV, determine the general conditions 
under which the sign change takes place 
and then perform a detailed analysis of 
the transitions between the different flavour regimes.
To make the discussion as transparent as possible 
we investigate in Section \ref{sec:N3dec} 
the behaviour of the baryon asymmetry 
under the general conditions under which 
the sign change of the baryon asymmetry $\eta_B$ 
in the 1-to-2 flavour regimes is possible in 
the case of decoupled $N_3$, 
in which the number of parameters 
is significantly smaller than in the general case 
with three heavy Majorana neutrinos.
The thorough analysis is performed in this Section 
with the CP violation necessary for the generation of the 
baryon asymmetry provided by the low-energy 
Dirac or/and Majorana phases present in the PMNS neutrino mixing matrix.
We conclude in Section \ref{sec:concs} with a  
summary of our results.

%
%
 \section{The Framework}
\label{Notation}
%
%
\subsection{Neutrino Masses and Neutrino (Lepton) Mixing}
%

 Throughout the present study we employ the reference
3-neutrino mixing scheme (see, e.g., \cite{Tanabashi:2018oca}):
\be
 \nu_{\alpha L}(x)=\sum_{j=1}^3 U_{\alpha j} \nu_{j L}(x),
\ee
%
 where  $\nu_{\alpha L}(x)$, $\alpha=e,\mu,\tau$, is the left-handed (LH) flavour 
neutrino field (which enters into the expression of the weak interaction 
Lagrangian), $\nu_{j L}(x)$, $j=1,2,3$, is the LH component 
of the field of a light neutrino $\nu_j$ with mass $m_j$,  
and  $U$ is the $3\times 3$ unitary Pontecorvo-Maki-Nakagawa-Sakata (PMNS) 
neutrino (lepton) mixing matrix. 
  We consider the case of  
light massive neutrinos $\nu_j$ being Majorana particles and
will use in our analysis the standard parametrisation 
of the PMNS matrix \cite{Tanabashi:2018oca} in this case:  
\begin{equation}
\label{PMNS}
U = \begin{pmatrix}
c_{12}c_{13}&s_{12}c_{13}&s_{13}\text{e}^{-i\delta}\\
-s_{12}c_{23}-c_{12}s_{23}s_{13}\text{e}^{i\delta}&c_{12}c_{23}-s_{12}s_{23}s_{13}\text{e}^{i\delta}&s_{23}c_{13}\\
s_{12}s_{23}-c_{12}c_{23}s_{13}\text{e}^{i\delta}&-c_{12}s_{23}-s_{12}c_{23}s_{13}\text{e}^{i\delta}&c_{23}c_{13}
\end{pmatrix}\times
\begin{pmatrix}
1&0&0\\
0&\text{e}^{\frac{i\alpha_{21}}{2}}&0\\
0&0&\text{e}^{\frac{i\alpha_{31}}{2}}
\end{pmatrix}\,.
\end{equation}
%
 Here $c_{ij} \equiv \cos\theta_{ij}$, $s_{ij} \equiv \sin\theta_{ij}$, 
the angles $\theta_{ij} = [0,\pi/2]$,
$\delta=[0,2\pi)$ is the Dirac CP violation (CPV) phase, 
and $\alpha_{21}$ and $\alpha_{31}$ are the two Majorana 
CPV phases \cite{Bilenky:1980cx}, 
$\alpha_{21 (31)} = [0,4\pi]$~
\footnote{Within the type I seesaw mechanism of neutrino 
mass generation we will consider the mass-eigenstate neutrinos 
to be Majorana fermions and it proves convenient to work with this extended
range of possible values of the two Majorana phases $\alpha_{21,31}$ 
\cite{Molinaro:2008rg} (see further).
}.
The Dirac and Majorana phases can be sources
of low-energy leptonic CP violation. 
In the case of CP invariance, 
we have $\delta =0,~\pi$ and 
$\alpha_{21 (31)} = k_{2(3)1}\pi$, $k_{2(3)1}=0,1,2,3,4$.

 In what concerns the light neutrinos masses $m_{1,2,3}$,
we use the ``standard'' convention of numbering 
the neutrino mass eigenstates in which 
$\Delta m^2_{21} \equiv m^2_2 - m^2_1 > 0$ and
$\Delta m^2_{31(32)} \equiv m^2_3 - m^2_{1(2)}$ 
are associated, together respectively with 
the angles $\theta_{12}$ and $\theta_{23}$, 
with the observed flavour conversion of solar 
(electron) neutrinos $\nu_e$ and the dominant oscillations   
of atmospheric muon neutrinos and antineutrinos,
$\nu_\mu$ and $\bar{\nu}_\mu$, 
while the angle $\theta_{13}$, together 
with  $\Delta m^2_{31(32)}$, 
is associated with the reactor $\bar{\nu}_e$ oscillations 
observed in the Daya Bay, RENO and Double 
Chooz experiments 
\cite{Tanabashi:2018oca}.
The enormous amount of neutrino oscillation data
accumulated over many years of research 
(see, e.g., \cite{Zyla:2020zbs})
made it possible to determine 
$\Delta m^2_{21}$, $\sin^2\theta_{12}$,  
$|\Delta m^2_{31}|$ ($|\Delta m^2_{32}|$),
$\sin^2\theta_{23}$ and $\sin^2\theta_{13}$ 
with remarkably high precision 
 (see, e.g., \cite{Capozzi_2020,Esteban_2020}).
\begin{table}\centering
\begin{tabular}{c|cccccc}\toprule
\rowcolor[gray]{.95}
         & $\theta_{12}$   & $\theta_{13}$  &$\theta_{23}$ &$\delta$ &
        $\Delta m_\odot^2$ & $\Delta m_\text{atm}^2$ \\ 
        \rowcolor[gray]{.95}
        \multirow{-2}{*}{\textbf{Ordering}}&(${}^\circ$) & (${}^\circ$) & (${}^\circ$) & (${}^\circ$) & ($10^{-5}\text{ eV}^2$) & ($10^{-3}\text{ eV}^2$) \\ 
        \hline
        \midrule
 \textbf{NO}& $33.44^{+0.77}_{-0.74}$ & $8.57^{+0.12}_{-0.12}$ & $49.2^{+0.9}_{-1.2}$ & $197^{+27}_{-24}$ & $7.42^{+0.21}_{-0.20}$ & $2.517^{+0.026}_{-0.028}$  \\ 
 \midrule
 \textbf{IO} & $33.45^{+0.78}_{-0.75}$ & $8.60^{+0.12}_{-0.12}$ & $49.3^{+0.9}_{-1.1}$ & $282^{+26}_{-30}$ & $7.42^{+0.21}_{-0.20}$ & $-2.498^{+0.028}_{-0.028}$ \\
 \bottomrule
\end{tabular}
\caption{
 Best-fit values and $1\sigma$ allowed ranges of the neutrino mixing
angles  $\theta_{12}$, $\theta_{13}$, $\theta_{23}$, and of the 
$\Delta m_\odot^2\equiv \Delta m^2_{21}$  and 
$\Delta m_\text{atm}^2 \equiv \Delta m^2_{31}$ 
($\Delta m_\text{atm}^2 \equiv \Delta m^2_{32}$) 
in the case of NO (IO) light neutrino mass spectrum, 
obtained in \cite{Esteban_2020}. 
We quote also the best-fit value and $1\sigma$ allowed ranges
of the Dirac CPV phase $\delta$ from \cite{Esteban_2020}.
However, these data on $\delta$ are not used in our analyses.
}
\label{Tab::BestFit}
\end{table}
%
 We report in Table \ref{Tab::BestFit} the best-fit values and $1\sigma$ 
ranges of the three neutrino mixing (or PMNS) angles and 
the two neutrino mass squared differences obtained from the global neutrino 
oscillation data analysis in \cite{Esteban_2020}.
In the numerical analyses we will perform, 
we will use the best-fit values of the three neutrino mixing angles
and of the two neutrino mass squared differences 
quoted in Table \ref{Tab::BestFit}.

The existing neutrino data, as is well known, 
do not allow to determine the sign of $\Delta m^2_{31(32)}$ 
and the two values of ${\rm sgn}(\Delta m^2_{31(32)})$ 
correspond to two possible types of light neutrino mass 
spectrum --  with normal ordering (NO) and inverted ordering (IO), 
which is reflected in Table  \ref{Tab::BestFit}. 
In the adopted convention, the two spectra read:
\begin{itemize}
\item {\bf Normal Ordering (NO):} $m_1< m_2< m_3$,
$\Delta m_{31}^2 \equiv \Delta m_\text{atm}^2>0$;
\item {\bf Inverted Ordering (IO):} $m_3< m_1< m_2$,
$\Delta m_{32}^2 \equiv \Delta m_\text{atm}^2<0$.
\end{itemize}
 Depending on the value of the lightest neutrinos mass, 
the neutrino mass spectrum can also be: 
\begin{itemize}
\item{\bf Normal Hierarchical (NH):}  $0\simeq m_1 \ll m_2 < m_3$, with 
$m_2 \simeq \sqrt{\Delta m_{21}^2}$ and 
$m_3 \simeq \sqrt {\Delta m_{31}^2}$; 
\item{\bf Inverted Hierarchical (IH):} $0\simeq m_3 \ll m_1 < m_2$, 
with $m_1 \simeq \sqrt{|\Delta m_{32}^2| - \Delta m_{21}^2}$ and 
$m_2 \simeq \sqrt{|\Delta m_{32}^2|}$;
\item{\bf Quasi Degenerate (QD):} $m_1\simeq m_2\simeq m_3$, 
with $m^2_{1,2.3}\gg |\Delta m^2_{31(32)}|$. 
\end{itemize}
All considered spectra are compatible with the existing data 
on the light neutrino masses \cite{Tanabashi:2018oca}.
The IO spectrum is disfavoured at approximately $2.7\sigma$ C.L. 
with respect to the NO spectrum by the global neutrino 
oscillation data \cite{Esteban_2020}. 
In our further analyses we will mostly be interested in 
the NH and IH spectra. 

 A few comments relevant for our further discussion 
are in order. As it follows from Table  \ref{Tab::BestFit},
we have  $\Delta m^2_{21} \ll |\Delta m^2_{31(32)}|$,
$\Delta m^2_{21}/|\Delta m^2_{31(32)}| \cong 1/30$.
Apart from some hints 
from the data of the T2K and NO$\nu$A 
experiments \cite{Abe:2019vii,NOvANu2020}
that the Dirac phase $\delta \sim 3\pi/2$,  
no other experimental information on
the Dirac and Majorana CPV phases in 
the PMNS matrix is available at present. 
The values of $\delta$ obtained in the global 
analyses \cite{Capozzi_2020,Esteban_2020}
have relatively large uncertainties. 
In view of this we will treat 
both the Dirac phase $\delta$ and the 
Majorana phases $\alpha_{21}$ and $\alpha_{31}$ 
as free parameters in our study.
 We recall that with $\theta_{13} \cong 0.15$,
the Dirac phase $\delta$ can generate
CP violating effects in neutrino 
oscillations \cite{Cabibbo:1977nk,Bilenky:1980cx,Barger:1980jm},
i.e., a difference between the probabilities of the 
$\nu_\alpha \rightarrow \nu_{\beta}$ and
$\bar{\nu}_\alpha \rightarrow \bar{\nu}_{\beta}$
oscillations, $\alpha\neq \beta =e,\mu,\tau$.
The magnitude of CP violation in
$\nu_\alpha \rightarrow \nu_{\beta}$ and
$\bar{\nu}_\alpha \rightarrow \bar{\nu}_{\beta}$
oscillations ($\alpha\neq \beta$)
is determined by \cite{Krastev:1988yu}
the rephasing invariant 
\be
J_{\rm CP} =
{\rm Im}\, \left (U_{\mu 3}\,U^*_{e3}\,U_{e2}\,U^*_{\mu 2}\right )\,,
\label{JCP02}
\ee
%
which 
\footnote{The $J_{\rm CP}$ factor in Eq. \eqref{JCP02}
is analogous to the rephasing invariant
associated with the Dirac 
CPV phase in the quark mixing matrix \cite{Jarlskog:1985cw}.
}
in the standard parametrisation 
of the PMNS matrix has the form:
\be
J_{CP} \equiv
{\rm Im}\,(U_{\mu 3}\,U^*_{e3}\,U_{e2}\,U^*_{\mu 2}) =
\frac{1}{8}\,\cos\theta_{13}
\sin 2\theta_{12}\,\sin 2\theta_{23}\,\sin 2\theta_{13}\,\sin \delta\,.
\ee
%
If the hints that 
$\delta$ has a value close to $3\pi/2$  
are confirmed by future more precise data 
one would have $J_{CP}\cong -\,0.03$, implying that 
the CP violating effects in neutrino 
oscillations would be relatively large 
and observable in currently running and/or future 
neutrino oscillation experiments 
(T2K, NO$\nu$A, T2HK, DUNE, see, e.g., 
\cite{Tanabashi:2018oca,Zyla:2020zbs}).

In what concerns the Majorana CPV phases 
in the PMNS matrix, 
the flavour neutrino oscillation probabilities
$P(\nu_\alpha \rightarrow \nu_{\beta})$ and
$P(\bar{\nu}_\alpha \rightarrow \bar{\nu}_{\beta})$,
$\alpha,\beta =e,\mu,\tau$, do not depend on
these phases \cite{Bilenky:1980cx,Langacker:1986jv}.
The Majorana phases can play
important role, e.g., in $|\Delta L| = 2$
processes like neutrinoless double beta 
($(\beta\beta)_{0\nu}$-)
decay $(A,Z) \rightarrow (A,Z+2) + e^- + e^-$,
$L$ being the total lepton charge,
in which the Majorana nature of
massive neutrinos $\nu_i$ manifests itself
(see, e.g., Refs.~\cite{Bilenky:1987ty,Bilenky:2001rz, 
Petcov_2020}).

Our interest in the Dirac and Majorana CPV
phases present in the neutrino mixing matrix
is stimulated also by the intriguing possibility
that the Dirac phase and/or the Majorana phases in the PMNS matrix 
$U$ can provide the CP violation
necessary for the generation of the observed
baryon asymmetry of the 
Universe \cite{Pascoli:2006ie, Pascoli:2006ci}.
In the present article we continue   
to explore this intriguing and very appealing 
possibility.

 Finally we comment briefly on the current limits 
on the absolute scale of light neutrino masses 
(or equivalently on the lightest neutrino mass). 
Using the existing best lower bounds on the 
$(\beta\beta)_{0\nu}$-decay half-lives of 
$^{136}$Xe  \cite{KamLAND-Zen:2016pfg}  
and $^{76}$Ge \cite{Agostini:2020xta} 
one can obtain the following 
``conservative'' upper limit on 
the light Majorana neutrino masses, 
which is in the range of the QD spectrum 
\cite{Penedo:2018kpc}:
$m_{1,2,3} \lesssim 0.58$ eV.

 The most stringent upper limit on the light neutrino 
masses, which does not depend on the nature of massive neutrinos, 
was obtained in KATRIN experiment by measuring the spectrum
of electrons near the end point in $^3$H $\beta$-decay 
\cite{Aker:2019uuj,Aker:2021gma}:
$m_{1,2,3} < 0.8$ eV (90\% C.L.).

  The Cosmic Microwave Background (CMB)
data of the WMAP and PLANCK experiments, combined with
supernovae and other cosmological and astrophysical data 
can be used to obtain information in the form of an upper limit on the sum of
neutrino masses. Depending on the model complexity and the input data used 
one typically finds
\cite{LVinZyla:2020zbs} (see also \cite{Capozzi_2020}): 
$\sum_j m_j <  (0.11 - 0.54)$ eV (95\% CL). 

%
\subsection{The Seesaw Mechanism}
%

 The leptogenesis we are going to discuss in the present article is 
based on the type I seesaw mechanism of neutrino mass generation
\cite{Minkowski:1977sc,Yanagida:1979as,GellMann:1980vs,Glashow:1979nm,Mohapatra:1979ia}. This rather simple mechanism is realised, as is well known, 
by extending the Standard Model (SM) with $n\geq 2$ right-handed (RH) 
neutrinos $\nu_{lR}$ ( RH neutrino fields $\nu_{lR}(x)$),
that are singlets under $SU(2)_L\times U(1)_{Y_W}$,
possess a Majorana mass term and couple through 
a Yuakawa-type interaction to the SM lepton and Higgs doublets, 
$(\psi_{\alpha L}(x))^T = (\nu^T_{\alpha L}(x)~~\ell^T_{\alpha L}(x))$, 
with $\alpha=e,\,\mu,\,\tau$, and $(\Phi(x))^T = (\Phi^{(+)}(x)~\Phi^{(0)}(x))$.  
The minimal type I seesaw scheme in which leptogenesis can be realised
is with $n=2$ RH neutrinos. In this scenario the lightest neutrino --  
$\nu_1$ ($\nu_3$) for NO (IO) neutrino mass spectrum -- 
is massless at tree and one-loop level. 
We will consider leptogenesis with both 
$n=3$ and $n=2$ RH neutrinos.

 Without loss of generality, we work in the basis in which 
i) the Majorana mass matrix of RH neutrinos, $M$, is diagonal and positive, 
$M=\text{diag}(M_1,\,M_2,\,M_3)$ with $M_i > 0$,
and ii) the charged lepton Yukawa couplings are flavour diagonal. 
In the chosen  basis, the neutrino Yukawa and the RH neutrino Majorana 
mass terms are given by:
\be
\label{Ynu}
{\cal L}_{\rm Y,M}(x) =
-\,\left (Y_{\alpha i} \overline{\psi_{\alpha L}}(x)\,i\tau_2\,\Phi^*(x)\,N_{iR}(x)
+ \hbox{h.c.} \right )
-\,\frac{1}{2}\,M_{i}\,\overline{N_i}(x)\, N_i(x)\,,
\ee
%
where $Y_{\alpha i}$ is the matrix of the neutrino Yukawa coupling 
and $N_i (x) = N_{i R}(x) + N_{i L}^c(x) = C (\overline{N_{i}}(x))^T$, with 
$N_{i L}^c(x) \equiv C (\overline{N_{i R}}(x))^T$ and $C$ being 
the charge-conjugation matrix. 
 The fields $N_{1,2,3}(x)$ 
correspond to Majorana neutrinos $N_{1,2,3}$ with 
masses $M_{1,2,3} > 0$
which in high scale leptogenesis can have values 
$M_{1,2,3}\sim (10^6 - 10^{14})$ GeV,  
and so we will refer to $N_{1,2,3}$ further on as ``heavy Majorana neutrinos'' 
or just ``heavy neutrinos''. 
After the spontaneous breaking of the electroweak symmetry, the neutral 
component of the Higgs doublet acquires a non-vanishing vacuum 
expectation value (VEV) $v = 246$ GeV, generating a neutrino Dirac 
mass term which, together with
the $N_{1,2,3}$ mass term,  
can be cast in the form:
\be
\label{Lnum}
\mathcal{L}^m_\nu= -\,\frac{1}{2}\begin{pmatrix}
\overline{\nu_{\alpha L}}&\overline{N^c_{i L}}
\end{pmatrix}
\begin{pmatrix}\mathbb{O}_{\alpha\beta}&\frac{v}{\sqrt{2}}Y_{\alpha j}
\\\frac{v}{\sqrt{2}}(Y^T)_{i \beta}&M_i\delta_{ij}
\end{pmatrix}
\begin{pmatrix}
\nu^c_{\beta R}\\N_{j R}
\end{pmatrix}+h.c.,
\ee
%
where $\alpha,\,\beta=e,\,\mu,\,\tau$ ($i,\,j=1,\,2,\,3$) 
and $\nu^c_{\beta R}(x)\equiv C (\overline{\nu_{\beta L}}(x))^T$.
 This mass term can be diagonalised by means of the Takagi 
transformation, 
which, at leading order in the seesaw expansion 
parameter $|vY_{\alpha i}|/M_i \ll 1$, leads to the 
well known expression for the tree-level 
light neutrino mass matrix $m_\nu^\text{tree}$: 
\be
 \left(m_\nu^\text{tree}\right)_{\alpha \beta}
 \cong -\, \frac{v^2}{2}Y_{\alpha i}~M_i^{-1}\left(Y^{T}\right)_{i\beta}.
\label{seesawnuMajM}
\ee

In the version of the high scale leptogenesis 
with heavy Majorana neutrino masses which are not hierarchical 
and have relatively low values, $M_1 \sim 10^6$ GeV,
$M_2 \cong 3M_1$, $M_3\cong 3M_2$,   
the 1-loop radiative correction to the light neutrino mass 
matrix can be non-negligible \cite{Moffat:2018wke,Moffat:2018smo}
\footnote{The higher-order corrections to the 
light neutrino mass matrix were shown  
to be suppressed with respect to the tree-level and 
one-loop contributions 
\cite{Lopez-Pavon:2015cga,Moffat:2018wke}.
}. 
We will be interested in high scale leptogenesis 
with hierarchical masses of the heavy Majorana neutrinos, 
$M_1 \ll M_2 \ll M_3$, in which the CP violation is provided 
exclusively by the Dirac and/or Majorana CPV phases present 
in the PMNS matrix. In this case successful leptogenesis 
is possible for $M_1 \gtrsim 10^{10}$ GeV 
\cite{Pascoli:2006ci,Moffat:2018smo}.
Under these conditions the 1-loop contribution 
to the light neutrino mass matrix $m_\nu$, 
as our numerical study has shown, is sub-leading 
and amounts to  
$\sim$(10\% - 20\%) effects.
Nevertheless, we have included it 
in our calculations. 
This contribution  
is given by \cite{Pilaftsis_1992,Aristizabal_Sierra_2011,Lopez_Pavon_2013} 
(see also, e.g., \cite{Grimus_2002}):  
\be
(m_\nu^\text{1-loop})_{\alpha \beta} = 
Y_{\alpha i}~ \frac{M_i}{32\pi^2}\left(\frac{\log{\left(M_i^2/m_H^2\right)}}
{M_i^2/m_H^2 - 1}+3\frac{\log{\left(M_i^2/m_Z^2\right)}}
{M_i^2/m_Z^2 - 1}\right)\left(Y^T\right)_{i \beta},
\ee
%
where $m_H = 125$ GeV and $m_Z = 91.2$ GeV are the Higgs and Z boson 
masses, respectively.  The light neutrino mass matrix 
including the one-loop correction reads:
\be
\left(m_\nu\right)_{\alpha \beta} \equiv 
 \left(m_\nu^\text{tree} + m_\nu^\text{1-loop}\right)_{\alpha\beta}
= -\,\frac{v^2}{2}\,Y_{\alpha i}~f(M_i)\left(Y^T\right)_{i \beta}\,,
\label{eq:mnutot}
\ee
%
with 
\be
f(M_i) \equiv M_i^{-1} - \frac{M_i}{16\pi^2v^2}\left(\frac{\log{\left(M_i^2/m_H^2\right)}}{M_i^2/m_H^2 - 1}+3\frac{\log{\left(M_i^2/m_Z^2\right)}}{M_i^2/m_Z^2 - 1}\right)\,.
\label{eq:fMi}
\ee
%
 It can be diagonalised as:
\be 
\label{eq:mnuDiag}
\hat{m}_\nu = U^\dagger m_\nu U^*\,,
\ee
%
where $\hat{m}_\nu \equiv \text{diag}(m_1,\,m_2,\,m_3)$. 
 With $m_\nu$ given by Eq. \eqref{eq:mnutot}, 
the Casas-Ibarra parametrisation \cite{Casas:2001sr} 
of the neutrino Yukawa couplings takes the form 
\cite{Lopez-Pavon:2015cga}:
\be
\label{eq:CasasIbarra}
Y_{\alpha j} = 
\pm i \frac{\sqrt{2}}{v} U_{\alpha a}\sqrt{m_a}R_{ja}\sqrt{f^{-1}(M_j)}\,,
\ee
%
where $R$ is a complex orthogonal matrix. 
In the present study we adopt the following parametrisation of the 
$R$-matrix:
\be
R = 
\begin{pmatrix}
1&0&0\\
0&c_1&s_1\\
0&-s_1&c_1
\end{pmatrix}
\begin{pmatrix}
c_2&0&s_2\\
0&1&0\\
-s_2&0&c_2
\end{pmatrix}
\begin{pmatrix}
c_3&s_3&0\\
-s_3&c_3&0\\
0&0&1
\end{pmatrix},
\ee
%
where $c_j \equiv \cos(x_j + i y_j)$ and $s_j \equiv \sin(x_j + i y_j)$, 
$x_j$ and $y_j$ being free real parameters ($j = 1,\,2,\,3$).

%
\subsection{The Baryon Asymmetry of the Universe and Flavoured Leptogenesis}
%

The baryon asymmetry of the Universe (BAU) can be parametrised by the 
baryon-to-photon ratio
\be
\eta_B \equiv \frac{n_B-n_{\bar{B}}}{n_\gamma},
\ee
%
where $n_B$, $n_{\bar{B}}$ and $n_\gamma$ are the number densities of baryons, 
anti-baryons and photons, respectively. Alternatively, it can be expressed 
in terms of the baryonic density parameter
\be
\Omega_B h^2 = \eta_B \frac{m_p n_\gamma}{\rho_c h^{-2}}\cong \frac{\eta_B}{2.73\times 10^{-8}}\,,
\ee 
%
where $m_p$ is the proton mass, 
$\rho_c$ is the critical density of the Universe
and $h$ is the Hubble expansion rate of the Universe ($H$) per 
unit of 100 (km/s)/Mpc (the numerical values of the constants are 
taken from \cite{10.1093/ptep/ptaa104}).
The present BAU has been determined independently from the estimates 
of the Big Bang Nucleosynthesis (BBN) and, with higher precision, from 
the measurements of the Cosmic Microwave Background (CMB) anisotropies 
made by the Planck observatory. 
The following results have been reported (at $68\%$ C.L.) 
in \cite{Cooke_2018, Planck2018}
\footnote{More precisely, the BBN estimate is taken from Eq. (14) of 
\cite{Cooke_2018}, while that from CMB corresponds to the one reported 
in the last column of Table 2 in \cite{Planck2018}.
}:
\begin{eqnarray}
\Omega_Bh^2 = 0.02235 \pm 0.00049 &\text{(BBN),}&\\
\Omega_Bh^2 = 0.02242 \pm 0.00014 &\text{(CMB).}&
\end{eqnarray}
%
From both we obtain the best-fit value of
\be
\label{eq:etaBfit}
\eta_B \cong 6.1 \times 10^{-10},
\ee
%
that is going to be our reference value in the further analyses.

 The generation of a matter-antimatter asymmetry in the expanding Universe 
can naturally be accomplished within the type I seesaw framework through 
thermal leptogenesis. 
Provided the Yukawa couplings in Eq. \eqref{Ynu} are 
CP violating, the out-of-equilibrium decays of the heavy Majorana neutrinos 
$N_i$ to leptons and Higgs doublets in the early Universe generate 
CPV asymmetries in the individual lepton flavour charges $L_\alpha$, 
as well as in the total lepton charge $L$. The so generated lepton 
asymmetry is then translated into an asymmetry in the baryon charge 
$B$ by the SM $(B+L)$-violating, but $(B-L)$-conserving,
sphaleron processes, which are effective at temperatures 
$T\cong 132  - 10^{12}$ GeV.

 In this work we will concentrate on the case in which the heavy 
neutrinos $N_{1,2,3}$ have hierarchical masses, 
namely $M_1 \ll M_2 \ll M_3$. In this case generically only the CPV 
decays of $N_1$ contribute to the generation of the CPV lepton asymmetry.  
We shall report in this section the relevant equations 
for one decaying heavy neutrino that will be used in our analysis.

The charged lepton final states
in the decays of the heavy neutrino $N_i$, 
$N_i\to \Phi^+ \psi_i$ and $N_i\to \Phi^- \overline{\psi_i}$,
are a superposition of the charged lepton flavour states, namely, 
\begin{eqnarray}
\label{eq:psi}
|\psi_i\rangle &=& \sum_{\alpha=e,\,\mu,\,\tau} C_{i\alpha} |\psi_\alpha\rangle\,,\\
\label{eq:psibar}
|\overline{\psi_i}\rangle &=& \sum_{\alpha=e,\,\mu,\,\tau} \overline{C}_{i\alpha} |\overline{\psi_\alpha}\rangle\,,
\end{eqnarray}
%
with the coefficients  $C_{i\alpha}$ and  $\overline{C}_{i\alpha}$  given by
\be
\label{eq:Cia}
C_{i\alpha}=\overline{C}_{i\alpha}=\frac{Y_{\alpha i}}{\sqrt{(Y^\dagger Y)_{ii}}}\,.
\ee
%
We are interested  in the decays of $N_1$, 
$N_1\to \Phi^+ \psi_1$ and $N_1\to \Phi^- \overline{\psi_1}$,
so index $i$ should be replaced with 1 in 
Eqs. \eqref{eq:psi} - \eqref{eq:Cia}.

If it were not for the SM charged lepton Yukawa interactions, 
the quantum states $|\psi_1\rangle$ and $|\overline{\psi_1}\rangle$ 
would be coherent superpositions of the charged lepton 
flavour states. However, when 
these interactions are  in thermal equilibrium, i.e.,
their rates are larger than the expansion rate 
of the Universe,
given the difference between the charged lepton Yukawa couplings, 
$h_e$, $h_\mu$, $h_\tau$, the flavour states become distinguishable and 
each flavour state experiences a different time-evolution -- actually, 
it is enough for the SM $\tau$- and $\mu$-Yukawa interactions to be in 
 equilibrium for the three lepton flavours to be distinguishable.
 If the SM charged lepton Yukawa interactions are faster than the 
process of the heavy neutrino decay into (anti)leptons,
 then the coherence in $|\psi_1\rangle$ ($|\overline{\psi_{1}}\rangle$) 
is efficiently destroyed \cite{Blanchet_2007} 
(see, e.g., also \cite{Dev_2018}) -- in this sense these are 
\textit{decoherence effects}.  The relevant processes are the interchanges 
between the LH leptons with their respective RH components
and vice verse through scattering processes involving the Higgs doublet.
By means of the optical theorem, 
the rates of these processes involving the tauon and the muon, 
$\Gamma_{\tau}$ and $\Gamma_{\mu}$, are
given by the imaginary part of 
the $\tau$, $\mu$ thermal self-energy and read 
\cite{Blanchet_2013, Moffat:2018wke} (see also, e.g., \cite{Davidson_2008} 
and references therein): 
$\Gamma_{\tau,\,\mu} \cong 8 \times 10^{-3} h_{\tau,\,\mu}^2~T$.
The comparison of $\Gamma_\tau$ and $\Gamma_\mu$ with the 
Hubble expansion rate $H$ gives 
\footnote{ The $\tau$- and $\mu$-Yukawa couplings are given 
by $h_{\tau} = \sqrt{2} m_{\tau} / v \cong 1.02\times10^{-2}$ and 
$h_\mu = \sqrt{2} m_\mu/ v \cong 6.08\times 10^{-4}$,  
where $m_{\tau}$ and $m_\mu$ are the $\tau^\pm$  
and $\mu^\pm$ masses, respectively, and $v = 246$ GeV. 
Given the smallness of the $e$-Yukawa coupling 
$h_e = \sqrt{2}m_e/v \cong 2.94\times 10^{-6}$, 
$m_e$ being the $e^\mp$ mass,  
the $e$-Yukawa interactions come into thermal equilibrium 
only at $T \lesssim 10^5$ GeV, being therefore ineffective 
at the temperatures of interest in the present work.
}:
\begin{eqnarray}\label{eq:Gammatau}
    \frac{\Gamma_\tau}{H}&\cong&  \frac{M_P}{T} ~ 4.85\times 10^{-8}\, \cong \left(\frac{1 \text{ GeV}} {T}\right) ~ 5.92\times 10^{11} \,,\\
    \label{eq:Gammamu}
    \frac{\Gamma_\mu}{H} &\cong&  \frac{M_P}{T} ~ 1.72\times 10^{-10}\cong \left(\frac{1 \text{ GeV}} {T}\right) ~ 2.10\times 10^{9}\,,
\end{eqnarray}
%
where $M_P \cong 1.22\times 10^{19}$ GeV is the Planck mass.
At $T\gg 10^{12}$ GeV, the rates of the $\tau$- and $\mu$-Yukawa interactions are much 
smaller than the expansion rate
of the Universe as $\Gamma_{\tau,\,\mu} \ll 1$. 
As a consequence, the flavour states are indistinguishable and the 
(anti)leptons produced via the $N_1$'s decay are always found in 
the coherent superposition defined in Eq. \eqref{eq:psi} (\eqref{eq:psibar}).
 This is the \textit{unflavoured} 
or \textit{single-flavour} \textit{regime}. 
For $M\gg 10^{12}$ GeV, leptogenesis 
proceeds in the 
unflavoured regime for its entire duration and is usually 
studied within the \textit{single-flavour approximation}, under which the 
$\mu$- and $\tau$-decoherence effects are neglected. 
Correspondingly, this scenario is typically dubbed 
\textit{unflavoured or single-flavoured leptogenesis}.
In the single-flavour approximation, the time-evolution of the number 
densities of $N_1$ and $B-L$ charge can be described by the set 
of semi-classical \textit{single-flavoured Boltzmann equations} (1BE1F):
\begin{eqnarray}
    \label{1BE1F:N}
    \frac{d N_{N_1}}{dz} &=& - D_1\left(N_{N_1} - N_{N_1}^{eq}\right)\,,\\
    \label{1BE1F:BL}
    \frac{d N_{B-L}}{dz} &=& \epsilon^{(1)} D_1\left(N_{N_1} - N_{N_1}^{eq}\right) - W_1 N_{B-L}\,,
\end{eqnarray}
%
where $z\equiv M_1 / T$.  The quantities $N_{N_1}$ and $N_{B-L}$ 
are respectively the number of heavy neutrinos $N_1$ and 
$B-L$ asymmetry in a comoving volume.  In the present work 
the comoving volume is normalised as in 
\cite{Moffat:2018wke,Moffat:2018smo,Brivio:2019hrj} 
so that it contains one photon at z = 0, i.e., $N^{eq}_{N_1}(0) = 3/4$. 
This normalisation within the Boltzmann statistics is equivalent to using 
$N^{eq}_{N_1}(z) = \frac{3}{8} z^2 K_2(z)$, where $K_n(z)$, 
$n = 1\,,2\,,\,...$, is the modified $n^\text{th}$ Bessel function 
of the second kind.

 The decay parameter $D_1$ is given by:
\be
\label{eq:D1}
D_1(z) = \kappa_1 z \frac{K_1(z)}{K_2(z)},
\ee
%
where $\kappa_1$ is defined as the ratio between the total decay rate 
of $N_1$ at zero temperature, 
$\Gamma_{N_1}^{(0)} = (Y^\dagger Y)_{11} M_1/8\pi$,
and the Hubble expansion 
rate $H$ at $z = 1$. It proves convenient to write $\kappa_1$ 
in the following form:
\be
\label{eq:kappa1}
\kappa_1 = \frac{\tilde{m}_1}{m_*},
\ee
%
 where
\be
\label{eq:m1mstar}
\tilde{m}_1 \equiv (Y^\dagger Y)_{11} v^2 / 2M_1\,,~~ 
m_* \equiv (8\pi^2v^2/3M_P)\sqrt{g_*\pi/5} \approx 10^{-3}~{\rm eV}\,,
\ee
%
 $g_* = 106.75$ being the number of relativistic degrees of freedom at $z=1$. 
The wash-out parameter $W_1$ reads:
\be
\label{eq:W1}
W_1(z) = \frac{1}{2N_\ell^{eq}}\,D_1(z)\,N_{N_1}^{eq}(z)\,,
\ee
%
where $N_\ell^{eq}$ is the equilibrium number density of leptons 
at $z=0$, which, within the adopted normalisation, is given by $N_\ell^{eq} = N_{N_1}^{eq}(0) = 3/4$ 
\footnote{ A detailed derivation of Eqs. \eqref{eq:D1} - \eqref{eq:W1} 
is given, e.g., in \cite{BUCHMULLER2005305}.
}. 

Finally, the CPV-asymmetry parameter $\epsilon^{(1)}$ is given by 
\cite{COVI1996169,Covi:1996fm,Buchmuller:1997yu} 
\footnote{We work with the same sign convention used in \cite{Moffat:2018smo}, 
so the CP-asymmetry has an opposite sign with respect to that defined in 
\cite{COVI1996169}. We note a wrong sign typo in the last expression in 
Eq. (2.44) in \cite{Moffat:2018smo} -- this can be checked by summing 
Eq. (2.53) of the same article over the flavour indices.
}:
\be
\label{eq:eps}
\epsilon^{(1)} = \frac{3}{16\pi (Y^\dagger Y)_{11}}\sum_{j\neq 1} \Im\left[(Y^\dagger Y)^2_{1j}\right]\frac{\xi(x_j)}{\sqrt{x_j}}\,,
\ee
%
with $x_j \equiv M_j^2/M_1^2$ and
\be
\xi(x) \equiv \frac{2}{3}x\left[(1+x)\log\left(1+\frac{1}{x}\right)-\frac{2-x}{1-x}\right]\,.    
\ee
%
We note that for large $x$,
$\xi(x) = 1 ~ + ~\mathcal{O}(1/x)$,
so that, in the hierarchical limit $M_1\ll M_2 \ll M_3$, 
$\epsilon^{(1)}\propto M_1$ 
 since $1/\sqrt{x_j} = M_1/M_j$ and in this limit 
$(Y^\dagger Y)_{11}\propto M_1$ and 
$\Im\left[(Y^\dagger Y)^2_{1j}\right]\propto M_1M_j$
\footnote{To be more precise, in $\epsilon^{(1)}$ also factors of the 
form $f^{-1}(M_{2,3})/M_{2,3}$,  given in Eq. \eqref{eq:fMi}, appear 
inside the summation. 
However, the mass dependence of these factors is logarithmic and, 
in the mass range 
$10^9 \lesssim M/ \text{GeV} \lesssim 10^{14}$ of 
interest to us, $f^{-1}(M)/M$ 
changes only by a factor of $1.1$
taking values in the interval $1.1 - 1.3$.
}.

 Since the Yukawas enter in $\epsilon^{(1)}$ only through the 
product $Y^\dagger Y$, there is no dependence on the PMNS matrix. 
There is therefore no contribution to $\epsilon^{(1)}$ from the 
CPV Dirac and Majorana phases in the PMNS matrix.

As the mass scale of leptogenesis is lowered to $M_1 \sim 10^{12}$ GeV, 
the single-flavour approximation becomes inaccurate since the SM 
$\tau$-Yukawa interactions enter in equilibrium during the generation 
of the lepton asymmetry, i.e. $\Gamma_\tau / Hz \sim 1$. This is a transition 
regime, which we will refer to as \textit{1-to-2 flavour transition}, 
where the $\tau$-decoherence effects cannot be neglected.
Moreover, as was noticed in \cite{Moffat:2018smo}, 
when the requisite CP violation in leptogenesis is provided 
exclusively by the Dirac and/or Majorana CPV phases 
of the PMNS matrix, the 1-to-2 flavour transition 
proceeds with an unusual behaviour of the baryon asymmetry $\eta_B$, 
which extends into the region of the unflavoured regime 
at $M > 10^{12}$ GeV. This unusual behaviour will be investigated
in detail in our work. Here it suffices to mention that  
due to CP violating quantum decoherence 
effects caused by the SM $\tau$-Yukawa interactions,
in which CP violation is provided by the low-energy 
leptonic CPV phases, the generation of BAU in the 
single-flavour approximation as described by 
Eqs. \eqref{1BE1F:N} and \eqref{1BE1F:BL}
fails and that the observed BAU can still be generated 
at $M_1 > 10^{12}$ GeV even if $\epsilon^{(1)} = 0$ and 
the  $\tau$-Yukawa interactions are not in full thermal 
equilibrium.
 
 For $10^{9} \ll T/\text{GeV}\ll 10^{12}$, the $\tau$-Yukawa interactions 
are in thermal equilibrium while that of $\mu$ are not, 
namely $\Gamma_\tau / H \gg 1$ while $\Gamma_\mu/H \ll 1$ . Correspondingly, 
the $\tau$-(anti)lepton state becomes distinguishable from the other 
flavour states and the coherence in 
$|\psi_1\rangle$ ($|\overline{\psi_1}\rangle$) gets eventually destroyed. 
 As a consequence, the CPV asymmetry in  
$L_\tau$ evolves differently with respect 
to the asymmetry in the sum of $L_e$ and $L_\mu$ charges,
$L_{\tau^\perp}\equiv L_{e+\mu}\equiv L_e + L_\mu$.
This corresponds to the \textit{two-flavour regime of leptogenesis 
or two-flavoured leptogenesis}.

 For $10^{9} \ll M_1/\text{GeV}\ll 10^{12}$, 
the $\tau$-Yukawa ($\mu$-Yukawa) 
interactions enter in thermal equilibrium at $z\ll 1$ ($z\gg 1$) and 
leptogenesis can be studied within the \textit{two-flavour approximation} 
under which only the $\mu$-decoherence effects are neglected.
If in addition the $\tau$-Yukawa interactions are assumed to be 
infinitely (=``sufficiently'') 
fast during the whole period of leptogenesis, 
the \textit{two-flavoured Boltzmann equations} (1BE2F) 
can be used to describe the time-evolution of the CPV asymmetries in the 
 $L_\tau$ and $L_{\tau^\perp}$ charges and of BAU.
The set of 1BE2F equations in the two-flavour approximation reads:
\begin{eqnarray}
\label{1BE2F:N}
\frac{dN_{N_1}}{dz} &=& - D_1\left(N_{N_1}-N_{N_1}^{eq}\right)\,,\\
\label{1BE2F:tau}
 \frac{dN_{\tau\tau}}{dz} &=& \epsilon_{\tau\tau}^{(1)} D_1 \left( N_{N_1}-N_{N_1}^{eq}\right) - W_1 p_{1\tau} N_{\tau\tau}\,,\\
 \label{1BE2F:taup}
 \frac{dN_{\tau^\perp\tau^\perp}}{dz} &=& \epsilon_{\tau^\perp\tau^\perp}^{(1)} D_1 \left( N_{N_1}-N_{N_1}^{eq}\right) - W_1 p_{1\tau^\perp} N_{\tau^\perp\tau^\perp}\,,
\end{eqnarray}
%
where $p_{1\tau} = |C_{1\tau}|^2$ and 
$p_{1\tau^\perp} = |C_{1e}|^2 + |C_{1\mu}|^2 = 1-p_{1\tau}$, 
while $N_{\tau\tau}$ and $N_{\tau^\perp\tau^\perp}$ are respectively 
the values of the asymmetries 
 in the charges
$\frac{1}{3}B - L_\tau$ and $\frac{2}{3}B - L_{\tau^\perp}$ 
in a comoving volume, 
so that $N_{B-L} = N_{\tau\tau} + N_{\tau^\perp\tau^\perp}$.
The expressions for the relevant CPV lepton asymmetries 
$\epsilon_{\tau \tau}^{(1)}$ and 
$\epsilon_{\tau^\perp\tau^\perp}^{(1)} = \epsilon_{ee}^{(1)}+\epsilon_{\mu\mu}^{(1)}$ 
will be given below.

 As the mass scale is lowered to $M_1 \sim 10^{9}$ GeV, 
leptogenesis approaches 
the \textit{2-to-3 flavour transition}, 
where the $\mu$-decoherence effects cannot be neglected
since the $\mu$-Yukawa interactions 
enter in equilibrium, 
$\Gamma_\mu / Hz \sim 1$. 
Therefore, the two-flavour approximation ceases to be accurate.
Actually, as we are going to show in the present study, 
there are choices of the parameters for which
the 1BE2F equations
are never accurate and cannot be used for the description of leptogenesis 
in the whole range $10^{9} \lesssim M_1/\text{GeV} \lesssim 10^{12}$. 
In addition, in certain regions of the parameter space,
the scale below which the 1BE2F set of 
equations starts to be valid
can be significantly lower than $\sim 10^{12}$ GeV.
 
 At $T\ll 10^9$ GeV, also the $\mu$-Yukawa interactions are in 
thermal equilibrium, i.e. $\Gamma_\mu/H \gg 1$. 
This is the \textit{three-flavour regime}: all the flavours are 
distinguishable, the coherent superposition in $|\psi_1\rangle$ ($|\overline{\psi_1}\rangle$) is fully destroyed and 
the CPV lepton asymmetries in each of the charges 
$L_\alpha$ ($\alpha = e,\,\mu,\,\tau$) evolve separately. 
At $M_1 \ll 10^{9}$ GeV, both the $\mu$- and $\tau$-Yukawa interactions enter 
in equilibrium at $z\ll 1$ corresponding to the 
\textit{three-flavoured leptogenesis} scenario. 
If the $\mu$- and $\tau$-Yukawa interactions are assumed to be 
infinitely ($\equiv$``sufficiently'') fast, then leptogenesis 
can be described by the 
\textit{three-flavoured Boltzmann equations} (1BE3F), namely:
\begin{eqnarray}
\label{1BE3F:N}
\frac{dN_{N_1}}{dz} &=& - D_1\left(N_{N_1}-N_{N_1}^{eq}\right)\,,\\
 \label{1BE3F:e}
 \frac{dN_{ee}}{dz} &=& \epsilon_{ee}^{(1)} D_1 \left( N_{N_1}-N_{N_1}^{eq}\right) - W_1 p_{1e} N_{ee}\,,\\
 \label{1BE3F:mu}
 \frac{dN_{\mu\mu}}{dz} &=& \epsilon_{\mu\mu}^{(1)} D_1 \left( N_{N_1}-N_{N_1}^{eq}\right) - W_1 p_{1\mu} N_{\mu\mu}\,,\\
\label{1BE3F:tau}
 \frac{dN_{\tau\tau}}{dz} &=& \epsilon_{\tau\tau}^{(1)} D_1 \left( N_{N_1}-N_{N_1}^{eq}\right) - W_1 p_{1\tau} N_{\tau\tau}\,,
\end{eqnarray}
%
where $p_{1\alpha} = |C_{1\alpha}|^2$, while $N_{\alpha\alpha}$ 
 is the value of the asymmetry in the charge
$\frac{1}{3}B - L_\alpha$ in a comoving volume, so 
that $N_{B-L} = \sum_{\alpha} N_{\alpha\alpha}$, with $\alpha = e,\,\mu,\,\tau$. 

 The CPV lepton asymmetries $\epsilon^{(1)}_{\alpha\alpha}$
 in the both set of equations 1BE2F and 1BE3F are given by 
\cite{COVI1996169,Covi:1996fm,Buchmuller:1997yu}  
\footnote{The need for a double flavour index will later be clarified as in the quantum treatment also the off-diagonal terms are relevant.
}:
\begin{equation}
\label{eq:eps1aa}
\begin{aligned}
\epsilon^{(1)}_{\alpha\alpha}  = 
\frac{3}{16\pi\left(Y^\dagger Y\right)_{11}}
\sum_{j\neq 1} & \Bigg\{\Im\left[ Y_{\alpha 1}^* Y_{\alpha j}(Y^{\dagger}Y)_{1j} \right] 
f_1\left(x_{j}\right) +
\Im\left[  Y_{\alpha 1}^*Y_{\alpha j}(Y^{\dagger}Y)_{j1} \right]f_2\left(x_{j}\right)
 \Bigg\}\,,
  \end{aligned}
\end{equation}
%
where
\be
f_1\left(x\right)
\equiv \frac{\xi\left(x\right)}{\sqrt{x}}\,,~~~ 
 f_2\left(x\right)
\equiv \frac{2}{3\left(x-1\right)}\,.
\ee
%
 In Eq. \eqref{eq:eps1aa},
 $\alpha = \tau^{\perp},\tau$ and $\alpha = e,\mu,\tau$ 
for the 1BE2F and 1BE3F equations \eqref{1BE2F:tau} - \eqref{1BE2F:taup} and 
\eqref{1BE3F:e} - \eqref{1BE3F:tau},  
respectively,
and $\epsilon_{\tau^\perp\tau^\perp}^{(1)} = \epsilon_{ee}^{(1)}+\epsilon_{\mu\mu}^{(1)}$. 
We have: $\sum_{\alpha} \epsilon_{\alpha\alpha}^{(1)} = \epsilon^{(1)}$, 
with $\epsilon^{(1)}$ as given in Eq. \eqref{eq:eps}.

To obtain a better description of the 
 physics of leptogenesis, the decoherence effects should always 
be included in the calculations. As already shown in, e.g., 
\cite{Simone_2007,Blanchet_2007,Blanchet_2013}, 
the \textit{density matrix equations} (DMEs) provide an accurate 
tool to study thermal leptogenesis accounting for quantum decoherence 
processes, especially when these are neither infinitely fast nor 
their effects negligible. The DMEs describe the time-evolution of 
the entries of the density matrix, which, in the three-flavour basis, 
is given by
\be
N = \sum_{\alpha,\beta}N_{\alpha\beta}|\psi_\alpha\rangle \langle \psi_\beta |
\ee
%
with $\alpha,\,\beta = e,\,\mu,\,\tau$. The diagonal 
entries $N_{\alpha\alpha}$ are the already defined number densities 
for the $\frac{1}{3}B-L_\alpha$ asymmetry, so 
that $N_{B-L} = \text{Tr}(N) = \sum_{\alpha} N_{\alpha\alpha}$. 
The off-diagonal elements $N_{\alpha\beta}$ describe the degree 
of coherence between the flavour states.
The DMEs in the three-flavour basis explicitly read 
\cite{Simone_2007, Blanchet_2007, Blanchet_2013}:
\begin{eqnarray}
\label{DME:N}
\frac{dN_{N_{1}}}{dz}&=&-D_{1}(N_{N_{1}}-N^\text{eq}_{N_{1}})\\
\label{DME:full3}
 \frac{dN_{\alpha\beta}}{dz} &=&
 \begin{aligned}[t]
 &\epsilon^{(1)}_{\alpha\beta}D_{1}(N^{}_{N_{1}}-N^\text{eq}_{N_{1}}) - 
\frac{1}{2}W_{1}\left\{P^{0(1)},N\right\}_{\alpha\beta} \\
-&\frac{\Gamma_\tau}{Hz}
\left[I_\tau,\left[I_\tau,N\right]\right]_{\alpha\beta}
-\,\frac{\Gamma_\mu}{Hz}\left[I_\mu,\left[I_\mu,N\right]\right]_{\alpha\beta}\,,
\end{aligned}
\end{eqnarray}
%
 where 
$I_{\tau}$ and $I_\mu$ are $3\times3$ matrices such that 
$(I_\tau)_{\alpha\beta} = \delta_{\alpha\tau}\delta_{\beta\tau}$ and 
$(I_{\mu})_{\alpha\beta} = \delta_{\alpha \mu}\delta_{\beta \mu}$, 
and
\begin{equation}
P^{0(1)}_{\alpha \beta} \equiv C_{1 \alpha} C_{1 \beta}^*,
\end{equation}
%
 are projection matrices
which generalise the notion of the projection probability. 
They appear in the anti-commutator structure, which explicitly reads:
\be
\left\{P^{0(1)},N\right\}_{\alpha\beta} = 
\sum_{\gamma = e,\,\mu,\,\tau}\left( C_{1\alpha}C_{1\gamma}^*N_{\gamma\beta} + 
C_{1\gamma} C_{1\beta}^*N_{\alpha\gamma}\right)\,.
\ee
%
The double-commutator 
structures in Eq. \eqref{DME:full3} give rise to an exponentially 
damping term proportional to $\Gamma_{\tau,\,\mu}/Hz$ for the 
equations describing the off-diagonal elements of $N$. 
If these terms are infinitely large, i.e., $\Gamma_{\tau,\,\mu} \to + \infty$,  
the density matrix is driven towards a diagonal form and the DMEs 
reduce to the three-flavoured set of Boltzmann equations 1BE3F. 
The CPV-asymmetry parameters are 
\cite{COVI1996169,Covi:1996fm,Buchmuller:1997yu,Abada:2006fw,Simone_2007,Blanchet_2013,Biondini_2018}:
\begin{equation}
\label{eq:eps1ab}
\begin{aligned}
\epsilon^{(1)}_{\alpha\beta}&=\frac{3}{32\pi\left(Y^{\dagger} Y\right)_{11}}
\sum_{j\neq 1}\Bigg\{ i\left[Y_{\alpha 1}Y^{*}_{\beta j}(Y^{\dagger}Y)_{j1}
- Y^{*}_{\beta 1}Y_{\alpha j}(Y^{\dagger}Y)_{1j}\right] f_1\left(x_j\right) \\
&+i\left[Y_{\alpha 1}Y^{*}_{\beta j}(Y^{\dagger}Y)_{1j}-Y^{*}_{\beta 1}Y_{\alpha j}(Y^{\dagger}Y)_{j1}\right] f_2\left(x_j\right) \Bigg\}.
 \end{aligned}
\end{equation}
%
Setting $\alpha = \beta$ in the above expression reproduces 
the asymmetry $\epsilon^{(1)}_{\alpha\alpha}$
defined in Eq. \eqref{eq:eps1aa}, 
while the trace coincides with the expression for $\epsilon^{(1)}$ 
given in Eq. \eqref{eq:eps}.

  In the numerical analyses that follow, we will use the ULYSSES Python 
package \cite{GRANELLI2021107813} to solve the sets of equations 
that we have introduced in the present section.
 The code computes, in particular, 
$N_{B-L} = N_{ee} + N_{\mu\mu} +  N_{\tau\tau}$, which is then converted into the 
baryon asymmetry of the Universe $\eta_B$ expressed in terms of the 
baryon-to-photon ratio using the following relation:
\begin{equation}
    \eta_B = \frac{28}{79}\frac{1}{27}
N_{B-L}\,,
\label{eq:etaBl}
\end{equation}
%
where $28/79$ is the SM sphaleron conversion coefficient and 
the $1/27$ factor comes from the dilution of the baryon asymmetry 
 due to the  change of the photon density  
between leptogenesis and recombination 
\cite{BUCHMULLER2005305}.

%
\section{The Baryon Asymmetry Sign Change}
\label{sec:sign}
%
%
  We first consider two-flavoured leptogenesis in 
the case of three heavy Majorana neutrinos $N_{1,2,3}$ 
with hierarchical masses, $M_1 \ll M_2\ll M_3$, 
 $M_1 \gtrsim 10^{9}$ GeV.
In this case generically 
only the CPV decays of the lighter Majorana neutrino $N_1$ 
contribute to the generation of CPV lepton asymmetry which is converted 
into a baryon asymmetry by the sphaleron effects. 
The \textit{density matrix equations} (DMEs) describing the 
evolution of 
 the number of $N_1$ in a comoving volume, $N_{N_1}$, 
and of the CPV asymmetries in the lepton charges $L_\tau$ and 
$L_{\tau^\perp} = L_{e +\mu} = L_e + L_\mu$ in the two-flavoured 
leptogenesis have the following form in the case of interest:
\begin{eqnarray}
\label{DMEN}
    \frac{dN_{N_1}}{dz} &=& - D_1 (N_{N_1}-N_{N_1}^{eq})\,, \\
\label{DMEtau}
    \frac{dN_{\tau\tau}}{dz}   &=&
     \epsilon_{\tau \tau}^{(1)} D_1 (N_{N_1}-N_{N_1}^{eq}) - 
    W_1\left(p_{1\tau} N_{\tau\tau} +
    \Re\left[C_{1\tau^\perp}C_{1\tau}^* N_{\tau\tau^\perp}\right]\right)\,,\\
\label{DMEtaup}
    \frac{dN_{\tau^\perp \tau^\perp}}{dz}   &=&
     \epsilon_{\tau^\perp \tau^\perp}^{(1)} D_1 (N_{N_1}-N_{N_1}^{eq}) -
    W_1\left(p_{1\tau^\perp} N_{\tau^\perp\tau^\perp} +
    \Re\left[ C_{1\tau^\perp}C_{1\tau}^*N_{\tau\tau^\perp}\right]\right) \,,
    \\
\label{DMEtautaup}
    \frac{dN_{\tau\tau^\perp}}{dz}   &=&
    \epsilon_{\tau\tau^\perp}^{(1)} D_1 (N_{N_1}-N_{N_1}^{eq}) - 
    \frac{1}{2}W_1\left(N_{\tau\tau^\perp}+C_{1\tau}C_{1\tau^\perp}^*N_{B-L}\right)-\frac{\Gamma_\tau}{Hz}N_{\tau\tau^\perp}\,.
\end{eqnarray}
%
The $B-L$ asymmetry is given by 
$N_{B-L} = N_{\tau\tau}+N_{\tau^\perp\tau^\perp}$. We find that 
(see Appendix A):
\be
\label{B-L_DME}
\begin{split}
N_{B-L}(z_f) = N_{B-L}^\text{1BE1F} (z_f)+ N_{B-L}^\text{decoh}(z_f)
\end{split}
\ee
%
where
\be
\label{1BE1F}
 N_{B-L}^\text{1BE1F}(z) \equiv \int_{z_0}^{z} e^{-\int_{z'}^{z_f}W_1(z'')dz''}\epsilon^{(1)}D_1(z')(N_{N_1}(z')-N_{N_1}^{eq}(z'))\,dz'\,,
\ee
\be
\label{decoh}
N_{B-L}^\text{decoh} (z)\equiv 
\int_{z_0}^{z} e^{-\int_{z'}^{z_f}W_1(z'')dz''} W_1(z')\lambda(z')\,dz'\,,
\ee
%
$z_0$ corresponding to the beginning of leptogenesis, which we set to $z_0=10^{-3}$ in all our numerical calculations, and 
\be
\label{lambda}
\lambda(z) \equiv 2 \int_{z_0}^{z} \Re\left[C_{1\tau}^*C_{1\tau^\perp}
N_{\tau\tau^\perp}(z')\frac{\Gamma_\tau}{Hz'}\right] dz'\,.
\ee
%

The term $N_{B-L}^\text{1BE1F}$ is the solution to the 
\textit{single-flavoured Boltzmann equations} (1BE1F) and vanishes if 
the CP violation in leptogenesis
is due only to the physical Dirac and/or Majorana CPV phases 
in the PMNS matrix since in that case  \cite{Pascoli:2006ci}
$\epsilon^{(1)} = \epsilon^{(1)}_{\tau\tau} + \epsilon^{(1)}_{\tau^\perp\tau^\perp}  
= \epsilon^{(1)}_{\tau\tau} + \epsilon^{(1)}_{ee} + \epsilon^{(1)}_{\mu\mu} =0$.  
The term $N_{B-L}^\text{decoh}$ incorporates the decoherence effects and 
one can have $N_{B-L}^\text{decoh}>(\gg)~N_{B-L}^\text{1BE1F}$.
As was shown in \cite{Moffat:2018smo}, $N_{B-L}^\text{decoh}$ 
can be the only source of CPV lepton asymmetry 
if the CP violation in leptogenesis is provided exclusively by 
the physical CPV phases in the PMNS matrix.   
In the discussion that follows we focus on this case.

The factor $\Lambda_\tau \equiv \Gamma_\tau / Hz = \text{const.} / M_1$ and 
can be taken out of the integration in \eqref{lambda}. In the high scale 
regime ($M \gtrsim 10^{12}$ GeV) we can work in the limit of 
$\Lambda_\tau \to 0$ and neglect all the terms of order 
$\mathcal{O}(\Lambda_\tau^2)$ in the lepton asymmetry. 
Since in our case $N_{B-L}^\text{decoh}$ is the only source 
of lepton asymmetry, at $M \gtrsim 10^{12}$ GeV we have 
$N_{B-L} = N_{B-L}^\text{decoh} = \mathcal{O}(\Lambda_\tau)$. 
Solving Eq. \eqref{DMEtautaup} at zero order in $\Lambda_\tau$ 
\footnote{The term $\propto N_{B-L}$ in Eq. \eqref{DMEtautaup} 
can be neglected since for
$\epsilon^{(1)} = 0$ 
it leads to correction $\mathcal{O}(\Lambda_\tau^2)$ 
in the asymmetry.}
with the integrating factor method we find:
\be
N_{\tau\tau^\perp}(z)=\int_{z_0}^{z} e^{-\frac{1}{2}\int_{z'}^z W_1(z'')\,dz''}\epsilon_{\tau\tau^\perp}^{(1)}D_1(N_{N_1}-N_{N_1}^{eq})\,dz'+ \mathcal{O}(\Lambda_\tau)\,.
\ee
%
Inserting this result in  Eq. \eqref{lambda} we get:
\be
\lambda(z) = 
\Lambda_\tau \mathcal{I}_1(\kappa_1;z)(p_{1\tau^\perp} \epsilon_{\tau\tau}^{(1)} + 
p_{1\tau} \epsilon_{\tau^\perp\tau^\perp}^{(1)})  + \mathcal{O}(\Lambda_\tau^2),\,
\ee
%
where
\be
\mathcal{I}_1(\kappa_1;z) \equiv\int_{z_0}^z dz' \int_{z_0}^{z'} dz'' 
e^{-\frac{1}{2}\int_{z_0}^{z'}W_1(\tilde{z})\,d\tilde{z}}D_1(z'')(N_{N_1}(z'') 
- N_{N_1}^{eq}(z''))\,,
\ee
%
and we have used the relation 
$2\Re[C_{1\tau}^*C_{1\tau^\perp}\epsilon_{\tau\tau^\perp}] = 
p_{1\tau^\perp} \epsilon_{\tau\tau}^{(1)} + 
p_{1\tau} \epsilon_{\tau^\perp\tau^\perp}^{(1)}$ 
(see Appendix A for a derivation of this relation).
To leading order in $\Lambda_\tau$ the final asymmetry  reads:
\be
\label{DecohAsym}
N_{B-L}(z_f) = 
\Lambda_\tau \mathcal{I}_2(\kappa_1;z_f)(p_{1\tau^\perp} \epsilon_{\tau\tau}^{(1)} + 
p_{1\tau} \epsilon_{\tau^\perp\tau^\perp}^{(1)})  + \mathcal{O}(\Lambda_\tau^2),\,
\ee
%
where $p_{1\tau^\perp} \epsilon_{\tau\tau}^{(1)} + 
p_{1\tau} \epsilon_{\tau^\perp\tau^\perp}^{(1)} 
= (1 - 2p_{1\tau}) \epsilon_{\tau\tau}^{(1)}$ and 
\be
\label{eq:I2}
\mathcal{I}_2(\kappa_1;z)\equiv\int_{z_0}^z 
e^{-\int_{z_0}^{z} W_1(\tilde{z})\,d\tilde{z}} W_1(z') \mathcal{I}_1(\kappa_1;z')\,dz'\,.
\ee
\begin{figure}
 \centering
    \begin{subfigure}[t]{\textwidth}
 \centering
\includegraphics[width=12cm]{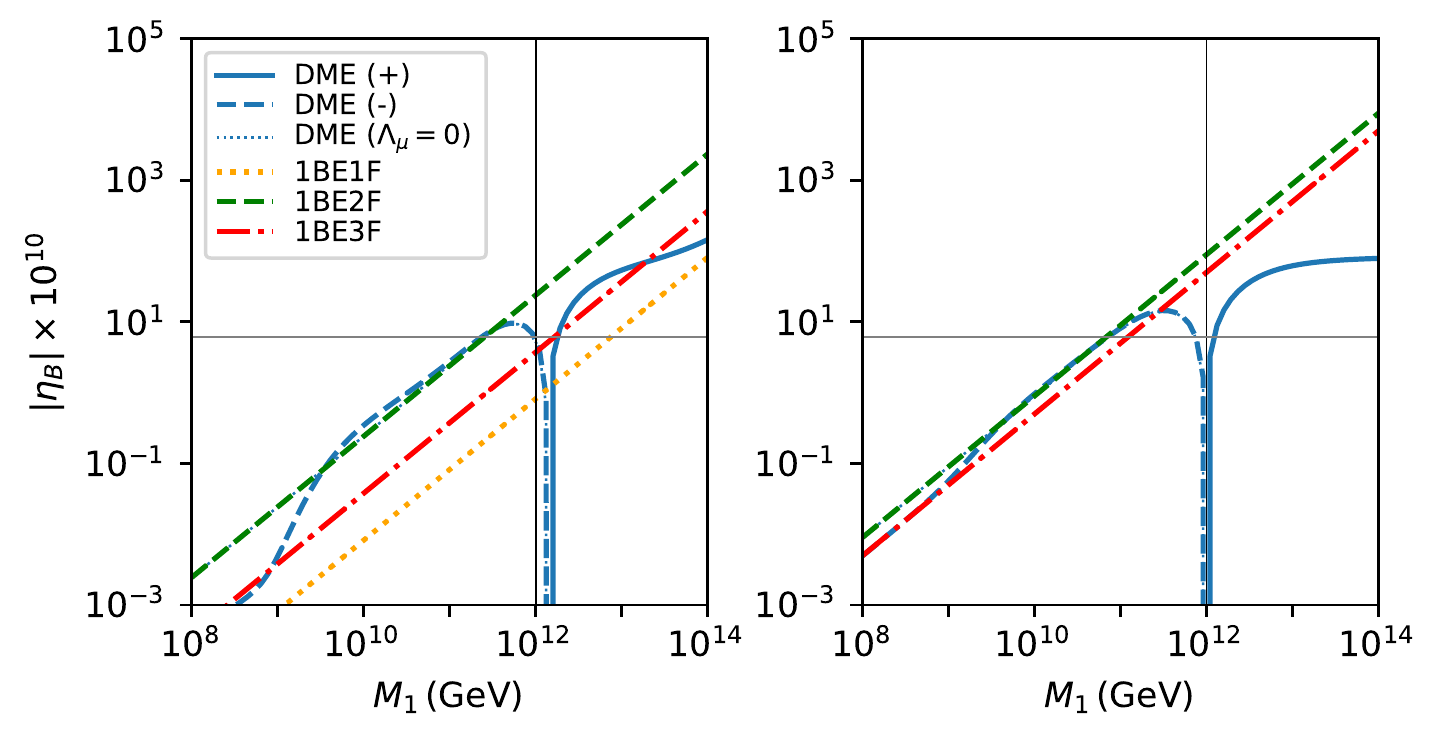}
    \end{subfigure}
    \begin{subfigure}[t]{\textwidth}
\centering
\includegraphics[width=12cm]{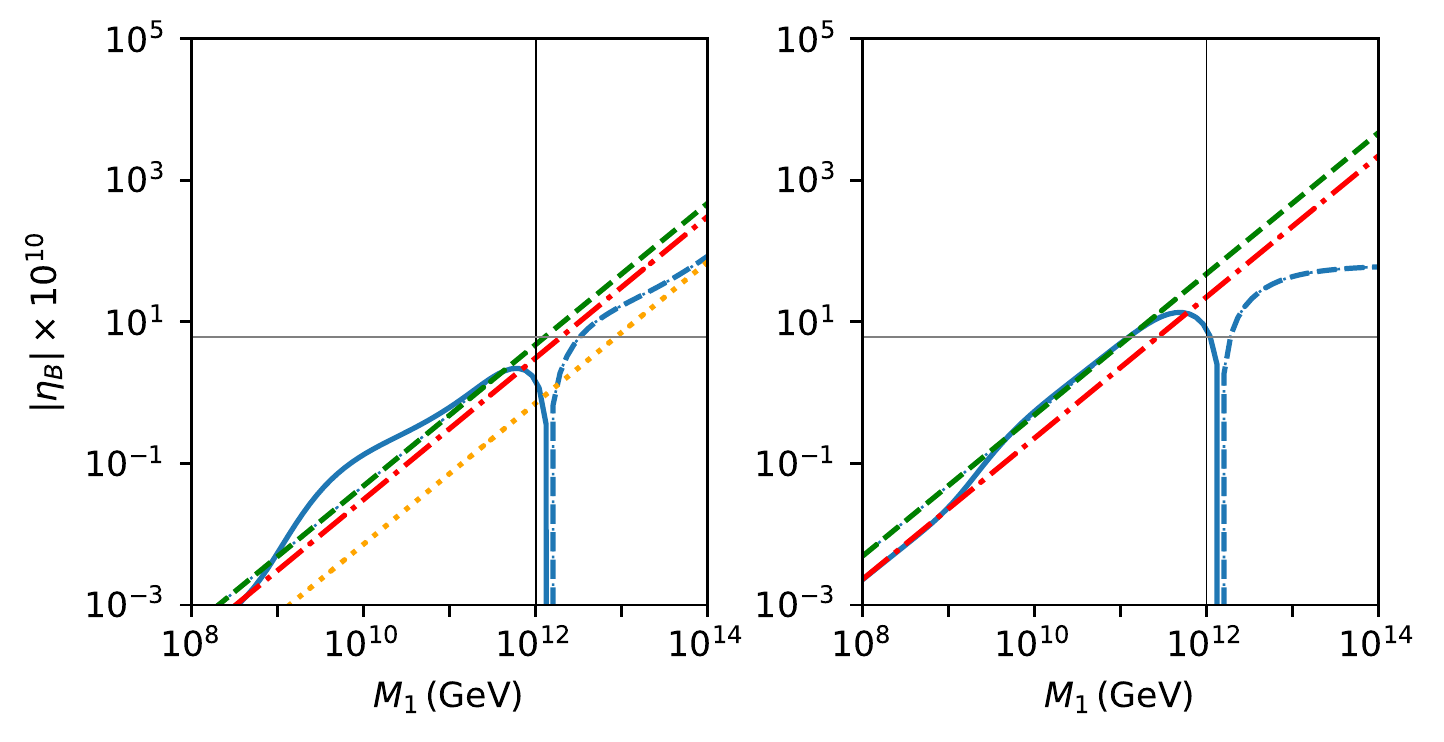}
    \end{subfigure}
    \caption{
The baryon asymmetry $|\eta_B|$ 
as a function of the mass scale $M_1$ 
 calculated with
DME (blue curve), and 1-, 2- and 3-flavoured
Boltzmann equations 1BE1F (orange dotted line), 1BE2F (green dashed line) 
and 1BE3F (red dash-dotted line). 
The solid (dashed) blue curve corresponds to 
$\eta_B > 0$ ($\eta_B < 0$),
 while the dotted blue curve is obtained for 
$\Lambda_\mu = 0$.
The lightest neutrino and the heavy Majorana neutrino masses, 
the Dirac and Majorana CPV phases and 
Casas-Ibarra parameters are set to: $m_1 = 0.0159$ eV, 
$M_3 = 5 M_2 = 50 M_1$, $\delta = 228^\circ$, 
$\alpha_{21} = 200^\circ$, $\alpha_{31} = 175^\circ$, 
$x_1 = -/+10^\circ$, $x_2=-/+ 20^\circ$, $x_3 = -/+ 10^\circ$ 
in the top/bottom panels and $y_1=y_2=0$ and $y_3= 30^\circ$/$0$ 
in the left/right panels, respectively. 
In the top-right and bottom-right panels the CP violation 
is due only to the CPV phases in the PMNS matrix ($y_{1,2,3} = 0$), 
i.e. $\epsilon^{(1)} = 0$, and 
the corresponding 1BE1F solution 
$\eta_B^{1BE1F}  = 0$.
The horizontal (vertical) grey (black) line corresponds to the observed value of 
$\eta_B$ (to $M_1 = 10^{12}$ GeV).
}
\label{fig:SignChange1}
\end{figure}
%
Since $\epsilon^{(1)}\propto M_1$ and $\Lambda_\tau\propto1/M_1$, 
the asymmetry given by Eq. \eqref{DecohAsym} is constant 
with the mass scale $M_1$. 
Thus, when the CP violation is provided by the 
CPV phases of the PMNS matrix, at $M_1 > 10^{12}$ GeV 
there should exist an interval of values of $M_1$ 
in which the baryon asymmetry $\eta_B$ is constant, 
i.e., does not change with $M_1$. 
Indeed, the numerical solutions of the DMEs show the existence of a 
plateau at values of $M_1 > 10^{12}$ GeV \cite{Moffat:2018smo},
as is illustrated in Fig. \ref{fig:SignChange1}, right panels. 

We note that if the CP violation in leptogenesis 
is due to the Casas-Ibarra matrix and thus
$\epsilon^{(1)} \neq 0$, 
$N_{B-L}^\text{1BE1F} \propto M_1$ eventually starts to dominate over 
$N_{B-L}^\text{decoh}$ as $M_1$ increases, 
recovering the single-flavour approximation 
as is clearly seen in Fig. \ref{fig:SignChange1}, left panels.

As $M_1$ decreases from $M_1 \sim 10^{12}$ GeV, 
$\Lambda_\tau$
increases and the solution of DMEs approaches 
the solution of the  
\textit{two-flavoured Boltzmann equations} (1BE2F), 
which we denote by  $N_{B-L}^\text{1BE2F}$.
In the mass range of $10^{9} \lesssim M_1/\text{GeV} \lesssim 10^{12}$, 
the asymmetry is approximately given 
by $N_{B-L}^\text{1BE2F}$.
However, the transition at $M \sim 10^{12}$ GeV may
take place with a sudden 
sign change of the baryon asymmetry,   
as  Fig. \ref{fig:SignChange1} shows. The sign change can happen if the 
solution given in Eq. \eqref{DecohAsym} has a 
different sign with respect to the solution $N_{B-L}^\text{1BE2F}$ 
\footnote{ A sign change of $\eta_B$ can occur 
also at $M \sim 10^{9}$ GeV, where the transition 
between the two-flavour and three-flavour regimes takes place.
If $\epsilon^{(1)} \neq 0$, another 
sign change can occur at a mass scale $M_1\gtrsim 10^{12}$ GeV
when $N_{B-L}^\text{1BE1F}$ starts dominating over $N_{B-L}^\text{decoh}$. 
However, investigating the conditions under which these 
sign changes of $\eta_B$ take place is beyond the scope of the 
present study.
}.
As Fig. 1 also indicates, we can have $\eta_B > 0$
either at  $M_1\gtrsim 10^{12}$ GeV or at $M_1\lesssim 10^{12}$ GeV. 
More generally, if we denote by $M_{10}$ the value of $M_1$ at which 
$\eta_B = 0$, we can have $\eta_B > 0$ and viable leptogenesis 
for certain values of $M_1$ lying either in the interval 
$M_1 > M_{10}$ or in the interval $M_1 < M_{10}$. 
Because of the change of the sign of $\eta_B$,
there is the possibility of finding the 
predicted $|\eta_B|$ equal to the observed value 
of the baryon asymmetry but $\eta_B$ having 
the wrong (negative) sign and so no successful 
leptogenesis (Fig.\ref{fig:SignChange1}, bottom-left panel).   
In view of this, it is of crucial importance  
to understand the conditions under which 
$\eta_B$  changes sign as well as what determines the value(s) 
of $M_1$ at which $\eta_B = 0$. 

We discuss in the next subsections the 
circumstances under which the 
sign change of $\eta_B$ can take place 
in the cases of  \textit{strong} and 
\textit{weak} wash-out regimes, for which $\kappa_1 \gg 1$ and 
$\kappa_1 \ll 1$, respectively. We concentrate on the physically 
interesting possibility of the requisite CP violation  
provided only by the Dirac and/or Majorana CPV phases 
of the PMNS matrix,
 which leads also to the existence of a ``plateau'' at $M_1 > 10^{12}$ GeV 
where to a good approximation $\eta_B$ does not depend on $M_1$.

%
\subsection{Strong Wash-Out Regime}
%

\begin{figure}[t!]
\centering
\includegraphics[width=12cm]{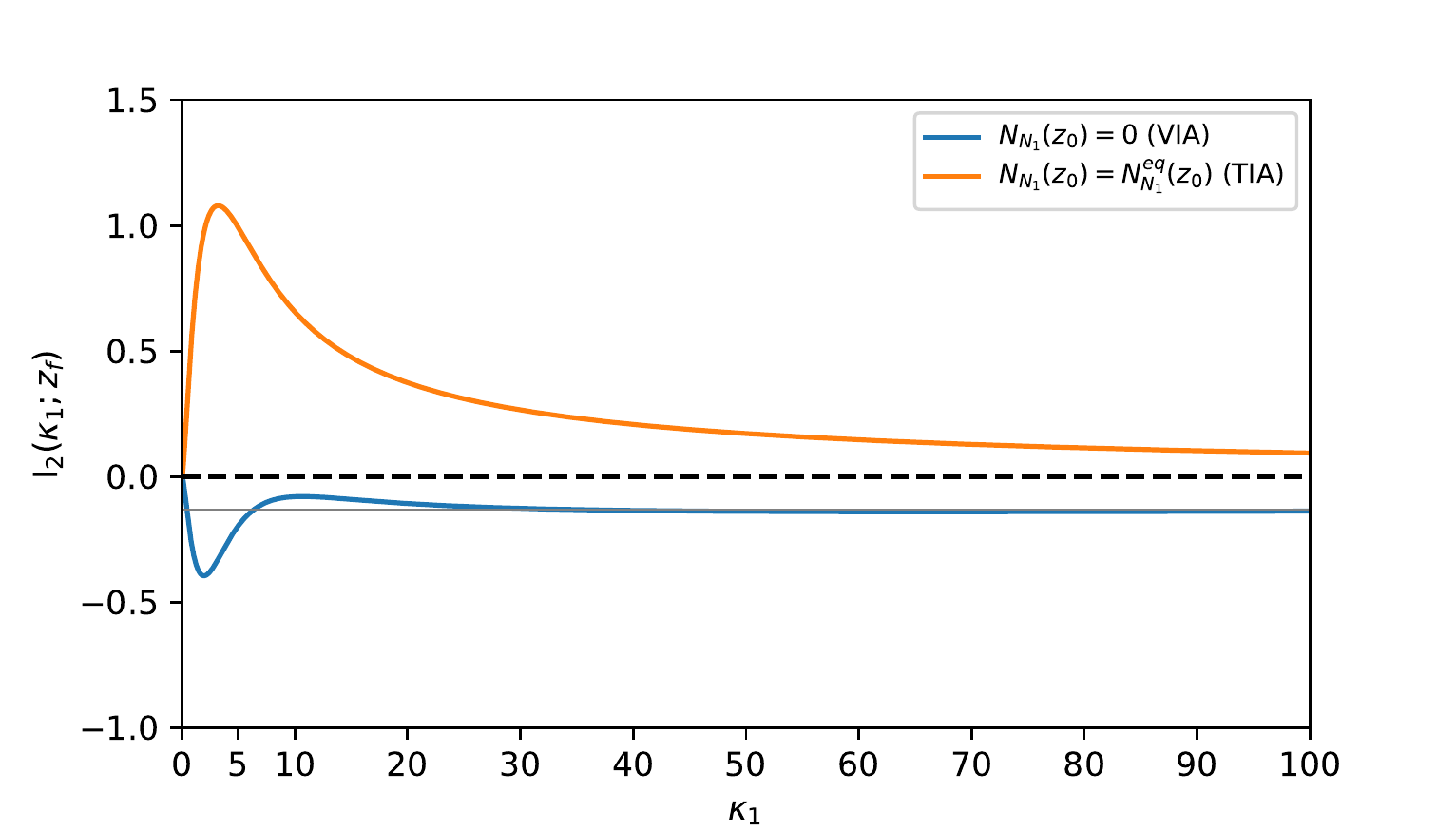}
\caption{
The dependence of the  function 
$\mathcal{I}_2(\kappa_1;z_f)$ (defined 
in Eq. \eqref{eq:I2}) 
on $\kappa_1$, computed numerically 
at $z_f = 1000$ in the cases of VIA (blue) and TIA (orange) initial 
abundances of $N_1$. Note that in the strong wash-out regime
in the VIA case for $k\gtrsim 10$ we have 
to a good approximation $\mathcal{I}_2\approx -0.13$ 
-- the value marked by the horizontal grey line. 
See text for further details.
}
\label{fig:I2}
\end{figure}

 In the strong wash-out regime the solution to the 1BE2F does not depend 
on the initial conditions since any initially generated asymmetry 
is erased by the strong wash-out processes. 
 A sufficiently accurate analytic expression of the solution 
to the 1BE2F, valid in the strong wash-out regime, is given by 
(see Appendix \ref{App:SW}):
\be
N_{B-L}^\text{1BE2F}(z_f) \simeq \frac{2 N_{N_1}^{eq}(0)}{\kappa_1 z_d(\kappa_1)}
\,\frac{p_{1\tau^\perp} \epsilon_{\tau\tau}^{(1)} 
+ p_{1\tau} \epsilon_{\tau^\perp\tau^\perp}^{(1)}}{p_{1\tau}p_{1\tau^\perp}}
=  \frac{2 N_{N_1}^{eq}(0)}{\kappa_1 z_d(\kappa_1)}
\,\frac{(1 - 2 p_{1\tau}) \epsilon_{\tau\tau}^{(1)}}{p_{1\tau}(1 - p_{1\tau})}\,.
\ee
%
 Since $p_{1\tau}(1 - p_{1\tau}) > 0$, 
a difference in sign between the solution $N_{B-L}^\text{1BE2F}(z_f)$
given above and the solution of 
Eq. \eqref{DecohAsym} occurs when $\mathcal{I}_2(\kappa_1;z_f)$ 
is negative. We show in Fig. \ref{fig:I2} the behaviour of 
$\mathcal{I}_2(\kappa_1;z_f)$ for $z_f = 1000$ computed numerically 
with the ULYSSES Python package \cite{GRANELLI2021107813} from 
Eq. \eqref{eq:I2} 
\footnote{We recall that $N_{B-L}(z)$, and therefore also 
$\mathcal{I}_2(\kappa_1;z)$, is frozen and kept constant after the wash-out 
processes become ineffective at $z_d(\kappa_1) \ll z_f$.
}
in the cases of vanishing initial abundance (VIA) of $N_1$, 
$N_{N_1}(z_0) = 0$, and thermal initial abundance (TIA) of $N_1$,
$N_{N_1}(z_0) = N_{N_1}^{eq}(z_0)$.
As follows from the behaviour of $\mathcal{I}_2(\kappa_1;z_f)$ shown in 
Fig. \ref{fig:I2}, a sign change of $\eta_B$ at $M_1 \sim 10^{12}$ GeV in 
the strong wash-out regime always happens for VIA, but never for TIA. 

%
\subsection{Weak Wash-Out Regime}
%

In the weak wash-out regime we need to 
consider separately the cases of 
the two different initial conditions -- 
VIA ($N_{N_1}(z_0) = 0$)
and TIA ($N_{N_1}(z_0) = N_{N_1}^{eq}(z_0)$).
%
\subsubsection{Vanishing Initial Abundance}
%
%
In the VIA case the asymmetry of interest 
in the two-flavoured
leptogenesis reads (see Appendix \ref{App:WWVIA}):
\be
\begin{split}
N^\text{1BE2F}_{B-L}(z_f) &
= \frac{81\pi^2}{1024N_\ell^{eq}}\kappa_1^2 (\epsilon_{\tau\tau}^{(1)}p_{1\tau} 
+ \epsilon_{\tau^\perp\tau^\perp}^{(1)} p_{1\tau^\perp})\\
&= -\, \frac{81\pi^2}{1024N_\ell^{eq}}\,
\kappa_1^2 \epsilon^{(1)}_{\tau\tau}(p_{1\tau^\perp} - p_{1\tau}).
\end{split}
\ee
%
where to get  the last 
equation we have used the fact 
that $\epsilon^{(1)} = 
\epsilon_{\tau\tau}^{(1)} +  \epsilon_{\tau^\perp\tau^\perp}^{(1)} = 0$. 
Using this condition also in Eq. \eqref{DecohAsym} we find:
\be
\label{B-L_DME2}
N_{B-L}(z_f) = \Lambda_\tau\mathcal{I}_2(\kappa_1;z_f) 
\epsilon_{\tau\tau}^{(1)}(p_{1\tau^\perp}-p_{1\tau}) + \mathcal{O}(\Lambda^2_\tau).
\ee
%
It is then clear from the comparison of the last two equations 
that $\mathcal{I}_2(\kappa_1;z_f)$ needs to be positive in order 
for a sign change of $\eta_B$ 
to occur at $M_1\sim 10^{12}$ GeV. 
Since, as shown in Fig. \ref{fig:I2}, 
$\mathcal{I}_2(\kappa_1;z_f)$ is always negative if $N_{N_1}(z_0) = 0$, 
the transition at $M\sim 10^{12}$ GeV  
in the weak wash-out regime in the VIA case always takes place 
without a sign change.

%
\subsubsection{Thermal Initial Abundance}
%
%
In the case of TIA, for which $N_{N_1}(z_0) = N_{N_1}^{eq}(z_0)$, and 
CP violation due only to the CPV phases in the PMNS matrix, 
the asymmetry of interest in the two-flavoured leptogenesis  
is described by the following analytic expression 
(see Appendix \ref{App:WWTIA}):
\be
\label{weakTIA}
N_{B-L}^\text{1BE2F}(z_f) = 
\epsilon_{\tau\tau}^{(1)}\,(p_{1\tau^\perp} - p_{1\tau})\mathcal{A}(\kappa_1;z_f),
\ee
%
where
\be 
\mathcal{A}(\kappa_1;z) \equiv 
\int_{z_D}^{z}\,dz'D_1(z')N_{N_1}(z')\int_{z'}^z \,dz'' W_1(z'')\,dz' > 0.
\ee
%
The comparison of \eqref{weakTIA} with Eq. \eqref{B-L_DME2} tells us 
that no sign change of $\eta_B$ should occur 
in the 1-to-2 flavour transition also in this case, given the fact that 
$\mathcal{I}_2(\kappa_1;z_f)$ is always positive for TIA.

%
\subsection{Transitions Between Different Flavour Regimes: Detailed Analysis}
%
%
The discussion and the results obtained 
in the preceding subsections
led to the important conclusion that 
we  should expect a sign change of the baryon asymmetry 
at the transition between the single- and two-flavoured 
leptogenesis in the case of VIA and strong wash-out regime 
of baryon asymmetry generation. 
However, certain important points 
could not be addressed within the approach
used in the discussion leading to this conclusion.
For example, the intermediate cases in which 
the asymmetries in different flavours 
$N_{\tau\tau}$ and $N_{\tau^\perp\tau^\perp}$
are generated in different wash-out regimes -- 
strong and weak --
could not and have not been considered.
The analysis performed by us also does not 
allow to determine the mass scale $M_{10}$
at which $\eta_B = 0$ and the transition between the 
two different flavour regimes considered takes place. 
Clearly, having  different wash-out 
regimes for the different flavour asymmetries 
and having a value of $M_{10}$ which differs 
significantly from $\sim 10^{12}$ GeV,  
might be possible, in principle, 
for choices of the parameters, 
namely the R-matrix angles $x_1 +i y_1,\,x_2 + iy_2,\,x_3 +i y_3$, the PMNS 
phases $\delta,\,\alpha_{21},\,\alpha_{31}$ 
and the mass of the lightest 
neutrino $m_1$, which differ from the choices 
considered by us. To address, in particular, 
the aforementioned points, we use an alternative approach 
to the problem of interest.

 We start from the following equation for $N_{B-L}$ in the case of 
$\epsilon^{(1)} = 0$  
(a detailed derivation of this equation 
is provided in Appendix \ref{FromDMEs}), 
which is valid as long as $M_1\gtrsim 10^{9}$ GeV 
where $\Lambda_\mu$ can be safely neglected:
\be
\label{dNB-Ldz}
\frac{dN_{B-L}}{dz} = - W_1(z)\left(N_{B-L}(z) - \lambda(z)\right).
\ee
%
The functions $W_1(z)$ and  $\lambda(z)$ are
given in Eqs. \eqref{eq:W1} and \eqref{lambda}, respectively.
Within this apparently simple equation for $N_{B-L}$, we have encoded all 
the decoherence effects on the system in the term $W_1\,\lambda$, which 
in particular contains both a source and a wash-out term, 
as will be clarified later on. The term $W_1 N_{B-L}$ is the usual wash-out term 
which tends to cancel any initially generated asymmetry.

Taking into account the expression for 
$N_{\tau\tau^\perp}$ given in 
Eq. \eqref{eq:Soltautaup} of Appendix \ref{FromDMEs},
the function $\lambda(z)$ 
can be cast in the form:
\be
\begin{split}
\lambda(z) =& \Lambda_\tau\,\epsilon_{\tau\tau}^{(1)}
(p_{1\tau^\perp} - p_{1\tau})
\int_{z_0}^z dz'\int_{z_0}^{z'}dz''\, D_1(z'') 
(N_{N_1}(z'') - N_{N_1}^{eq}(z'')) 
e^{-\Lambda_\tau(z'-z'')}e^{-\frac{1}{2}\int_{z''}^{z'} d\tilde{z}\,W_1(\tilde{z})} \,\\  
&-\,\Lambda_\tau p_{1\tau}p_{1\tau^\perp}\int_{z_0}^z dz'\int_{z_0}^{z'}dz''\, W_1(z'')N_{B-L}(z'')e^{-\Lambda_\tau(z'-z'')}e^{-\frac{1}{2}\int_{z''}^{z'}d\tilde{z}\, W_1(\tilde{z})}.
\end{split}
\ee
%
The first term in this expression for  $\lambda(z)$
is the only source of $B-L$ asymmetry, while the second is 
an \textit{integrated} wash-out term.
In the limit of $\Lambda_\tau \to 0$, i.e., 
for $M\gg 10^{12}$ GeV, the first term scales as 
$\Lambda_\tau$, while the second term scales as 
$\Lambda_\tau^2$ and can be neglected.
We note also that the integrated wash-out term 
can be suppressed by a small value of $p_{1\tau}p_{1\tau^\perp}$ as well.
Given that the source term is proportional to 
$\Lambda_\tau\epsilon_{\tau\tau}^{(1)}(p_{1\tau^\perp} - p_{1\tau})$, 
also the $B-L$ asymmetry will be proportional to it at any $z$:
\be
\label{eq:NNtilde}
N_{B-L}(z) \equiv 
\Lambda_\tau\epsilon_{\tau\tau}^{(1)}(p_{1\tau^\perp}-p_{1\tau}) \tilde{N}_{B-L}(z),
\ee
%
where $\tilde{N}_{B-L}$ can only depend, apart from $z$, 
on $\kappa_1$, $\Lambda_\tau$ and $p_{1\tau}$ 
(through the product $p_{1\tau}p_{1\tau^\perp}$).
This means that the function $\lambda(z)$ can be written as:
\be
\label{eq:lambdaS}
\lambda(z) = \Lambda_\tau\, \epsilon_{\tau\tau}^{(1)}(p_{1\tau^\perp}-p_{1\tau})  
\mathcal{S}(p_{1\tau},\kappa_1,\Lambda_\tau;z)\,,
\ee
%
with 
\be
\label{eq:SDEF}
\begin{split}
\mathcal{S}(p_{1\tau},\kappa_1,\Lambda_\tau;z) \equiv& 
\int_{z_0}^z dz'\int_{z_0}^{z'}dz''\, 
D_1(z'') (N_{N_1}(z'') - N_{N_1}^{eq}(z''))
e^{-\Lambda_\tau(z'-z'')}e^{-\frac{1}{2}\int_{z''}^{z'} d\tilde{z}\,W_1(\tilde{z})}\\ 
&-\, \Lambda_\tau p_{1\tau}(1 - p_{1\tau})
\int_{z_0}^z dz'\int_{z_0}^{z'}dz''\, W_1(z'')\tilde{N}_{B-L}(z'')
e^{-\Lambda_\tau(z'-z'')}e^{-\frac{1}{2}\int_{z''}^{z'}d\tilde{z}\, W_1(\tilde{z})}.
\end{split}
\ee
%
It follows from Eqs. \eqref{dNB-Ldz} -- \eqref{eq:SDEF}  that 
in the VIA case of interest we have: 
\be
\tilde{N}_{B-L}(z_f) = \int_{z_0}^{z_f}W_1(z)
\mathcal{S}(p_{1\tau},\kappa_1,\Lambda_\tau;z)\, e^{-\int_{z}^{z_f}W_1(z')\,dz'}\,dz.
\label{eq:tNBmL}
\ee
%
The last equation, combined with Eq. \eqref{eq:NNtilde}, 
cannot be used to compute the final asymmetry because 
inside $\mathcal{S}$ a dependence on $\tilde{N}_{B-L}$ is ``hidden''. 
 However, it is clear from the last expression 
and Eqs. \eqref{eq:NNtilde} and \eqref{eq:lambdaS} that,
given the signs of $\epsilon_{\tau\tau}^{(1)}$ and of 
$(p_{1\tau^\perp}-p_{1\tau}) = (1 - 2\,p_{1\tau})$,
the sign of  the asymmetry $N_{B-L}$
depends on the sign evolution of $\mathcal{S}$. 
We therefore analyze the behaviour of the function $\mathcal{S}$ to better 
understand the sign change at the 1-to-2 flavour transition. 
We construct the function $\mathcal{S}$ by first solving numerically the full 
set of DMEs with the ULYSSES Python package \cite{GRANELLI2021107813} and 
then we compute explicitly $\mathcal{S}$ using the definition of 
$\lambda(z)$ in Eq. \eqref{lambda} together with 
Eqs. \eqref{eq:lambdaS} and \eqref{eq:NtautaupDEF}
\footnote{The ULYSSES Python code allows to calculate 
numerically $N_{B-L}$, $N_{\tau\tau^\perp}$ and $\lambda(z)$, 
which then can be used to obtain $\mathcal{S}$.
}.
 
 We consider the case of heavy Majorana neutrinos 
having a vanishing initial abundance (VIA), i.e., $N_{N_1}(z_0) = 0$. 
At the beginning of leptogenesis at $z > z_0$, but $z$ relatively 
close to $z_0$, both the term 
involving $N_{N_1}(z'')$ and the integrated wash-out term in 
Eq. \eqref{eq:SDEF} are much smaller than the term  
involving $N_{N_1}^{eq}(z'')$
\footnote{The integrated wash-out term is negligible because 
both $W_1(z'')$ and $\tilde{N}_{B-L}(z'')$ are strongly suppressed.}, 
so that $\mathcal{S}$ starts its evolution with a negative sign. 
As $z$ increases,  $\mathcal{S}$ receives contributions from both 
terms in Eq.  \eqref{eq:SDEF}. At values of $z > z_{eq}$, where 
$z_{eq}$ corresponds to the time of evolution at which 
$N_{N_1}= N_{N_1}^{eq}$, we have, as our numerical analysis shows,  
$N_{N_1}(z) - N_{N_1}^{eq}(z) > 0$. As $z$ increases, the 
source term in Eq.  \eqref{eq:SDEF} goes through zero and becomes positive.
Let us call $\tilde{z}_\Lambda$ the value of $z$ at which $\mathcal{S} = 0$, 
so that at $z<\tilde{z}_\Lambda$ ($z>\tilde{z}_\Lambda$) 
we have $\mathcal{S}< 0$ ($\mathcal{S}> 0$).

It should be clear from Eq. \eqref{eq:tNBmL} that 
for $z< \tilde{z}_\Lambda$, $\tilde{N}_{B-L} <0$
and therefore also the second (integrated wash-out) term in 
Eq. \eqref{eq:SDEF} is positive. 
However, it is significantly smaller than the absolute value of  
the negative source term involving  $N_{N_1}^{eq}$.
At $z = \tilde{z}_\Lambda$, this negative term is compensated 
by the sum of the source term involving $N_{N_1}^{eq}(z'')$ 
and the integrated wash-out term.
At $z > \tilde{z}_\Lambda$, $\mathcal{S}$ is positive and 
remains so as $z$ increases.
 
    In the TIA case, as our numerical analysis shows,
$\tilde{z}_\Lambda$ does not exist since, in particular,
$N_{N_1}(z)>N_{N_1}^{eq}(z)$ for $z>z_0$ 
and correspondingly the function $\mathcal{S}$ 
has a positive sign for the entire period of leptogenesis.
This explains why no sign change can be present in the 
TIA case, as proven in the previous section. 
In what follows we focus our discussion on the VIA case only.

If in the VIA case the $B-L$ asymmetry is frozen at 
$z_f<\tilde{z}_\Lambda$, then, as we have discussed, 
 $\mathcal{S} < 0$ and therefore $\tilde{N}_{B-L}(z_f)<0$
(see Eq. \eqref{eq:tNBmL}).
Thus, we can have $\tilde{N}_{B-L}(z_f) > 0$ only 
if $z_f>\tilde{z}_\Lambda$.

To highlight this behaviour we focus for a moment on the strong wash-out 
regime. Suppose that $\kappa_1 \gg 1$, so that there exist two moments 
$z_{in}$ and $z_{out}$ for which $W_1(z_{in}<z<z_{out}) \gtrsim 1$. 
Then, for $z_{in} < z < z_{out}$ to a good approximation we have 
$dN_{B-L}/dz \cong 0$~
\footnote{We have checked numerically that 
$dN_{B-L}/dz \cong 0$ is indeed a sufficiently 
good approximation within the analysis performed by us.
} 
(see, e.g., \cite{Fong:2010up}),
and following the same steps as in Appendix \ref{Approximations}, 
from  Eq. \eqref{dNB-Ldz} we get:
\be
N_{B-L}(z) \simeq \lambda(z). 
\ee
%
After $z_{out}$ the asymmetry gets frozen so that:
\be
N_{B-L}(\infty) \simeq \lambda(z_{out}). 
\ee
%
Hence, as follows from Eq. \eqref{eq:lambdaS},
the sign of the final asymmetry reads:
\be
\label{eq:CasesStrong}
\text{sgn}(N_{B-L}(\infty))=
\begin{cases}
-\text{sgn}(\epsilon_{\tau\tau}^{(1)})\text{sgn}(p_{1\tau^\perp}-p_{1\tau})\,, & \text{if }z_{out}<\tilde{z}_\Lambda~~(\mathcal{S}<0)\,; \\
\text{sgn}(\epsilon_{\tau\tau}^{(1)})\text{sgn}(p_{1\tau^\perp}-p_{1\tau})\,, &\text{if }z_{out}>\tilde{z}_\Lambda~~(\mathcal{S}>0)\,. 
\end{cases}.
\ee
%
At $z_{out}=\tilde{z}_\Lambda$ we have $\mathcal{S}=0$ and therefore 
$N_{B-L}(\infty) = 0$.
Analytic expression for both $z_{in}$ and $z_{out}$ are 
given in \cite{BUCHMULLER2005305}:
\be
z_{in}(\kappa_1)\simeq \frac{2}{\sqrt{\kappa_1}}\,,\qquad 
z_{out} \simeq 1.25\log(25\kappa_1)\,.
\ee
%

In the weak wash-out regime the analysis is
more complicated as the asymmetry 
may freeze at $z_f \neq z_{out}$ and we do not have 
any analytic expression for this case. 
However, on the basis of the numerical analysis we did,
we expect leptogenesis to end at $z_f$ smaller 
than a few tens.

 We note that $\tilde{z}_\Lambda$ depends only on 
$p_{1\tau},\,\kappa_1$ and $\Lambda_\tau$, i.e.
$\tilde{z}_\Lambda = \tilde{z}_\Lambda\left(p_{1\tau},\kappa_1,\Lambda_\tau\right)$.
If we neglect the weak dependence on the mass scale $M_1$ of 
$\kappa_1$, which comes from the loop contribution to the light neutrino 
masses \cite{Moffat:2018wke}, the only dependence of  $\tilde{z}_\Lambda$ on 
$M_1$ is inside $\Lambda_\tau \propto 1/M_1$. 
Therefore we have 
\footnote{Approximate analytic expression for the value of 
$M_{10}$ at which $\tilde{z}_\Lambda = 0$ and the 1-to-2 
flavour transition takes place is derived in Appendix C.
} 
$\tilde{z}_\Lambda \simeq 
\tilde{z}_\Lambda\left(p_{1\tau},\kappa_1,M_1\right)$.
In addition, in the limit of 
$\Lambda_\tau\to 0$, i.e., at $M_1\gg 10^{12}$ GeV, 
the integrated wash-out term -- 
the second term in Eq. \eqref{eq:SDEF} -- 
can be neglected so that the dependence of  $\mathcal{S}$, 
and therefore of $\tilde{z}_\Lambda$,
on $p_{1\tau}$ drops off, i.e.,
$\tilde{z}_\Lambda \simeq \tilde{z}_\Lambda(\kappa_1,M_1)$.
In terms of the Casas-Ibarra parametrisation this means that 
$\tilde{z}_\Lambda$ does not depend on the PMNS phases. 
As the  mass scale $M_1$ decreases,
the integrated wash-out term becomes 
non-negligible
activating a dependence on the PMNS phases through the 
product $p_{1\tau}p_{1\tau^\perp}$.

In the general case of three heavy Majorana neutrinos having 
non-generate but also non-hierarchical masses
(e.g., $M_3 = 3M_2$, $M_2 = 3 M_1$),   
the discussion is rather complicated due to the large number 
of parameters present in the Casas-Ibarra parametrisation. 
 To make the discussion as transparent as possible we consider  
the case of decoupled $N_3$, in which the number of parameters 
is significantly smaller than in the general case.

%
\section{The Case of Decoupled \texorpdfstring{$N_3$}{N3}}
\label{sec:N3dec}
%
%
In the case of decoupled  heavy 
Majorana neutrino $N_3$ ($M_1 \ll M_2 \ll M_3$), 
the lightest neutrino, as is well known,
 is massless at tree and one loop level, 
i.e., $m_1 \cong 0$ ($m_3 \cong 0$), 
and the light neutrino mass spectrum is 
normal (inverted) hierarchical, denoted as NH (IH).
The set of parameters relevant for our discussion 
includes: the masses of the two heavy Majorana neutrinos $M_1$ and $M_2$;
the three CPV phases $\delta,\,\alpha_{21},\,\alpha_{31}$ 
of the PMNS matrix; the real and imaginary parts 
$x$ and $y$ of the complex angle of the  
Casas-Ibarra orthogonal $R$-matrix. 
The $R$-matrix for the NH and IH light neutrino mass spectra 
of interest has the form:
\begin{eqnarray}
\label{NHR}
R^\text{(NH)}&
=&\begin{pmatrix}
0& R_{12} & R_{13}\\
0& R_{22} & R_{23} \\
1&0&0
\end{pmatrix}
=\begin{pmatrix}
0&\cos\theta & \sin\theta\\
0& -\sin  \theta&\cos\theta\\
1&0&0
\end{pmatrix}\,,\\
\label{IHR}
R^\text{(IH)}&
=&\begin{pmatrix}
R_{11}& R_{12}&0\\
R_{21}& R_{22} &0\\
0&0&1
\end{pmatrix}
=\begin{pmatrix}
\cos  \theta&\sin  \theta&0\\
-\sin  \theta&\cos  \theta&0\\
0&0&1
\end{pmatrix}\,,
\end{eqnarray}
%
with $\theta = x + i y$. 
Both $R$-matrices in Eqs. \eqref{NHR} 
and \eqref{IHR} have  det$(R) = 1$. 
In the literature, the  factor 
$\varphi = \pm 1$ is sometimes included in the definition of $R$ to allow 
for the both cases det$(R) = \pm 1$. We choose instead to 
work with matrices in  Eqs. \eqref{NHR} and \eqref{IHR} but
extend the range of the Majorana phases $\alpha_{21(31)}$ 
from  $[0, 2\pi]$ to $[0, 4\pi]$, 
which effectively accounts for the case of  det$(R) = -\,1$ 
\cite{Molinaro:2008rg}, 
so that the same full set of $R$ and Yukawa matrices is considered.

In what follows we will analyse the case of 
hierarchical mass spectrum of the two heavy Majorana neutrinos, 
$M_1 \ll M_2$. 
In the subsequent numerical analysis we use $M_2 = 10M_1$.

%
\subsection{CP Violation from Low-Energy CPV Phases of the PMNS Matrix }
%
%
We are interested in the scenario of leptogenesis in which the CP violation 
is due exclusively to
the low-energy CPV phases present in the PMNS matrix. 
Correspondingly, we should avoid contributions to CP violation 
in leptogenesis associated with the $R$-matrix.
To satisfy this requirement we can \cite{Pascoli:2006ci}
either set i) $y=0$ and $x\neq 0$, 
which corresponds to a real $R$-matrix; 
or 2) $x = k\pi$, $k=0,1,2$, and $y\neq 0$, so that
in the case of NH (IH) spectrum 
the product $R_{12}R_{13}$ ($R_{11}R_{12}$) of 
the $R$-matrix elements 
(see Eqs. \eqref{NHR} and \eqref{IHR}),
which enters into the expression for the CPV 
asymmetry $\epsilon_{\tau\tau}^{(1)}$, is purely imaginary.  

 We first report the expressions for $\kappa_1$, $p_{1\tau}$ and 
$\epsilon_{\tau\tau}^{(1)}$ in the NH case for $y=0$ (real  $R_{12}R_{13}$), 
relevant for our further analysis:
\begin{eqnarray}
\label{eq:k1NH}
    k_1 &=& \frac{1}{m_*}\frac{f^{-1}(M_1)}{M_1} 
\left(m_2 \cos^2x + m_3 \sin^2x\right)\,,\\
\label{eq:ptNH}
    p_{1\tau}&=&\frac{m_2 |U_{\tau 2}|^2 \cos^2x+m_3  |U_{\tau 3}|^2 
\sin^2x+\sqrt{m_2 m_3} \Re\left(U_{\tau 2}^* U_{\tau 3}\right) \sin2x}
{m_2 \cos^2x+m_3 \sin^2x}\,,\\
\label{eq:eNH}
    \epsilon_{\tau\tau}^{(1)} &=& \frac{3 M_1}{16\pi v^2}\frac{f^{-1}(M_2)}{M_2}
\frac{\sqrt{m_2 m_3} (m_3-m_2) \sin2x}
{m_2 \cos^2x+m_3 \sin^2x}\Im\left(U_{\tau2}^*U_{\tau3}\right) 
+ \mathcal{O}\left(\frac{M_1}{M_2}\right)\,,
\end{eqnarray}
%
where $m_*$ and $f(M_{1,2})$ are defined in 
Eqs. \eqref{eq:m1mstar} and  \eqref{eq:fMi}.
The corresponding expressions for the IH spectrum can formally be obtained 
from those given above by changing 
$m_{2(3)}\rightarrow m_{1(2)}$ and $U_{\tau2(\tau 3)}\rightarrow U_{\tau1(\tau 2)}$.

For the $M_1$ and $M_2$ mass ranges we are going to consider, 
namely $M_1 = (10^9 - 10^{13})$ GeV, $M_2 = 10M_1$, 
the factors $f^{-1}(M_1)/M_1$ and $f^{-1}(M_2)/M_2$ 
in the expressions for $k_1$ and  $\epsilon_{\tau\tau}^{(1)}$
vary slowly in the intervals $1.16 - 1.27$ 
and $1.19 - 1.30$, respectively, increasing from the minimal values 
as $M_1$ and $M_2$ increase.
\begin{figure}[t]
    \centering
\includegraphics[width=12cm]{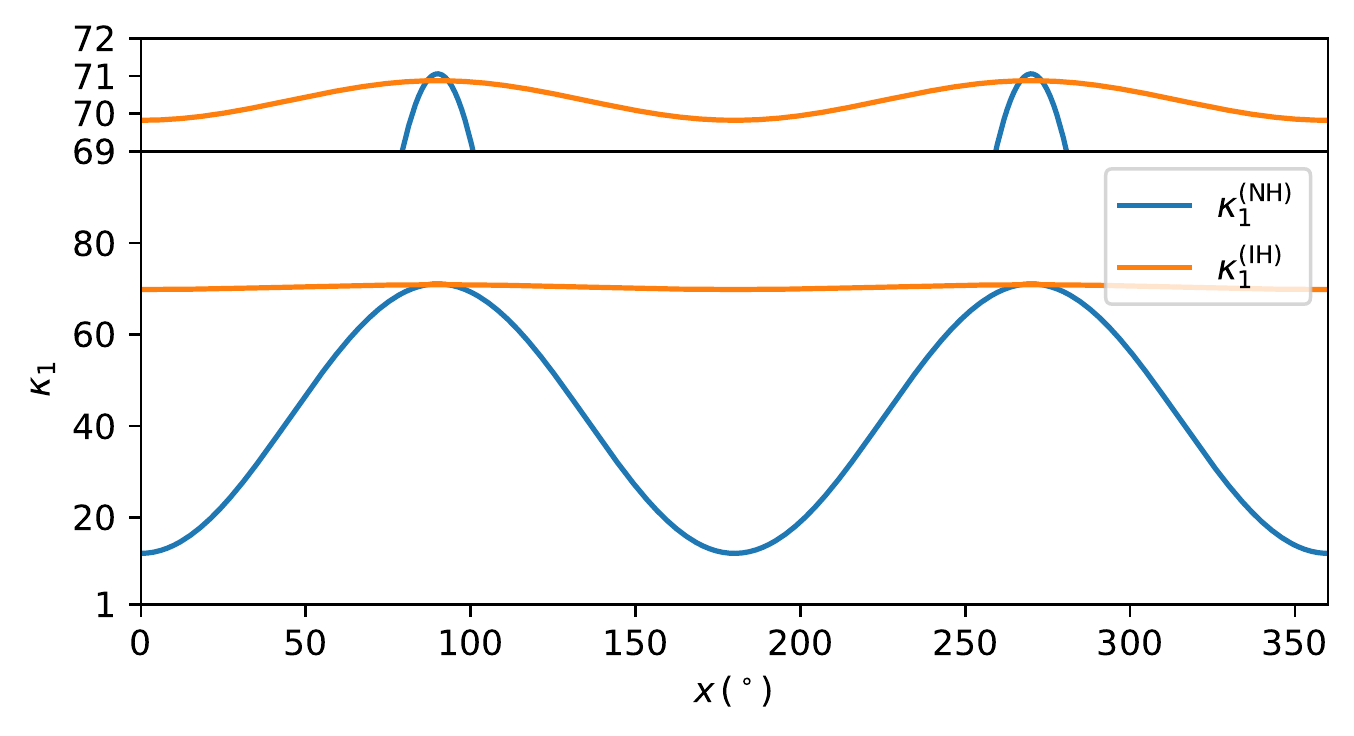}
    \caption{The decay parameter 
$\kappa_1$ versus $x$ for NH (blue) and IH (orange) light neutrino 
mass spectra with real $R$-matrix, i.e., $y = 0$. 
As the figure shows, $\kappa_1 \gg 1$, meaning that leptogenesis 
occurs always in the strong wash-out regime. The figure illustrates 
also  the periodicity of $\pi$ in the $\kappa_1$ dependence on $x$. 
The top panel illustrates the small oscillations of the IH curve.}
    \label{fig:k1}
\end{figure}
%

The combinations of the PMNS entries that appear in 
equations \eqref{eq:k1NH} - \eqref{eq:eNH} are given by:
\begin{eqnarray}
    |U_{\tau1}|^2                           &=& s_{12}^2 s_{23}^2 + c_{12}^2 c_{23}^2 s_{13}^2-2 s_{12}c_{12} s_{23}c_{23}  s_{13}  \cos\delta\,,\\
    |U_{\tau2}|^2                           &=& c_{12}^2 s_{23}^2+ s_{12}^2 c_{23}^2 s_{13}^2+2 s_{12}c_{12}s_{23} c_{23}  s_{13}  \cos\delta\,,\\
    |U_{\tau 3}|^2                          &=& c_{23}^2 c_{13}^2\,,\\
\label{eq:Imt2t3}    
\Im\left(U_{\tau 2}^* U_{\tau 3}\right) &=&c_{23} c_{13}  \left[c_{12} s_{23} \sin\left(\frac{\alpha_{21}-\alpha _{31}}{2}\right)+ s_{12}c_{23} s_{13} \sin \left(\frac{\alpha_{21}-\alpha_{31}}{2}+\delta \right)\right]\,,\\
\label{eq:Ret2t3}
    \Re\left(U_{\tau 2}^* U_{\tau 3}\right) &=&-c_{23} c_{13}  \left[c_{12} s_{23} \cos \left(\frac{\alpha_{21}-\alpha_{31}}{2}\right)+s_{12}c_{23}  s_{13} \cos \left(\frac{\alpha_{21}-\alpha_{31}}{2}+\delta \right)\right]\,,\\
 \label{eq:Imt1t2}       
 \Im\left(U_{\tau 1}^* U_{\tau 2}\right) &=&
    \begin{aligned}[t]
    &- s_{12}c_{12} \left(s_{23}^2-c_{23}^2 s_{13}^2\right)\sin \left(\frac{\alpha_{21}}{2}\right)+\\
    &- s_{23}c_{23} s_{13} \left[c_{12}^2 \sin \left(\delta -\frac{\alpha_{21}}{2}\right)+s_{12}^2 \sin \left(\frac{\alpha_{21}}{2}+\delta \right)\right]\,,
    \end{aligned}\\
   \label{eq:Ret1t2}
    \Re\left(U_{\tau 1}^* U_{\tau 2}\right) &=&
    \begin{aligned}[t]
    &-s_{12}c_{12} \left(s_{23}^2-c_{23}^2 s_{13}^2\right) \cos \left(\frac{\alpha_{21}}{2}\right) \\
    &+s_{23}c_{23} s_{13} \left[c_{12}^2 \cos \left(\delta -\frac{\alpha_{21}}{2}\right)-s_{12}^2 \cos \left(\frac{\alpha_{21}}{2}+\delta \right)\right]\,.
    \end{aligned}
\end{eqnarray}
%
In the NH (IH) case, only the Majorana CPV phase difference (phase) 
$\alpha_{21} - \alpha_{31}$ ($\alpha_{21}$) is physically relevant
\footnote{We will call the Majorana phase difference 
$\alpha_{23}$ simply ``Majorana phase'' 
and will denote it as $\alpha_{23}$ in what follows.
}. 
We note also that the dependence on $\delta$ is always 
suppressed by $\sin\theta_{13}$.   
Thus, for the NH (IH) neutrino mass spectrum 
$\tilde{z}_\Lambda$ is predominantly 
a function of the Majorana phase 
$\alpha_{23}$   ($\alpha_{21}$), 
of the real part of the $R$-matrix angle $x$ and 
of the mass scale $M_1$, exhibiting also subleading
dependence on $\delta$.

 For the CP conserving values of the Dirac and Majorana 
phases, $\delta = 0,\pi$, $\alpha_{21} = k_{21}\pi$ and 
$\alpha_{31}=k_{31}\pi$, $k_{21} = 0,1,2,...$, $k_{31} = 0,1,2,...$, 
with $\alpha_{23}\neq \pm 2n\pi$ ($\alpha_{21} \neq 2n\pi$), 
$n=0,\,1,\,2$, in the NH (IH) case, and real values of the 
elements of the $R$-matrix, $x \neq 0$, $y = 0$, the CP-symmetry is 
nevertheless violated in leptogenesis
due to the interplay of the CP conserving PMNS and real $R$ matrices
\cite{Pascoli:2006ci} and $\epsilon_{\tau\tau}^{(1)}\neq 0$, as also follows 
from Eqs. \eqref{eq:eNH}, \eqref{eq:Imt2t3} and \eqref{eq:Imt1t2}.

We present graphically in Fig. \ref{fig:k1} the dependence 
of the decay parameter $\kappa_1$ on $x$ for NH and IH  
neutrino mass spectra when $y = 0$.
As Fig. \ref{fig:k1} shows, in both the NH and IH cases leptogenesis 
occurs in the strong wash-out regime, i.e., $\kappa_1 \gg 1$ 
for any choice of $x$.
We can therefore rely on Eq. \eqref{eq:CasesStrong} to study the 
change of sign of $\eta_B$ 
in the 1-to-2 flavour transition 
 when $y=0$ and $x\neq 0$.
\begin{figure}[t]
    \centering
 \includegraphics[width=12cm]{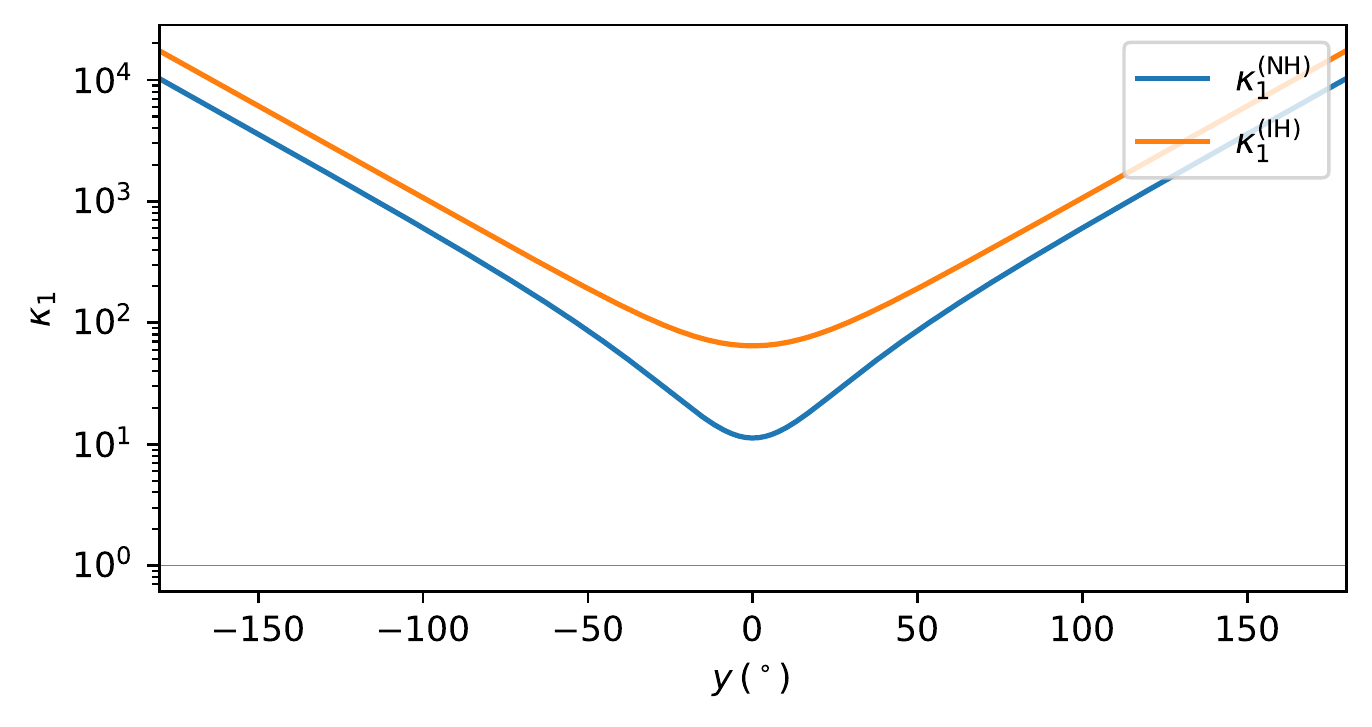}
    \caption{The decay 
parameter $\kappa_1$ as a function of $y$ for NH (blue) and
IH (orange) light neutrino mass spectra in the case of purely 
imaginary product $R_{12}R_{13}$ ($R_{11}R_{12}$), i.e., $x=0,\,\pi,\,...$ . 
As the figure shows, $\kappa_1 \gg 1$, meaning that leptogenesis 
always takes place in the strong wash-out regime.
}\label{fig:k1y}
\end{figure}

  For $y = 0$ and $x\neq 0$, i.e., for real elements of the $R$-matrix, 
the final baryon asymmetry 
$\eta_B$ is always suppressed in the IH case 
with respect to that in the NH case \cite{Pascoli:2006ci}. 
This is a consequence of the fact that 
the CPV asymmetry $\epsilon_{\tau\tau}^{(1)}$ 
in the NH and IH cases, $\epsilon^\text{(NH)}$ and 
$\epsilon^\text{(IH)}$, 
are proportional respectively to
 $m_3 - m_2$ and $m_2 - m_1$ (see Eq. \eqref{eq:eNH} and the 
subsequent discussion), 
$m_{3,2} \equiv m_{3,2}$(NH) and 
$m_{2,1} \equiv m_{2,1}$(IH) being the 
corresponding light neutrino masses of the two spectra 
(see section 2.1), so that
\be\label{eq:epsratio}
\frac{\left|\epsilon^\text{(IH)}\right|}
{\left|\epsilon^\text{(NH)}\right|}\propto
\frac{1}{2}\left(\frac{\Delta m^2_{21}}
{\Delta m^2_\text{atm}}\right)^{3/4}\approx \frac{1}{30}.
\ee
%
As a consequence, it is impossible to have a  
successful leptogenesis for IH neutrino mass spectrum
with CP violation provided 
only by the CPV phases in the PMNS matrix and real $R$-matrix 
for $M_1 \lesssim 10^{13}$ GeV.
The suppression can be avoided in the considered scenario 
if the product $R_{11}R_{12}$ of the $R$-matrix elements
is purely imaginary \cite{Pascoli:2006ci},
i.e., if $x = k\pi$, $k=0,1,2$, and $y\neq 0$.
Under the conditions $x = k\pi$ and $y\neq 0$,
the expressions for $\kappa_1$, $p_{1\tau}$ and $\epsilon_{\tau\tau}^{(1)}$ 
in the IH case take the form:
\begin{eqnarray}
\label{eq:k1IHy}
    k_1 &=& \frac{1}{m_*}\frac{f^{-1}(M_1)}{M_1} 
\left(m_1 \cosh^2y + m_2 \sinh^2y\right)\,,\\
\label{eq:ptIHy}
    p_{1\tau}&=&\frac{m_1 |U_{\tau 1}|^2 \cosh^2y+m_2  |U_{\tau 2}|^2 
\sinh^2y-\sqrt{m_1 m_2} \Im\left(U_{\tau 1}^* U_{\tau 2}\right) 
\sinh2y}{m_1 \cosh^2y+m_2 \sinh^2y}\,,\\
\label{eq:eIHy}
    \epsilon_{\tau\tau}^{(1)} &=& -\frac{3 M_1}{16\pi v^2}\frac{f^{-1}(M_2)}{M_2}
\frac{\sqrt{m_1 m_2} (m_2+m_1) \sinh2y}
{m_1 \cosh^2y+m_2 \sinh^2y}\Re\left(U_{\tau 1}^*U_{\tau 2}\right) 
+ \mathcal{O}\left(\frac{M_1}{M_2}\right)\,.
\end{eqnarray}
%
The corresponding expressions for the NH spectrum can
formally be obtained from those given above  by changing  
$m_{1(2)}\rightarrow m_{2(3)}$ and $U_{\tau1(\tau 2)}\rightarrow U_{\tau2(\tau 3)}$.

The suppression of $\eta_B$
in the IH case is now avoided due 
to the presence of the factor $(m_1 + m_2)$ in the CPV-asymmetry 
$\epsilon_{\tau\tau}^{(1)}$. 

 For  $x = k\pi$, $k=0,1,2$, and $y\neq 0$, the strong 
 wash-out condition $\kappa_1\gg 1$ is always satisfied, 
as is shown in  Fig. \ref{fig:k1y}.  
 Thus, also in this case we can rely
 on Eq. \eqref{eq:CasesStrong} to discuss the 
 the sign change of $\eta_B$   
at the 1-to-2 flavour transition.

For the $R$-matrix corresponding to the NH (IH) spectrum 
with elements $R_{12}$ and $R_{13}$ 
($R_{11}$ and $R_{12}$) 
whose product is purely imaginary, i.e.,  $x = k\pi$ and $y\neq 0$, 
the CP-symmetry can also be violated in leptogenesis 
due to the interplay of the $R$-matrix 
and the CP conserving PMNS matrix  
\cite{Pascoli:2006ci}.  
This possibility is realised  
for the CP conserving values of the Dirac and Majorana 
phases, $\delta = 0,\pi$, $\alpha_{21} = k_{21}\pi$ and 
$\alpha_{31}=k_{31}\pi$, $k_{21} = 0,1,2,...$, $k_{31} = 0,1,2,...$, 
with $\alpha_{23}\neq \pm (2n+1)\pi$ 
($\alpha_{21} \neq (2n+1)\pi$), $n=0,\,1$, in the 
NH (IH) case. Under these conditions we have 
$\epsilon_{\tau\tau}^{(1)}\neq 0$,
as also follows from Eqs. \eqref{eq:eIHy}, \eqref{eq:Ret2t3} 
and \eqref{eq:Ret1t2}.

 Given that the asymmetry $\eta_B$  is 
approximately constant in the plateau region at 
$M_1 \gtrsim 3\times 10^{12}$ GeV and 
decreases with the mass scale (see, e.g., Fig. \ref{fig:SignChange1} and 
the discussion in Sec. \ref{sec:sign}), the condition  
ensuring that $\eta_B$ is greater than or equal to the present 
BAU corresponds to the region of the parameter space for which we can 
have successful leptogenesis at $M_1 \gtrsim 3\times 10^{12}$ GeV.
Using Eqs. \eqref{eq:etaBfit}, \eqref{eq:etaBl} 
and \eqref{DecohAsym}, together with the fact that 
$\mathcal{I}_2(k_1;z_f)\approx -0.13$ in the strong wash-out regime in the 
VIA case (see Fig. \ref{fig:I2}), we get for the condition of interest:
\be
\label{eq:SuccessLG}
\eta_B = -\,0.13~\frac{28}{79}\frac{1}{27}\Lambda_{\tau}
\epsilon_{\tau\tau}^{(1)}(1-2p_{1\tau}) \gtrsim 6.1\times10^{-10}.
\ee
%
\begin{figure}[t]
    \centering
    \includegraphics[width=12cm]{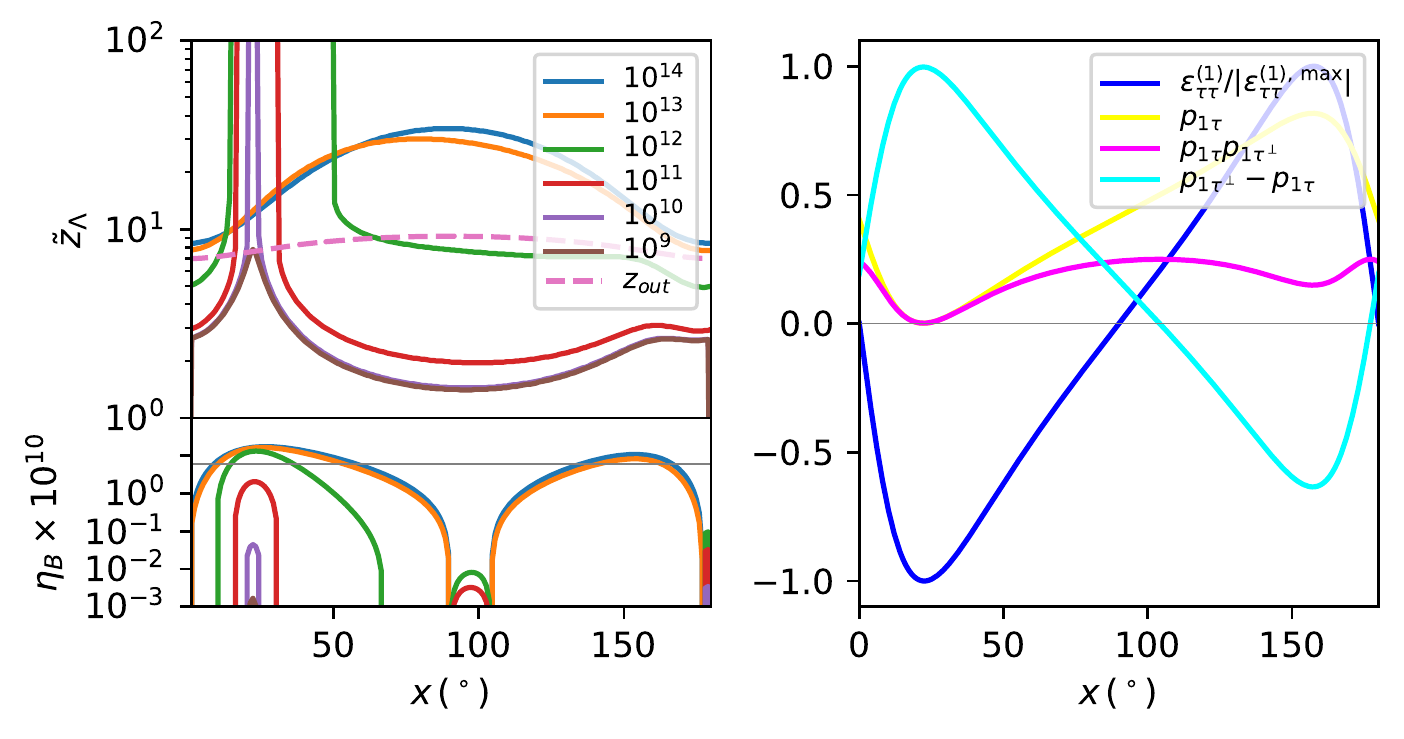}
    \caption{
    Top-left panel: $\tilde{z}_\Lambda$ versus $x$
for different mass scales $M_1 = 10^{14}$ (blue), $10^{13}$ (orange), 
$10^{12}$ (green), $10^{11}$ (red), $10^{10}$ (purple), $10^{9}$ (brown) GeV.
The dashed pink line corresponds to $z_{out}$, 
when the generated $\eta_B$ gets frozen. 
Bottom-left panel: the final baryon asymmetry 
(with the correct positive sign) $\eta_B$ 
versus $x$ for the set of values of $M_1$ 
used to obtain the top panel. 
The horizontal grey line marks the present BAU at $6.1\times 10^{-10}$. 
The right panel shows $\epsilon_{\tau\tau}^{(1)}$ normalised to 
its maximal value (dark blue), $p_{1\tau}$ (yellow), $p_{1\tau}p_{1\tau^\perp}$ 
(magenta) and $p_{1\tau^\perp}-p_{1\tau}$ (cyan). 
The sign of $\epsilon_{\tau\tau}^{(1)}(p_{1\tau^\perp}-p_{1\tau})$ 
is related to the sign of 
$\eta_B$ via Eq. \eqref{eq:CasesStrong}. 
The plots are obtained for $M_2 = 10M_1$, 
$\delta = 3\pi/2$, $\alpha_{21} = \alpha_{31} = 0$ and NH spectrum.
See text for further details.
}
\label{fig:zptNH_d_270_a21_0_a31_0}
\end{figure}
%
%
\subsection{CP Violation due to the Dirac Phase}
%

We consider in this subsection the scenario of leptogenesis with decoupled 
$N_3$ and CP violation due only to the Dirac phase $\delta$. To this end  
 the Majorana phase $\alpha_{23}$ ($\alpha_{21}$)
is set in the NH (IH) case to the following CP conserving values: 
i) $\pm 2n\pi $ ($ 2n\pi$), $n=0,\,1,\,2$, when $x\neq 0$ and $y=0$; 
ii) $\pm (2n+1)\pi $ ($(2n+1)\pi$), $n=0,\,1$, when $x=k\pi$, $k=0,1,2$, 
and $y \neq 0$.  With the choices of the values of  
$\alpha_{23}$ ($\alpha_{21}$) made we avoid the 
situation in which one of the sources of CP violation 
in leptogenesis is the interplay of CP conserving 
Majorana phases and $R$-matrix elements \cite{Pascoli:2006ci} 
that could be generated by 
the first term in the r.h.s. of Eq. \eqref{eq:Imt2t3} 
(Eq. \eqref{eq:Imt1t2}).

%
\subsubsection{The Case of Real \texorpdfstring{$R$}{R}-Matrix (\texorpdfstring{$x\neq 0$}{x} and \texorpdfstring{$y=0$}{y})}
%

 We show in the top-left panel of Fig. \ref{fig:zptNH_d_270_a21_0_a31_0} 
the curves of $\tilde{z}_\Lambda$ versus the angle $x$ 
for different mass scales $M_1$ assuming NH mass spectrum. 
The other parameters are set to:
$\delta =3\pi/2$, $\alpha_{23}=0$ and $y=0$.
The interpretation of the figure 
in this panel on the basis of Eq. \eqref{eq:CasesStrong}
should be the following. The values of $M_1$ and $x$ for 
which the $\tilde{z}_\Lambda$ curve lies 
above the $z_{out}$ curve correspond to $\text{sgn}(N_{B-L}(\infty)) 
= -\text{sgn}(\epsilon_{\tau\tau}^{(1)})\text{sgn}(p_{1\tau^\perp}-p_{1\tau})$. 
Alternatively, if the  $\tilde{z}_\Lambda$  curve lies below the $z_{out}$ one,  
then 
$\text{sgn}(N_{B-L}(\infty)) = 
\text{sgn}(\epsilon_{\tau\tau}^{(1)})\text{sgn}(p_{1\tau^\perp}-p_{1\tau})$. 
The intersection points correspond to a vanishing $\eta_B$  and 
mark the 1-to-2 flavour transition. 
In the bottom-left panel the generated baryon asymmetry 
$\eta_B$ (with the correct sign)
at different mass scales $M_1$ 
is also depicted. In the right panel we show the behaviour with 
$x$ of the relevant quantities: $\epsilon_{\tau\tau}^{(1)}$ 
(normalised to its absolute
maximal value), 
$p_{1\tau}$, $p_{1\tau}p_{1\tau^\perp}$ and $p_{1\tau^\perp}-p_{1\tau}$.
\begin{figure}
    \begin{subfigure}[t]{\textwidth}
    \centering
    \includegraphics[width=12cm]{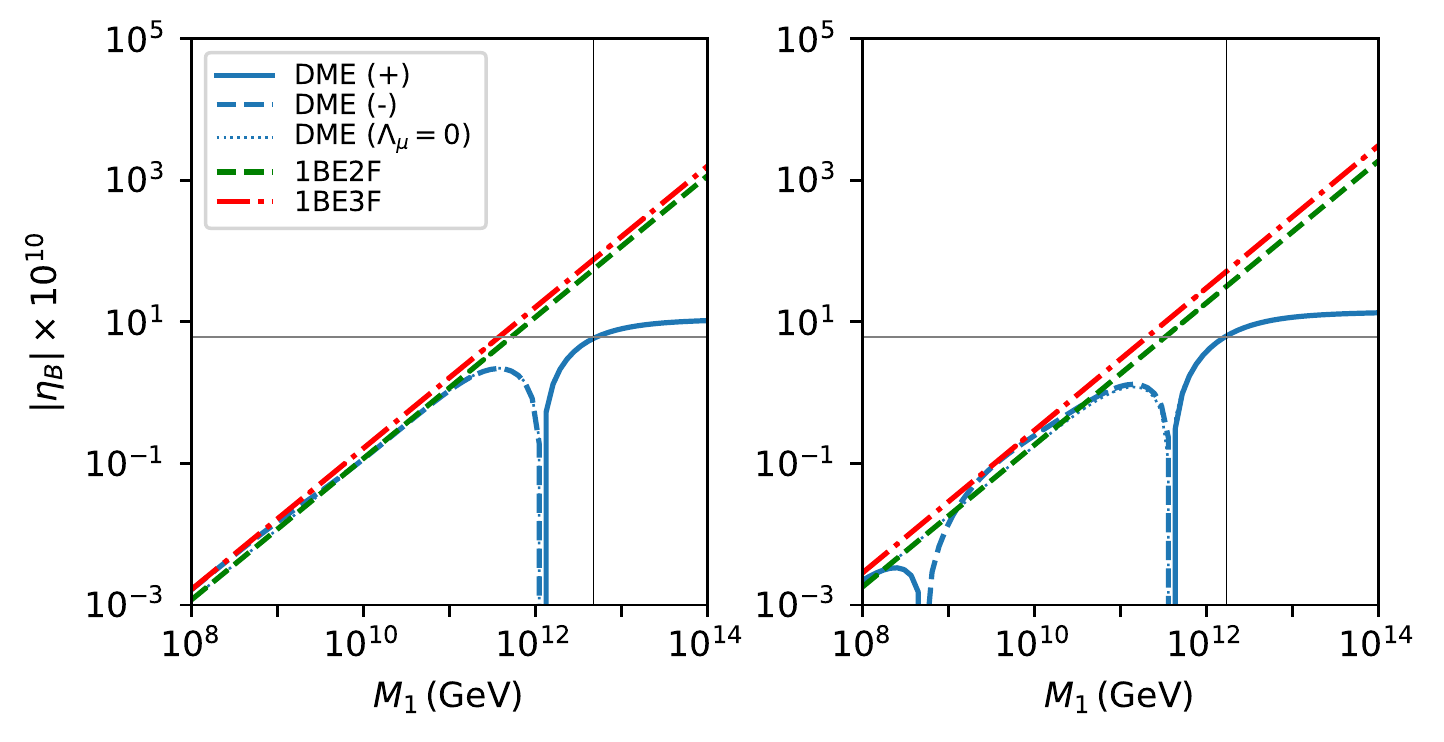}
    \end{subfigure}
    \begin{subfigure}[t]{\textwidth}
    \centering
    \includegraphics[width=12cm]{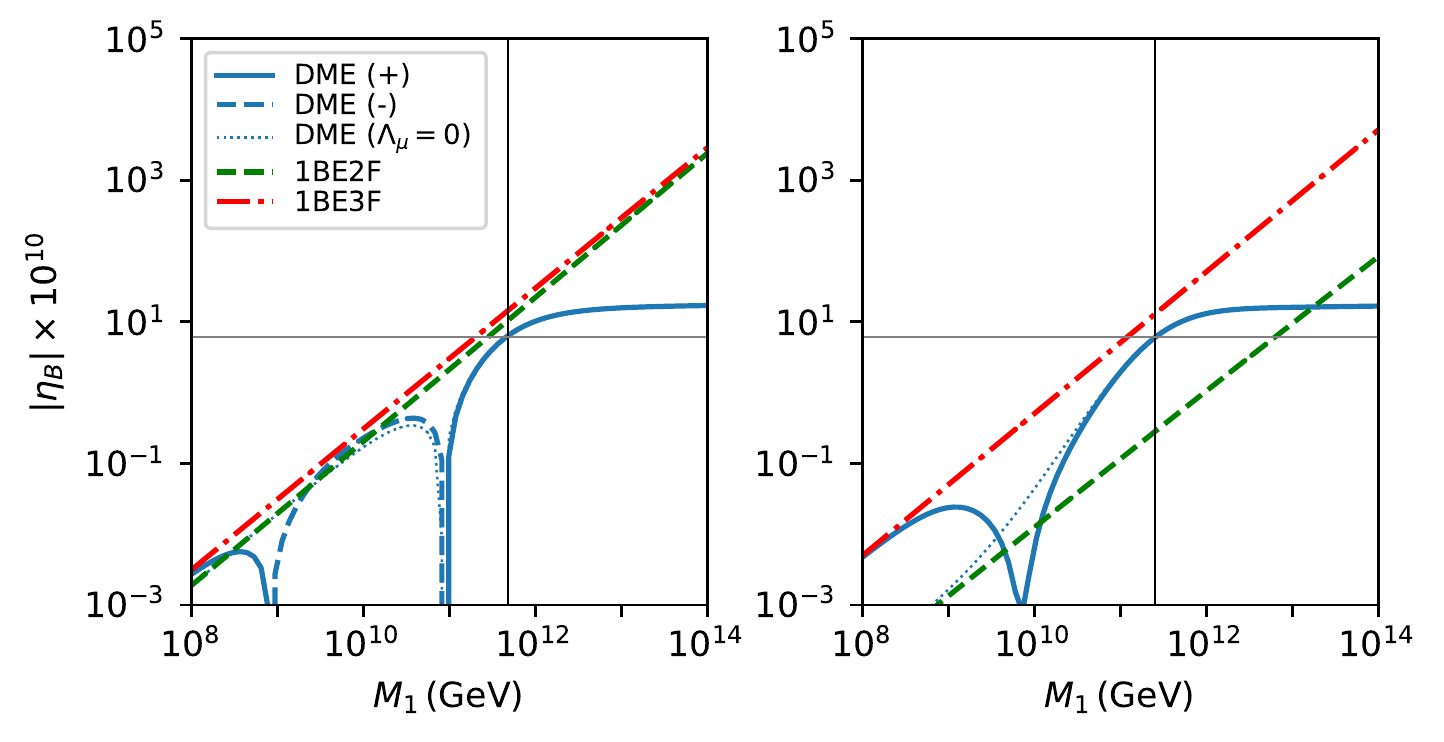}
    \end{subfigure}
    \caption{
The four plots show the absolute value of the 
baryon asymmetry $\eta_B$ versus $M_1$ for NH spectrum,
obtained with different sets of equations: 
DME (blue), 1BE2F (green), 1BE3F (red). 
The PMNS phases are chosen as in Fig. \ref{fig:zptNH_d_270_a21_0_a31_0}, 
$\delta = 3\pi/2$, $\alpha_{23} = 0$.
The top-left, top-right, bottom-left and bottom-right panels are 
obtained for $x=150^\circ$, $40^\circ$,
$30^\circ$ and $22.2^\circ$, respectively. 
The solid (dashed) blue line corresponds to 
$\eta_B > 0$ ($\eta_B < 0$); the dotted blue line 
is obtained for $\Lambda_\mu = 0$.
The horizontal grey line marks the present BAU at 
$6.1\times 10^{-10}$ and is reproduced with the
solution of DMEs at the minimal mass scales 
marked by the vertical black line, namely 
at $M_1\simeq 4.7\,(1.7) \times 10^{12}$ GeV top-left (top-right) panel 
and $4.8\,(2.5) \times 10^{11}$ GeV bottom-left (bottom-right) panel.
See text for further details.
}
    \label{fig:SCNH_d_270_a21_0_a31_0}
\end{figure}
%

For the considered choice of the parameters, depending on $x$ we can 
have different scenarios. At $x\lesssim 10^\circ$ and $x\gtrsim 60^\circ$, 
the $10^{13,\,14}$ GeV curves lie above the $z_{out}$ line while 
the  $10^{12,\,11\,,10\,,9}$ GeV curves lie
below (with the $10^{12}$ GeV curve lying near the $z_{out}$ line).
We can conclude that, for $x\lesssim 10^\circ$ and $x\gtrsim 60^\circ$, 
the 1-to-2 flavour transition occurs 
at values of $M_1$ slightly larger than 
$10^{12}$ GeV and with a sign change. 
 According to the bottom-left panel, 
in the indicated ranges of $x$
we can have successful leptogenesis 
for values of $x\approx 150^\circ$ 
at $M\gtrsim 10^{13}$ GeV.
At $x^\text{min} \simeq 22.2^\circ$ the magenta curve 
in the right panel of Fig. \ref{fig:zptNH_d_270_a21_0_a31_0}, 
corresponding to $p_{1\tau}p_{1\tau^\perp}$, reaches an absolute minimum 
for the chosen value of  $\alpha_{23} = 0$.
We then note that in the range $10^\circ\lesssim x \lesssim 60^\circ$, 
the transition occurs at $M\lesssim 10^{12}$ GeV, and the more the range 
of $x$ is squeezed around $x^\text{min}$, the lower is the mass scale of 
the transition. At $x^\text{min}$ no sign change occurs at the transition, 
given that all the $\tilde{z}_\Lambda$ curves 
obtained for $M_1 > 10^9$ GeV lie above the $z_{out}$ line.
As is shown in the bottom-left panel 
of Fig. \ref{fig:zptNH_d_270_a21_0_a31_0}, 
at $x^\text{min}$ the final baryon asymmetry 
$\eta_B$ is positive 
at any mass scale  and reaches the observed value 
at $10^{11}\lesssim M_1/$GeV $\lesssim10^{12}$. 

In Fig. \ref{fig:SCNH_d_270_a21_0_a31_0} we plot 
$\eta_B$ as a function of $M_1$, calculated using the density 
matrix equations (DMEs), two-flavoured (1BEF2) and three-flavoured (1BEF3)
Boltzmann equations, for different values of $x$, 
namely $x=150^\circ$ (top-left panel), $40^\circ$ (top-right panel),
$30^\circ$ (bottom-left panel) and $22.2^\circ$ (bottom-right panel). 
The PMNS phases and $M_2$ are the same as 
in Fig. \ref{fig:zptNH_d_270_a21_0_a31_0}.
The solid (dashed) blue curve corresponds to $\eta_B > 0$ ($\eta_B < 0$) 
and is obtained including the contributions from the $\mu$-Yukawa 
interactions, thus allowing to account for the 2-to-3 flavour 
transition when using the DMEs.
The sign change of $\eta_B$ at the 1-to-2 (2-to-3) 
flavour transition is present in both the top and the bottom-left 
(in the top-right and bottom-left) panels.
The mass scale $M_1$ of the 1-to-2 (2-to-3) flavour transition 
decreases with $x$ as it approaches the value $x^\text{min}=22.2^\circ$
(stays essentially constant at $M_1\simeq 10^{9}$ GeV). 
In the bottom-right panel the two transitions overlap 
so that the solution to the DMEs never approaches the 1BE2F solution.

For  $x = 150^\circ$ corresponding to the  
top-left panel in Fig. \ref{fig:SCNH_d_270_a21_0_a31_0}, 
successful leptogenesis takes place at 
$M_1\simeq 4.0 \times 10^{12}$ GeV and the 1-to-2 flavour 
transition happens around $M_1 \simeq 10^{12}$ GeV, as one would have 
expected from the considerations made after Eq. \eqref{eq:Gammatau}. This panel 
shows an example of what we will call 
``standard'' scenario of a 1-to-2 flavour transition 
at $\sim 10^{12}$ GeV under the assumption made about the source 
of CP violation as well as the strong wash-out regime of $\eta_B$ generation. 
We note also that the 2-to-3 flavour transition happens 
at $M_1 \sim 10^{9}$ GeV with no sign change, 
as the DME solution (solid blue line) interpolates smoothly between 
the 1BE2F (green) line and  the 1BE3F (red) one.

The remaining three panels of Fig. \ref{fig:SCNH_d_270_a21_0_a31_0} 
represent  examples of ``non-standard'' scenarios
of the transition of interest, as the 1-to-2 flavour transition 
happens at a mass scale which decreases from $10^{12}$ GeV as $x$ 
decreases approaching $x^\text{min}$, the value of $x$ 
at which the sign change 
does not occur and, as we will discuss, $\epsilon^{(1)}_{\tau\tau}$ 
has an absolute maximum
(bottom-right panel). 
The scenarios in these panels are 
``non-standard'' for the following reasons.
Firstly, the product 
$p_{1\tau}p_{1\tau^\perp}\simeq p_{1\tau} 
\simeq 2.5\times10^{-2},\, (1.5\times10^{-3})$
for $x=30^\circ,\,(22.2^\circ)$ is so small that the integrated 
wash-out term in $\lambda$ is 
additionally strongly suppressed, allowing the plateau due 
the $\tau\tau^\perp$-decoherence contribution to extend 
below $10^{12}$ GeV. Secondly, since for $x=30^\circ\,(22.2^\circ)$ we have 
\be
p_{1\tau}\kappa_1\simeq 0.63,\, (2.8\times10^{-2}),
\quad p_{1\tau^\perp}\kappa_1\simeq\kappa_1\simeq 24,\,(19),
\ee
%
the CPV asymmetry in the $\tau^\perp$-flavour 
is generated in the strong wash-out regime, 
while the CPV asymmetry in the $\tau$-flavour is produced in the weak  
wash-out regime. This scenario could not and was
not considered in Sec. \ref{sec:sign}
~\footnote{Moreover, since $p_{1\tau}\kappa_1\gtrsim 10^{-2}$, 
the analytic approximation 
used in Sec. \ref{sec:sign}
in the weak wash-out regime is not sufficiently 
accurate \cite{BUCHMULLER2005305}.
}.
Finally, 
Fig. \ref{fig:SCNH_d_270_a21_0_a31_0} 
shows that the two-flavour approximation in the range 
$10^{9}\lesssim M_1/$GeV $\lesssim 10^{12}$ 
based on 1BE2F is not always accurate.  
For the case  considered in the bottom-right panel of the figure, 
the DME solution for the asymmetry $\eta_B$ is enhanced by a factor of 
$\sim 10$ with respect to the asymmetry obtained by
solving the Boltzmann equations in the two-flavour approximation. 
This leads, in particular, to successful leptogenesis at 
$M_1\gtrsim 2.5\times 10^{11}$ GeV.
Most remarkably, in the case shown 
in the bottom-left panel,
the asymmetry $\eta_B$ predicted
by the DMEs (blue solid curve) at  
$M_1 \gtrsim 4.8 \times 10^{11}$ GeV 
has the correct sign allowing for successful leptogenesis, 
while the 1BE2F solution (green curve) gives $\eta_B < 0$
 in the indicated range of $M_1$ 
and thus non-viable leptogenesis. 

We note also that, as the top-right and bottom-left panels in 
Fig. \ref{fig:SCNH_d_270_a21_0_a31_0} show, 
at the 2-to-3 flavour transitions 
at $M_1\cong 10^9$ GeV, the baryon asymmetry $\eta_B$ 
changes sign going through zero, 
in contrast to the behaviour of $\eta_B$ shown 
in the top-left and bottom-right panels.
For the chosen values of the CPV phases of the PMNS matrix
the presence of this zero in $\eta_B$, as 
Fig. \ref{fig:SCNH_d_270_a21_0_a31_0} indicates,
depends on the value of $x$. 
For $x=22.2^\circ$ (bottom-right panel), 
the 2-to-3 flavour transition takes place with 
$\eta_B$ not going through zero but only through 
a relatively shallow minimum at $M_1\cong 10^{10}$ GeV.

\vspace{0.4cm}
%
\leftline{{\bf Ranges of $M_1$ and $\delta$ for Viable Leptogenesis}}
%

\vspace{0.4cm}
 The case illustrated in the bottom-right panel of 
Fig. \ref{fig:SCNH_d_270_a21_0_a31_0} is interesting for the 
following additional reasons. As our scan of the relevant parameter 
space shows, it is the case in which 
successful leptogenesis with two hierarchical in mass heavy Majorana 
neutrinos and CP violation provided by the Dirac phases $\delta$ 
takes place for the minimal for the considered scenario 
value of $M_{1\text{min}} \cong 2.5\times 10^{11}$ GeV. 
The value of $x = x^\text{min}=22.2^\circ$
maximises the CPV asymmetry $\epsilon_{\tau\tau}^{(1)}$. 
 Indeed, 
$\epsilon_{\tau\tau}^{(1)}$ depends on $x$ through the factor 
\be 
f_{1\epsilon}(x) = \dfrac{\sqrt{a}\,\sin2x}{a\,\cos^2x + \sin^2x}\,,~~~
a \equiv m_2/m_3\,, 
\label{eq:f1e}
\ee
%
which has an absolute maximum at 
\footnote{To be more precise, $f_{\epsilon}(x)$
has an absolute maximum at 
$x^\text{max} = 0.5\arccos((m_3 - m_2)/(m_3 + m_2)) = 22.5^\circ$, 
where we have made use of $m_3 = \sqrt{\Delta m^2_{31}}$, 
$m_2 = \sqrt{\Delta m^2_{21}}$ and the best-fit values of 
$\Delta m^2_{31}$ and $\Delta m^2_{21}$ given in Table 
\ref{Tab::BestFit}.
However, as can be easily checked, 
$f_{\epsilon}(x^\text{max}) - f_{\epsilon}(x^\text{min}) \cong 10^{-4}$.
}
$x^\text{min}=22.2^\circ$: $f_{\epsilon}(x^\text{min}) \cong 1.00$.
The chosen value of $\delta$ also maximises $|\epsilon_{\tau\tau}^{(1)}|$.
 As $M_1$ increases from the value $M_1 \cong 2.5\times 10^{11}$ GeV,
$\eta_B$ also increases from $\eta_B = 6.1\times 10^{-10}$
and, as  Fig. \ref{fig:SCNH_d_270_a21_0_a31_0} bottom-right panel shows, 
for $x =x^\text{min}$ and $\delta = 3\pi/2$ reaches a plateau at 
$M_1 = 2.7\times 10^{12}$ GeV, where $\eta_B = 1.60\times 10^{-9}$ and 
\footnote{As $M_1$ increases beyond  $2.7\times 10^{12}$ GeV, 
$\eta_B$ continues to grow very slowly due to the dependence 
of $\epsilon_{\tau\tau}^{(1)}$ on $f^{-1}(M_2)/M_2$, and at 
$M_1 = 10^{14}$ GeV ($M_2 = 10M_1$) 
we have $\eta_B \cong 1.67\times 10^{-9}$.
} 
is larger than the observed value of $\eta_B$ by the factor 
$C_{P1}\cong 2.62$. For the value of $M_1 \gtrsim 2.7\times 10^{12}$ GeV 
of the plateau, we have $\eta_B \propto (-\,\epsilon_{\tau\tau}^{(1)})$ 
(see Eq. \eqref{eq:SuccessLG}). Thus, 
$\eta_B$ will be compatible with the observed value of BAU 
for smaller value of  $(-\,\epsilon_{\tau\tau}^{(1)}) > 0$, 
i.e., for smaller $(-\,f_\epsilon(x)\sin\delta) > 0$. 
The plateau value of $\eta_B$ corresponds to 
$x =x^\text{min}$ and $\delta = 3\pi/2$
for which  $(-\,f_\epsilon(x^\text{min}\sin(3\pi/2)) = 1$. 
Thus, fixing $x = x^\text{min}$ we get the minimal value of 
$(-\,\sin\delta) > 0$ for which we can have successful leptogenesis 
at $M_1\gtrsim 2.7\times 10^{12}$ GeV:  
\be
(-\,\sin\delta) \gtrsim C^{-1}_{P1} \cong 0.38\,,~~{\rm or}~~
202.4^\circ \lesssim \delta \lesssim 337.6^\circ
\label{eq:drange1}
\ee
%
The derived condition on $\delta$ is a necessary condition for 
successful leptogenesis within the considered scenario 
\footnote{In \cite{Pascoli:2006ci} in the same scenario 
the following condition for successful leptogeneis was obtained 
using the 1BE2F and assuming that the two-flavoured 
leptogenesis regime does not extend beyond 
$M_1 = 5\times 10^{11}$ GeV: 
$|\sin\theta_{13}\sin\delta| \gtrsim 0.090$.
In the same article  the minimal scale 
of viable leptogenesis was found to be 
$M_{1\text{min}}\cong 2.2\times 10^{11}$ GeV, to be compared with 
$M_{1\text{min}}\cong 2.5\times 10^{11}$ GeV found by us.
The lower limit on $(-\sin\delta)$ we have obtained in  
Eq. \eqref{eq:drange1} implies 
$|\sin\theta_{13}\sin\delta| \gtrsim 0.057$, where we have 
used the best  fit value of $\theta_{13}$ from Table \ref{Tab::BestFit}.
It is clear that our results based on the DME, in particular, 
extend the ranges of $\delta$ and $M_1$, for which we can 
have successful leptogenesis, derived in \cite{Pascoli:2006ci}. 
}.

 As $M_1$ decreases from $2.7\times 10^{12}$ GeV to 
$2.5\times 10^{11}$ GeV, $\eta_B$ decreases from the value at the plateau 
to the observed value and correspondingly, 
the interval of values of $\delta$ for which one can have 
viable leptogenesis also decreases shrinking to the point 
$\delta = 3\pi/2$ at $M_1 = 2.5\times 10^{11}$.
Clearly, there exists a correlation between the value of 
$\delta$ and the scale $M_1$ of viable leptogenesis. 
It follows from the preceding discussion also that 
in the considered scenario of CP violation provided by the Dirac CPV phase 
$\delta$ of the PMNS matrix, it is possible to reproduce the 
observed value of BAU for values of $M_1$ spanning at least three 
orders of magnitude, i.e., for 
$2.5\times 10^{11} \lesssim M_1/\text{GeV} \lesssim 10^{14}$.

 In the case we have considered with 
$\alpha_{23} = 0$ and $x = x^\text{min}=22.2^\circ$, 
we can have successful leptogenesis, as Eq. \eqref{eq:drange1} shows,
for $\delta$ lying in the interval $\pi < \delta < 2\pi$, 
where $\sin\delta < 0$. So, the sign of $\sin\delta$ 
is anticorrelated with the sign of the observed $\eta_B$.
This result holds also for the alternative possible values 
of $\alpha_{23} = \pm 2\pi$ and 
all possible value of $x$, 
for which we can have viable leptogenesis.
In other words, in the case under study 
there are no values of $\delta$ from the interval 
$0 < \delta < \pi$ where $\sin\delta >0$, for which 
it is possible to reproduce the observed value of BAU. 

Indeed, we note first that 
there is a periodicity of $\pi$ in the dependence on $x$, 
and of $4\pi$ in the dependence on $\alpha_{23}$,
of all relevant quantities on which the 
\begin{figure}[t]
\centering
\includegraphics[width=12cm]{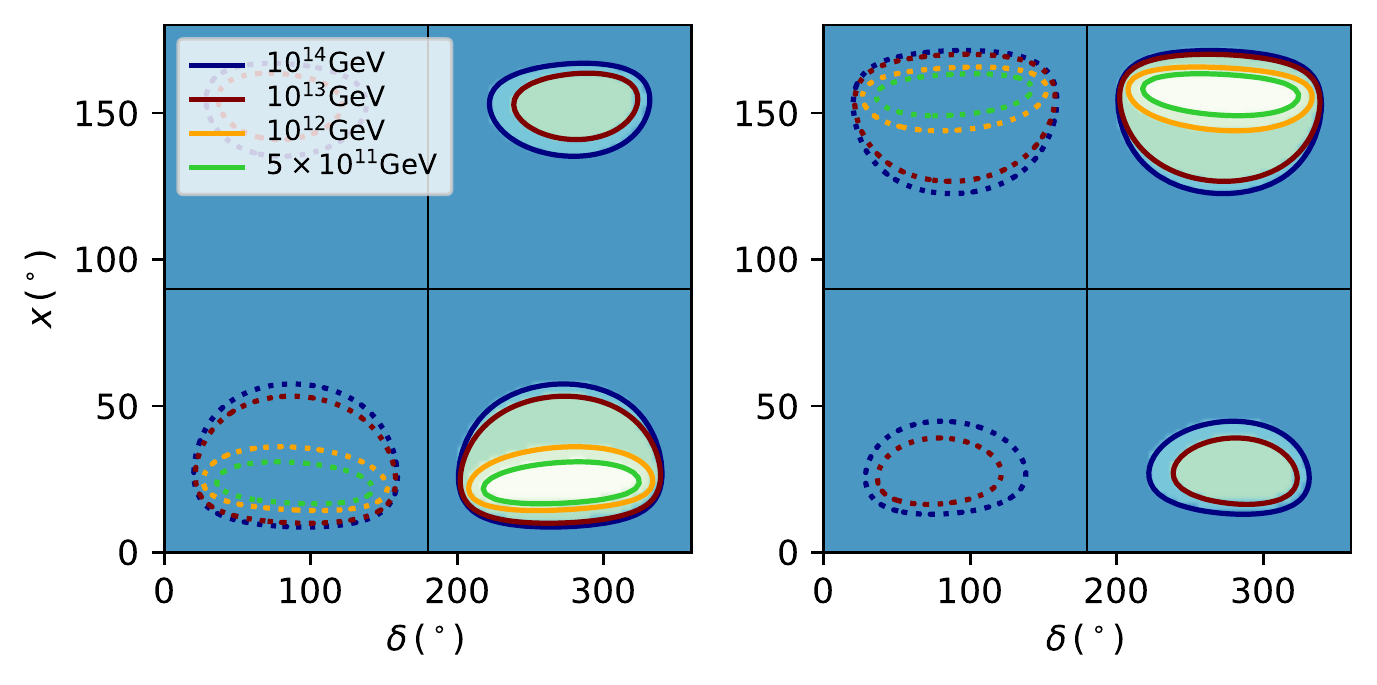}
\caption{Regions of viable leptogenesis 
in the $\delta-x$ half plane, $0\leq x \leq \pi$, 
for NH spectrum, real $R$-matrix, 
CP violation due to the Dirac phase $\delta$, 
$\alpha_{23} = 0$ (left panel) and $2\pi$ 
(right panel) and different values of $M_1$. 
The solid contours corresponding to 
fixed values of $M_1 $ surround the 
regions in which there 
is a combination of values of 
$\delta$ and $x$ for which $\eta_B = 6.1\times 10^{-10}$.
The dotted contours surround regions 
where one can have $|\eta_B| = 6.1\times 10^{-10}$ 
but $\eta_B < 0$. 
The predicted $\eta_B$
outside the contours  is always smaller in 
magnitude than the observed BAU.
The regions of viable leptogenesis 
in the half-plane $-\pi \leq x \leq 0$ 
(or $\pi \leq x \leq 2\pi$), which are not 
shown,  can be obtained from those in the figure 
by substituting $x$ with $x - \pi$. 
See text for further details.
}
\label{fig:NHscanxd_a21_0and360_a31_0}
\end{figure}
%
\noindent 
predicted sign of 
$\eta_B$ depends: $\epsilon^{(1)}_{\tau\tau}$, $k_1$ and 
$p_{1\tau}$ (see Eqs. \eqref{eq:k1NH} - \eqref{eq:Ret1t2}). 
Therefore one gets the 
same results for $x$ and $x -\pi$. In the example with 
$\alpha_{23} = 0$ and 
$x=22.2^\circ$ we have considered, 
we get the same result for $x = -\,157.8^\circ$ 
(or equivalently $x = 202.2^\circ$).
If we set $\alpha_{23} = 2\pi$, 
the results will be the same for 
$\alpha_{23} = -\,2\pi$.
Therefore the only possibility to have 
viable leptogenesis with $\sin\delta > 0$ 
is when  $\alpha_{23} = 2\pi$.
The quantities  
$\Re(U_{\tau 2}^* U_{\tau 3}) \sin2x \propto -\,\beta_\alpha\sin2x$
and  $\Im (U_{\tau 2}^* U_{\tau 3}) \sin2x 
\propto \beta_\alpha\,\sin\delta \sin2x$,
on which  respectively $p_{1\tau}$ and $\epsilon_{\tau\tau}^{(1)}$
depend, change sign when  $\alpha_{23}$ 
is changed from 0 to $2\pi$: $\beta_\alpha = 1~(-1)$ for 
$\alpha_{23} = 0~(2\pi)$.
In addition $\Re(U_{\tau 2}^* U_{\tau 3})$ and $p_{1\tau}$  
exhibit a sub-leading dependence on $\delta$ 
via terms proportional to $\sin\theta_{13}\cos\delta$. 
Thus, in what concerns the present discussion, changing the sign of 
$\sin\delta$ has negligible effect on  $p_{1\tau}$. 
We recall that for $\alpha_{23} = 0$,
$p_{1\tau}$ has a minimum at 
$x^\text{min}=22.2^\circ$ where $p_{1\tau} \ll 1$, 
so that we have  $(1 - 2p_{1\tau}) > 0$ 
for the quantity on which, in particular, 
the sign of $\eta_B$ depends.
The change of the sign of  $\Re(U_{\tau 2}^* U_{\tau 3})\sin2x$
leads to a significant change of the value of 
$p_{1\tau}$, leading for $x^\text{min}=22.2^\circ$ 
to  $(1 - 2p_{1\tau}) < 0$ and thus to 
non-viable leptogenesis for $\sin \delta > 0$.

It follows from the preceding considerations that 
the change of the signs of both $\Re(U_{\tau 2}^* U_{\tau 3})$
and $\Im (U_{\tau 2}^* U_{\tau 3})$ in the expressions for
$p_{1\tau}$ and $\epsilon_{\tau\tau}^{(1)}$
can only be compensated simultaneously by changing 
$x$ to $\pi - x$, i.e., by a change of 
the sign of $\sin2x$. This implies 
that in addition to the solutions we have found for 
$\alpha_{23} = 0$ for certain ranges of $x$ 
(e.g., for $0 < x < \pi/2$)
and of $\delta$ in the interval $\pi < \delta < 2\pi$,
for $\alpha_{23} = 2\pi$ 
we will have successful leptogenesis in the  
range of $\pi - x$ (e.g., for $\pi/2 < x < \pi$)
and for $\delta$ in the same interval.
Thus, in the case of  $\alpha_{23} = 2\pi$, 
a value of $\delta$ from the interval 
$0 < \delta < \pi$ with $\sin2x < 0$ ($\sin2x > 0$)
leads either to a wrong sign of $\eta_B$ 
due to the interplay of the signs of 
$p_{1\tau}$ and $\epsilon_{\tau\tau}^{(1)}$, 
or else to a value of  $\eta_B$ which is smaller than the observed one.

The conclusions of the preceding discussions 
are confirmed by the numerical scan of the parameters 
$\delta$ and $x$ in the case of $\alpha_{23} = 0$ and $2\pi$ 
and several fixed values of $M_1$, 
the results of which are shown in 
Fig. \ref{fig:NHscanxd_a21_0and360_a31_0}.
Thus, in the considered scenario there is a direct and unique 
relation between the sign of $\sin\delta$ and 
the sign of the baryon asymmetry of the Universe. 
If the measurement of $\delta$ in the low-energy 
neutrino oscillation experiments will show that $\delta$ 
lies in the interval 
$[0,\pi]$, the considered leptogenesis 
scenario will be ruled out.
If, however, $\delta$ will be found to 
lie in the lower half-plane, $\pi < \delta < 2\pi$, 
this will not only lend support for the discussed scenario, 
but also will allow to obtain constraints 
on the leptogenesis scale. 

\vspace{0.3cm} 
Given that for $x \neq 0$, $y = 0$  
and CP violation due only to the Dirac phase $\delta$
leptogenesis is unsuccessful at any mass scale in the IH case 
(see  Eq. \eqref{eq:epsratio} and the discussion related to it)
we do not consider this case.

%
\subsubsection{Purely imaginary \texorpdfstring{$R_{11}R_{12}$}{R11R12} (\texorpdfstring{$R_{12}R_{13}$}{R12R13}) 
(\texorpdfstring{$x = k\pi$}{xp}, \texorpdfstring{$k=0,1,2$}{k}, \texorpdfstring{$y\neq 0$)}{y}}
%

We discuss next the leptogenesis scenario in 
which CP violation is still provided by the Dirac phase only, 
but now $x = k\,\pi$, $k=0,1,2$, and $y\neq 0$ so that the product 
$R_{12}R_{13}$ ($R_{11}R_{12}$) is purely imaginary in the NH (IH) case 
and the suppression of the CPV asymmetry shown  
in  Eq. \eqref{eq:epsratio} is avoided.
\begin{figure}[t]
\centering
\includegraphics[width=12cm]{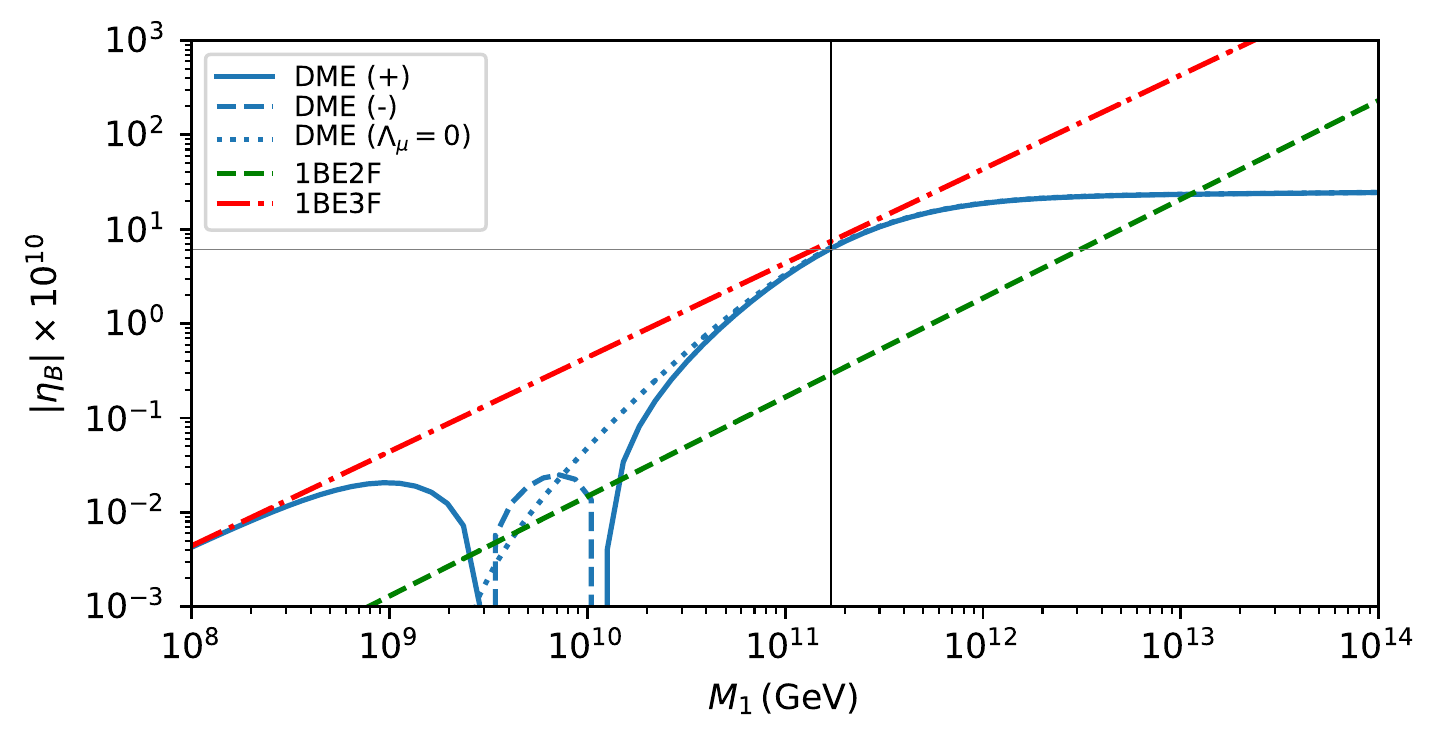}
\caption{The baryon asymmetry $\eta_B$ versus $M_1$ for NH spectrum,  
$x=0$,  $y=26^\circ~(-\,26^\circ)$, $\delta = \pi/2$
and $\alpha_{23} = \pi$ ($3\pi$), for which successful 
leptogenesis takes place for the minimal value of 
$M_1 \cong 1.7\times 10^{11}$ GeV (vertical black line). 
The requisite CP violation is provided by 
the Dirac phase $\delta$. The horizontal black line corresponds 
to the observed $\eta_B = 6.1\times 10^{-10}$. 
See text for further details.
}
\label{fig:SCNH_d_90_a21_180_a31_0_y26}
\end{figure}
%

\vspace{0.3cm}
%
\leftline{\bf NH Spectrum}
%

\vspace{0.3cm}
 The analysis is similar to that performed in the preceding subsection.
We report below the results on the ranges of $\delta$ and $M_1$ for which 
one can have successful leptogenesis. For the minimal value of 
$M_1$ we get $M_1 = 1.7\times 10^{11}$ GeV, which is obtained for 
$\delta = \pi/2$, $y=26^\circ$ ($y=-\,26^\circ$) and 
$\alpha_{23} = \pi$ or $-\,3\pi$ 
($3\pi$ or $-\pi$). This case is illustrated by  
Fig. \ref{fig:SCNH_d_90_a21_180_a31_0_y26}.
The value of $y$ maximises the factor 
\be 
f_{2\epsilon}(x) = \dfrac{\sqrt{a}\,\sinh 2y}{a\,\cosh^2 y + \sinh^2 y}\,,~~~
a \equiv m_2/m_3\,, 
\label{eq:f2e}
\ee
%
in the expression for $\epsilon_{\tau\tau}^{(1)}$, and thus 
maximises $\epsilon_{\tau\tau}^{(1)}$ with respect to $y$.
For $\alpha_{23} = \pi$ or $(-\,3\pi)$ 
the value of $\delta = \pi/2$ maximises  $\epsilon_{\tau\tau}^{(1)}$ 
which is proportional to $\sin\delta$. At the plateau which begins at 
$M_1\cong 2.1\times 10^{12}$ GeV we have 
$\eta_B \cong C_{P2}\,6.1\times 10^{-10}$ with $C_{P2}\cong 3.9$.
Correspondingly, at $M_1\gtrsim 2.1\times 10^{12}$ GeV we can have 
successful leptogenesis for 
\be
\sin\delta \gtrsim C^{-1}_{P2} \cong 0.25\,,~~{\rm or}~~
14.6^\circ \lesssim \delta \lesssim 165.4^\circ\,,
\label{eq:drange3}
\ee
%
 As $M_1$ decreases from $2.1\times 10^{12}$ GeV to $1.7\times 10^{11}$ GeV,
$\eta_B$ decreases from the value at the plateau to the observed value and 
the width of the intervals of values of $\delta$ 
in Eq. \eqref{eq:drange3} 
decreases. At $M_1 = 1.7\times 10^{11}$ GeV 
it shrinks to the point $\delta = \pi/2$.  
In what concerns $M_1$, successful leptogenesis is 
possible for values of $M_1 \gtrsim 1.7\times 10^{11}$ GeV, 
which span at least three orders of magnitude.

 It follows from the preceding discussion that 
for  $\alpha_{23} = \pi$ or $(-\,3\pi)$, 
one can have successful leptogenesis for value of $\delta$ 
from the interval $0 < \delta < \pi$ where 
$\sin \delta > 0$.
Performing an analysis similar to that 
in the preceding subsection, we 
\begin{figure}
\centering
\includegraphics[width=12cm]{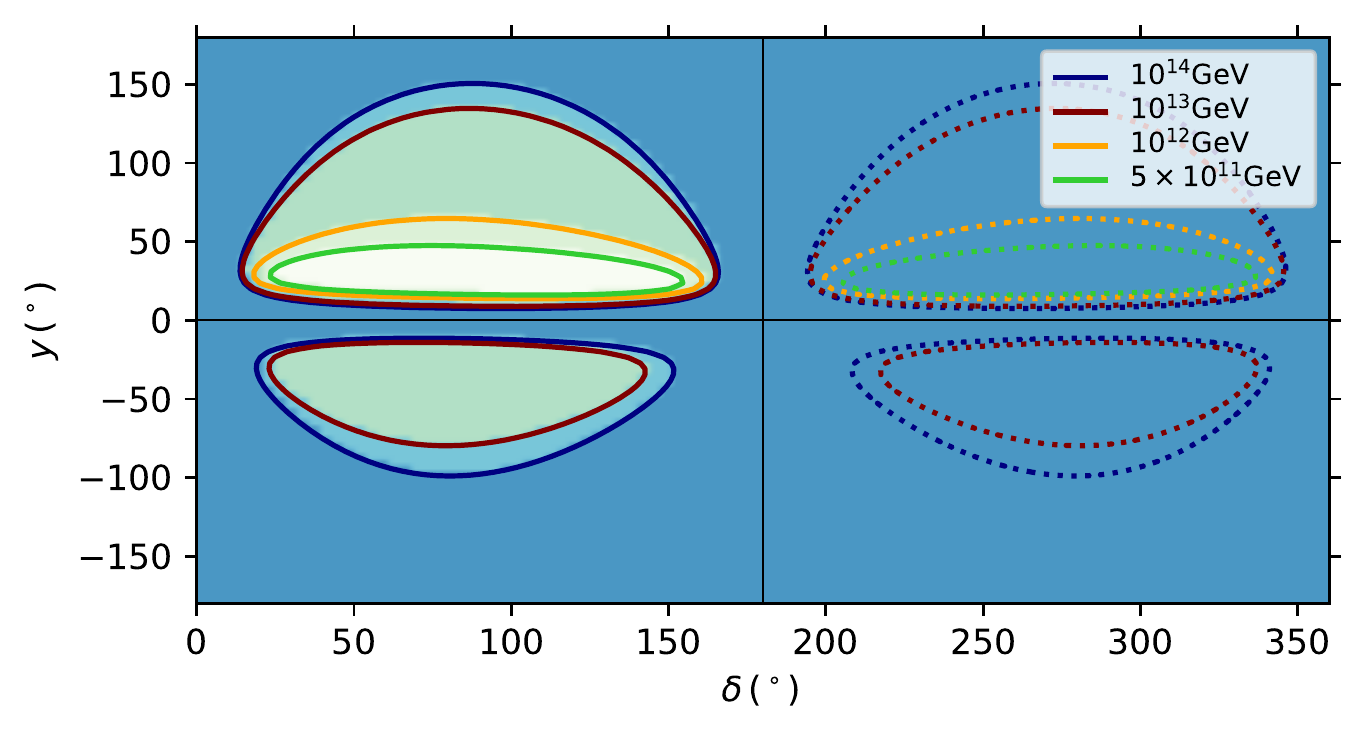}
\caption{Regions of viable leptogenesis 
in the $\delta - y$ plane for NH spectrum, $x = k\pi$, $k=0,1,2$, 
$\alpha_{23} = \pi$ and 
different $M_1$. The solid contours corresponding to 
fixed values of $M_1 $ surround the regions in which there 
is a combination of values of 
$\delta$ and $y$ for which $\eta_B = 6.1\times 10^{-10}$.
The dotted contours surround regions 
where one can have $|\eta_B| = 6.1\times 10^{-10}$ 
but $\eta_B < 0$.  
The predicted $\eta_B$
outside the contours is  smaller in 
magnitude than the present BAU.
Setting $\alpha_{23} = 3\pi$ leads 
to a figure which can be obtained from 
the present by changing $y$ to $-\,y$.
See text for further details.
}
\label{fig:SCNH_scanyd_a21_180_a31_0}
\end{figure}
%
\noindent 
find that also in this case 
there is a direct relation between the sign of $\sin\delta$ 
and the sign of the observed BAU in the sense that 
for the values of the parameters in the considered case, 
no region with viable leptogenesis exists for 
$\delta$ from the interval $\pi < \delta < 2\pi$,
where $\sin\delta < 0$. 
Changing  the value of $\alpha_{23}$ 
from $\pi$ to $3\pi$, for example, 
one finds that the viable regions of values of $y$ 
and $\delta$ from the interval $0 < \delta < \pi$, 
for which it is possible to reproduce the observed value of 
BAU, shift to the regions corresponding  
to $(-\,y)$ with $\delta$ remaining in the same  
interval $0 < \delta < \pi$. 
This is confirmed by the numerical scan of the 
$y-\delta$ parameter space for 
$\alpha_{23} = \pi$ and $3\pi$, 
the results of which are shown in 
Fig. \ref{fig:SCNH_scanyd_a21_180_a31_0}.

Obviously, the discussed leptogenesis scenario
will be ruled out if $\delta$ determined 
in neutrino oscillation experiments 
is found definitely to lie in the interval
$[\pi, 2\pi]$, while if $\delta$ is found to 
be in the upper half-plane, $0 < \delta < \pi$, 
the scenario will be proven viable and it will 
be possible to obtain also constraints on the 
leptogenesis scale of the scenario.

%
\vspace{0.3cm}
%
\leftline{\bf IH Spectrum}
%

\begin{figure}[t]
\centering
\includegraphics[width=12cm]{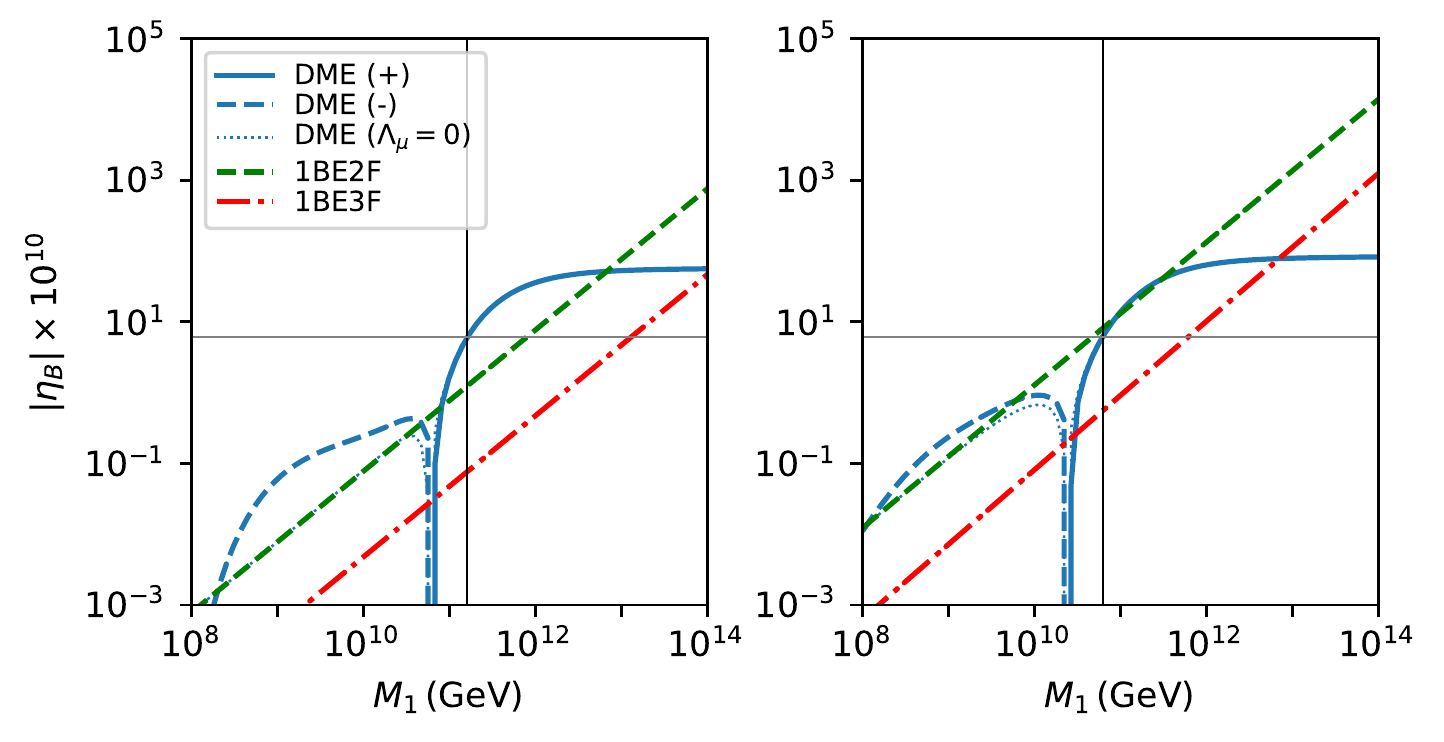}
\caption{The same as in Fig. \ref{fig:SCNH_d_270_a21_0_a31_0} 
but for $x=0$ and $y\neq 0$ and in the IH case.
The two panels correspond to CP violation from the Dirac phase only 
with $\delta = 3\pi/2$, $\alpha_{21} = \pi$, 
$y=-100^\circ$ (left panel) and $y=-46.5^\circ$ (right panel).
The vertical black lines correspond 
to $M_1 \simeq 1.6 \times 10^{11}$ GeV (left panel) and 
$M_1 \simeq 6.2 \times 10^{10}$ GeV (right panel). 
See text for further details.
}
\label{fig:SCIH_d_270_a21_180_a31_0}
\end{figure}

\vspace{0.3cm}
 We analyse in somewhat greater detail the case of IH spectrum.
We show in Fig. \ref{fig:SCIH_d_270_a21_180_a31_0} the
modulus of the baryon asymmetry $\eta_B$ versus 
$M_1$ for the IH spectrum with  
 $\delta = 3\pi/2$, $\alpha_{21} = \pi$,
$y=-100^\circ$ (left panel) and $y=-46.5^\circ$ (right panel).
 The dependence of $\eta_B$ on $M_1$ exhibits a number of 
interesting features. The 1-to-2 flavour transition 
described by DME takes place with a sign change of $\eta_B$.
At values of $M_1 < M_{10}$ ($M_1 > M_{10}$), $M_{10}$ being the 
value of $M_1$ at which $\eta_B = 0$, we have 
$\eta_B < 0$ ($\eta_B > 0$). In the case illustrated in 
Fig. \ref{fig:SCIH_d_270_a21_180_a31_0}, we have 
$M_{10} \ll 10^{12}$ GeV. When $y$ changes from 
$(-100^\circ)$ to $(-46.5^\circ)$, $M_{10}$ decreases from 
$6.0\times 10^{10}$  GeV to $2.4\times 10^{10}$ GeV. 
Most importantly, the minimal value of 
$M_1$ at which one can have successful leptogenesis 
also decreases from $M_1 = 1.6\times 10^{11}$ GeV to 
$M_1 = 6.2\times 10^{10}$, with both values being $\ll 10^{12}$ GeV.

 Further,  the DME solution for $\eta_B$ 
shown in the left (right) panel of Fig. \ref{fig:SCIH_d_270_a21_180_a31_0} 
is at $M_1 \leq 10^{12}$ GeV larger than (similar in magnitude to) 
$|\eta_B|$ found with 1BE2F, except in a narrow region around 
$M_{10}$. Still, $\eta_B$ obtained from the 1BE2F equations 
shown in both panels, in contrast 
to that derived from DME ones,
has a wrong sign, i.e., predicts $\eta_B < 0$ and thus 
non-viable leptogenesis.
The value of $|\eta_B|$ obtained with 1BE3F is in both cases, as the panels 
show,  significantly smaller than those found with DME and 1BE2F. 
Moreover, the 2-to-3 flavour transition described by the DME solution  
takes place at $M_1 \lesssim 10^{8}$ GeV, with the $\mu$-Yukawa interaction 
having the effect of enhancing the DME solution for $|\eta_B|$ 
in the interval $10^8 \lesssim M_1 /\text{GeV} \lesssim 10^{10}$. 
Both these features are in the region of values of $M_1$ 
for which the calculated $|\eta_B|$ is significantly smaller that 
the observed $\eta_B$. However, they might be relevant in a leptogenesis 
scenario with three heavy Majorana neutrinos with 
non-hierarchical masses, e.g.,  with $M_3 \cong 3M_2\cong 9M_1$. 
%

\begin{figure}[t]
\centering
\includegraphics[width=12cm]{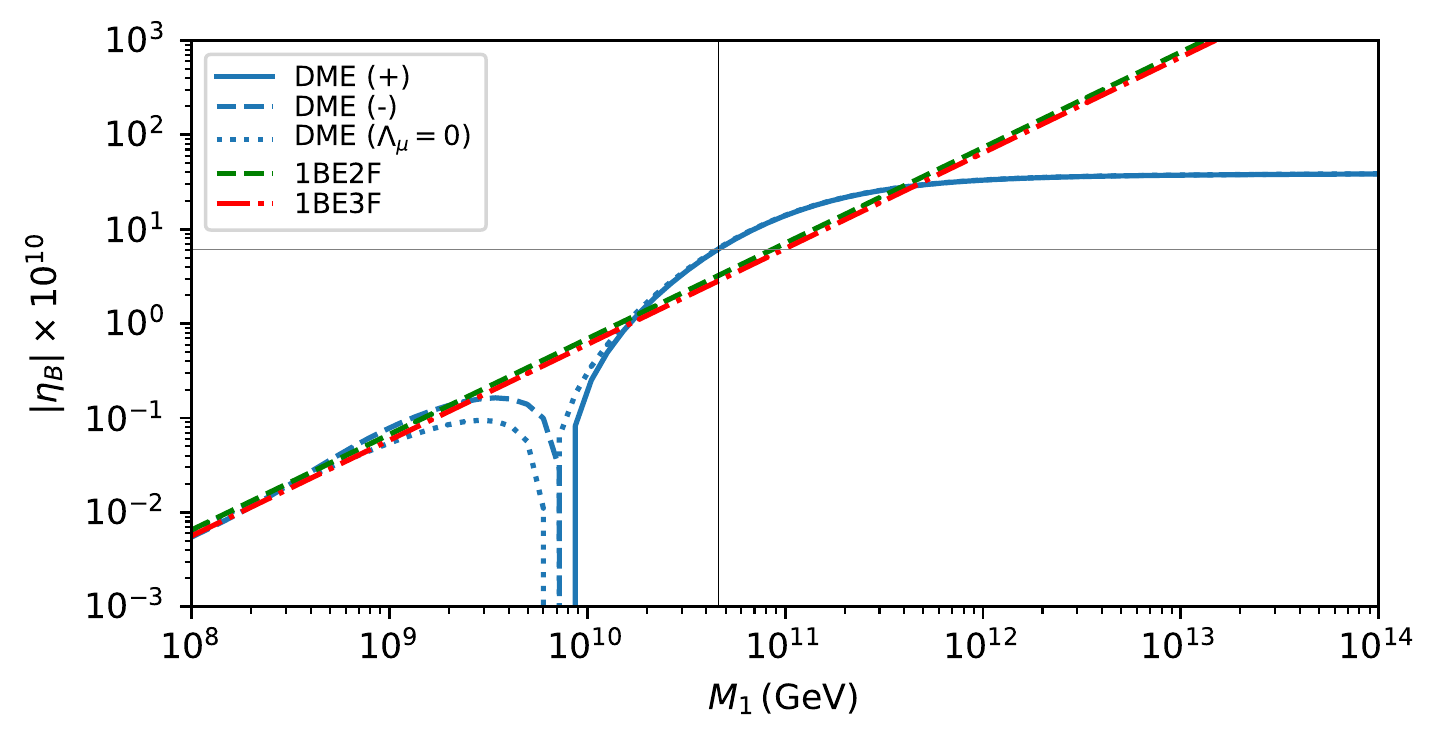}
\caption{The baryon asymmetry $\eta_B$ versus $M_1$ for IH spectrum,  
$x=0,\pi$,  $y =-\,73^\circ$ ($y = 73^\circ$), 
$\alpha_{21} = \pi$ ($3\pi$) and $\delta = 211^\circ$, 
for which successful leptogenesis 
takes place for the minimal value of $M_1 \cong 4.6\times 10^{10}$ GeV 
(vertical black line). The requisite CP violation is provided by 
the Dirac phase $\delta$. The horizontal black line corresponds 
to the observed $\eta_B = 6.1\times 10^{-10}$. 
See text for further details.
}
\label{fig:SCIH_d_211_a21_180_y-73}
\end{figure}

 In what concerns the range of $\delta$ and $M_1$ 
for which we have successful leptogenesis, we find that 
(see Fig. \ref{fig:SCIH_d_211_a21_180_y-73}):
i) the minimal value of $M_1$ is $M_1 \cong 4.6\times 10^{10}$ GeV 
and corresponds to the values of $\alpha_{21} = \pi~(3\pi)$, 
$y= -\,73^\circ$ ($y = +\,73^\circ$) and $\delta = 211^\circ$;
ii) the plateau of values of $\eta_B$ is present at 
$M_1 \gtrsim 1.2\times 10^{12}$ GeV;
iii) at the plateau 
$\eta_B \cong C_{P3}\,6.1\times 10^{-10}$ with 
$C_{P3}\cong 6.1$.
Correspondingly, at $M_1\gtrsim 1.2\times 10^{12}$ GeV 
successful leptogenesis is possible for 
\be
|\sin\delta| \gtrsim C^{-1}_{P3}\, |\sin(\delta = 211^\circ)| \cong 0.084\,,~~
{\rm or}~ 185^\circ \lesssim \delta \lesssim 355^\circ\,,
\label{eq:drange4}
\ee
%
where we have used the fact that the plateau value of $\eta_B$ 
corresponds to $\delta = 211^\circ$.
%

 We note that for the chosen values of  $\alpha_{21} = \pi$, 
$y= -\,73^\circ$,  
$\epsilon_{\tau\tau}^{(1)}$ 
is proportional to $\sin\delta$ and thus at 
$\delta = 211^\circ$ 
$|\epsilon_{\tau\tau}^{(1)}|$ is smaller by the factor 0.515 
than for $\delta = 3\pi/2$. However, due to the fact that,
as can be shown, the value 
of $p_{1\tau}$ at $\delta = 211^\circ$
is smaller approximately by a factor of 6 
than that at $\delta = 3\pi/2$,
the minimal $M_1$ at which we can have successful leptogenesis 
is also smaller than the one for $3\pi/2$ which reads  
$M_1 \cong 10^{11}$ GeV. At the same time, $\eta_B$ at the  plateau 
for $\delta = 3\pi/2$
is by a factor of approximately 1.9 times 
larger than the plateau value of $\eta_B$ for 
$\delta =  211^\circ$ 
and reads: 
$\eta_B =  7.3\times 10^{-9}$.

\begin{figure}[t]
\centering
\includegraphics[width=12cm]{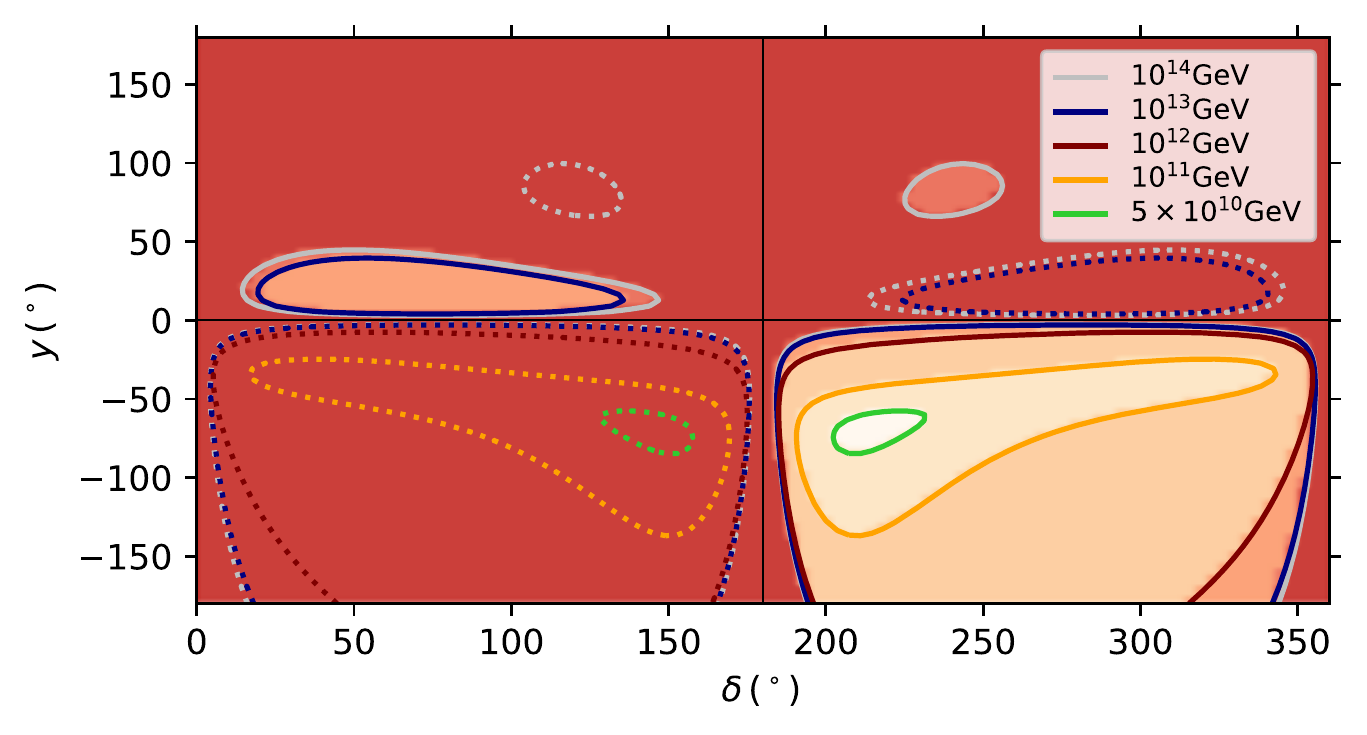}
\caption{Regions of viable leptogenesis 
in the $\delta - y$ plane for IH spectrum,  
$x = k\pi$, $k=0,1,2$, $\alpha_{21} = \pi$, 
corresponding to CP violation due to the Dirac phase $\delta$,
and different $M_1$. The solid contours corresponding to 
fixed values of $M_1 $ surround the regions in which there 
exists a combination of values of 
$\delta$ and $y$ for which $\eta_B = 6.1\times 10^{-10}$.
The dotted contours surround regions 
where one can have $|\eta_B| = 6.1\times 10^{-10}$ 
but $\eta_B < 0$. The predicted $\eta_B$
outside the contours  is smaller in 
magnitude than the present BAU.
Setting $\alpha_{21} = 3\pi$ leads 
to a figure which can be obtained from 
the present by changing $y$ to $-\,y$.
See text for further details.
}
\label{fig:SCIH_scanyd_a21_180_a31_0}
\end{figure}

 As we have seen, in the discussed case of Dirac CP violation 
and IH neutrino mass spectrum, successful leptogenesis is possible for 
values of $\delta$ from the interval $\pi < \delta < 2\pi$ 
where $\sin\delta < 0$. Performing a scan over 
$y$ and $\delta$ for the possible values 
of $\alpha_{21} = \pi$ and $3\pi$ with $x=k\pi$, $k=0,1,2$, 
we find that for $M_1 \gtrsim 10^{13}$ GeV one can have a 
successful leptogenesis also for values of $\delta$ from the interval 
$0 < \delta < \pi$, and a small range of values of 
$y$ from the interval  $0 < y < 50^\circ$. 
The appearance of this second region 
is related to the slow increase of $\epsilon_{\tau\tau}^{(1)}$ and thus of 
$\eta_B$ with $M_1$ due to the factor $f^{-1}(M_2)/M_2$.
The results of the scan are presented graphically in 
Fig. \ref{fig:SCIH_scanyd_a21_180_a31_0}. 
Thus, in this case we have a direct relation between the sign of 
$\sin\delta$ and the sign of the baryon asymmetry of the Universe 
only for $M_1 < 10^{13}$ GeV.

\begin{figure}
    \begin{subfigure}{\textwidth}
    \centering
    \includegraphics[width=0.8\textwidth]{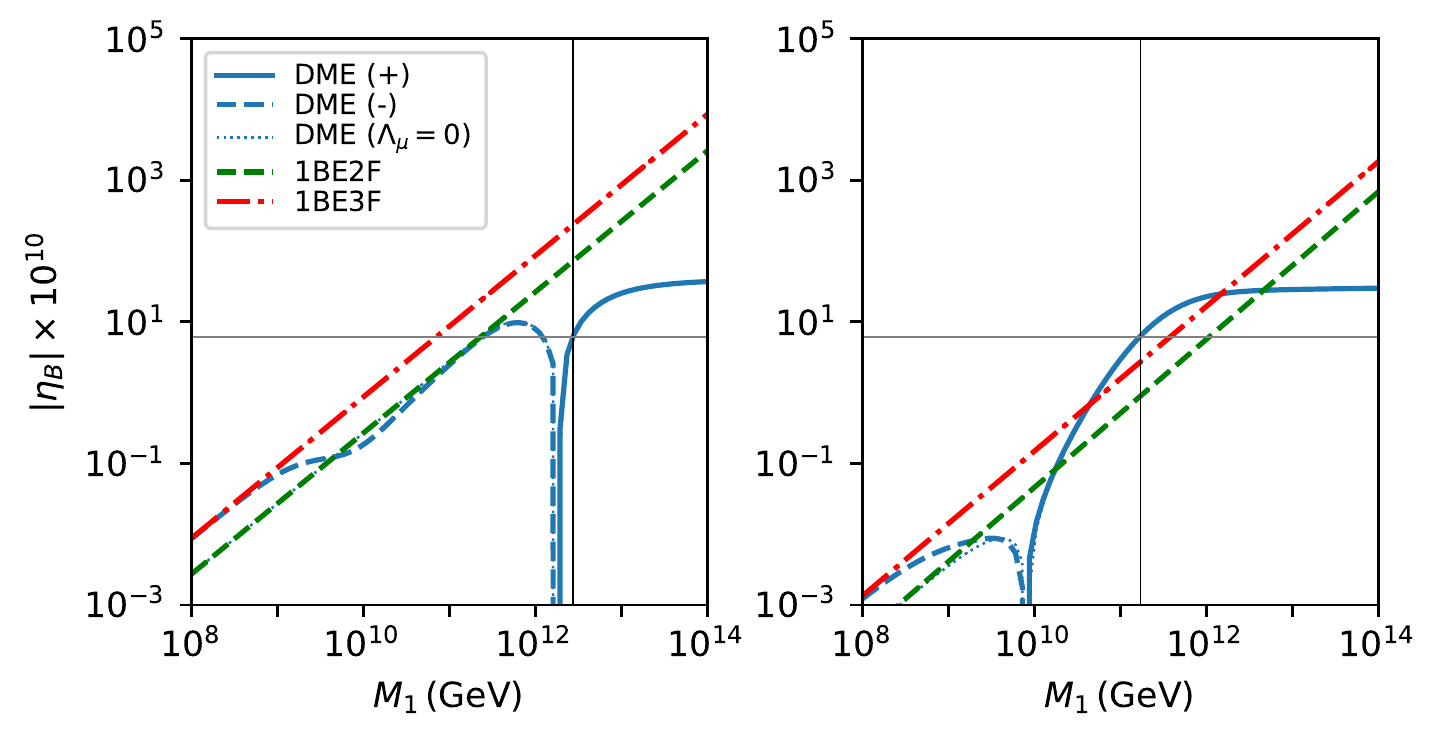}
        \end{subfigure}
\centering
\includegraphics[width=0.8\textwidth]{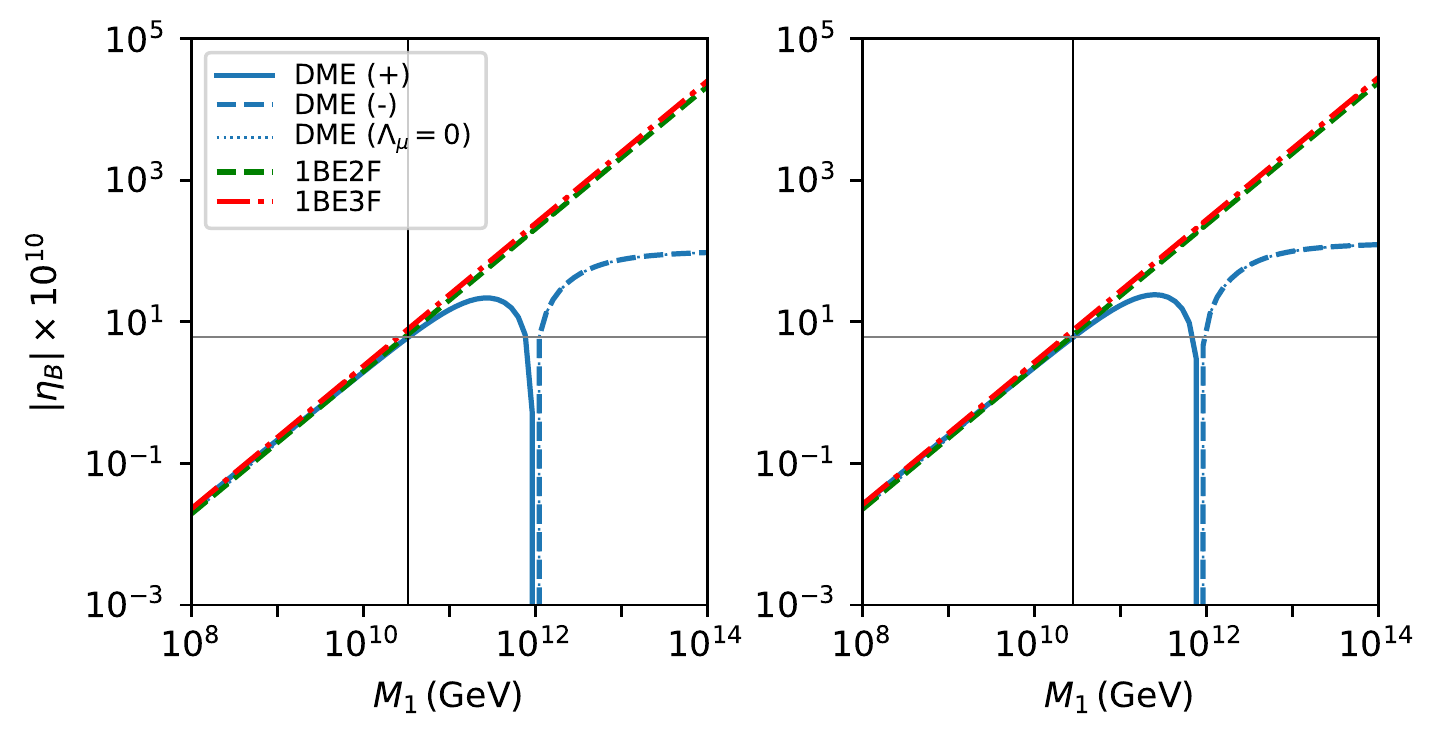}
\caption{Leptogenesis with CP violation due to the Majorana phases 
of the PMNS matrix in the case of NH spectrum. 
Top panels: Two examples of ``standard'' (left) and ``non-standard'' (right) 
behaviour of  $\eta_B$ as a function of $M_1$. 
The results shown are obtained as those  
in Fig. \ref{fig:SCNH_d_270_a21_0_a31_0}.  
The parameters in the left (right) panel are set to 
$\delta = \pi$, $y = 0$, $x = 30^\circ\,(20.4^\circ)$ and 
$\alpha_{23} = 3\pi/2\,(700^\circ)$. The vertical black lines are 
at $M_1 \simeq 2.7\times 10^{12}$ GeV (left) and 
$M_1 \simeq 1.7\times 10^{11}$ GeV (right) and intersect the horizontal 
grey line of the observed BAU at the points of successful leptogenesis.
Bottom panels: The left (right) panel show 
the dependence of $\eta_B$ on $M_1$ 
for real (purely imaginary) $R_{12}R_{13}$ 
and $\delta = \pi$ in the case of minimal $M_1$ for which 
it is possible to have successful leptogenesis with 
CP violation provided by the Majorana 
phase $\alpha_{23}$: 
the left (right) panel are obtained 
for $x = 20^\circ$, $y=0$ ($x = k\pi$, $k=0,1,2$,  
$y = 19^\circ$) and 
$\alpha_{23} = 102^\circ~(80^\circ)$. 
See text for further details.
}
\label{fig:plateau_SCNH_d_180_a21L270_a21R_700}
\end{figure}
%
\subsection{CP Violation due to the Majorana Phases}
%
%
 We investigate next the considered leptogenesis 
scenario with two heavy Majorana neutrinos with hierarchical 
masses in which the CP violation is provided only by the 
Majorana phases of the PMNS matrix. Thus, 
the Dirac phase and the $R$-matrix elements are chosen   
not to contribute to the CP violation necessary for 
the generation of BAU. We note that in the case of CP violation 
due to the Majorana phases, 
the additional CP violation due to the Dirac phase has 
sub-leading effects in leptogenesis as a consequence of the 
suppression by the factor $\sin\theta_{13}$. 
However, in certain cases these effects are non-negligible.

\vspace{0.2cm}
\leftline{{\bf NH Spectrum}}

\vspace{0.2cm}
For real  $R$-matrix it is impossible to have 
successful leptogenesis for IH light neutrino mass spectrum 
and CP violation originating from CPV phases of the PMNS matrix, 
as we have already discussed, so we consider only the case of 
NH spectrum. We show in Fig. \ref{fig:plateau_SCNH_d_180_a21L270_a21R_700}, 
top-left (top-right) panel, example of the 
behaviour of $\eta_B$ as a function of 
$M_1$ for  $y = 0$  (real $R$-matrix), $x = 30^\circ\,(20.4^\circ)$, 
$\delta = \pi$, and $\alpha_{23} = 3\pi/2\,(700^\circ)$. 
The vertical black lines are at $M_1 \simeq 2.7\times 10^{12}$ GeV (left panel) 
and $M_1 \simeq 1.7\times 10^{11}$ GeV (right panel) and intersect 
the horizontal grey line of the observed BAU at the points of 
successful leptogenesis. The scenarios illustrated 
in the top-left and top-right panels of 
Fig.  \ref{fig:plateau_SCNH_d_180_a21L270_a21R_700} correspond 
respectively to what we dubbed ``standard'' and ``non-standard'' 
behaviour of the baryon asymmetry $\eta_B$. 
The salient features of the behaviour of $\eta_B$ 
in the two cases are analogous to those discussed in 
detail in the two preceding subsections 
and we are not going to comment on them further.                   

 We show in the bottom-left (right) panel of 
Fig.  \ref{fig:plateau_SCNH_d_180_a21L270_a21R_700}
the dependence of $\eta_B$ on $M_1$ 
for real (purely imaginary) $R_{12}R_{13}$ 
and $\delta = \pi$ in the case in which 
it is possible to have successful leptogenesis with 
CP violation provided by the Majorana 
phase $\alpha_{23}$ for the  minimal in the 
considered scenario value of $M_1$. 
The other relevant parameters have the following values 
in the left (right) panels:  
$x = 20^\circ$, $y=0$ ($x = k\pi$, $k=0,1,2$,  
$y = 19^\circ$) and $\alpha_{23} = 102^\circ~(80^\circ)$. 
We will discuss in some detail in what follows first the case 
of $x = 20^\circ$, $y=0$ and $\alpha{23} = 102^\circ$, 
extending the discussion after that to the whole 
plane $0 < x < 180^\circ$.
 
The bottom-left panel of  
Fig.  \ref{fig:plateau_SCNH_d_180_a21L270_a21R_700}
illustrates one example of ranges of values of the Majorana phase 
$\alpha_{23}$ and $M_1$, 
for which one can have successful leptogenesis in the case of 
$0 < x < 90^\circ$, $y=0$, $\delta = \pi$ 
and CP violation provided by $\alpha_{23}$. 
As we have already remarked, this example corresponds to minimal $M_1$ 
for having viable leptogenesis in the case under 
study and is by far not exhaustive. We identify next 
qualitatively all the regions of  $\alpha_{23}$ in the interval 
$0 < \alpha_{23} < 720^\circ$ and of $M_1$ where 
it is possible to have viable leptogenesis with 
$0 < x < 90^\circ$, $y=0$ and $\delta = \pi$.

 We note first that in the case under discussion 
$\eta_B$ goes through zero and changes sign 
in the 1-to-2 flavour regime transition in any 
of the regions of the parameter space of interest  
and there is always a plateau of values of $|\eta_B|$ 
at $M_1 > M_{10}$, $M_{10}$ being the value of $M_1$ 
at which $\eta_B = 0$. 
We recall that ${\rm sgn}(\eta_B)$ at the plateau is determined 
by ${\rm sgn}(-\,(1-2p_{1\tau})\epsilon_{\tau\tau}^{(1)})$ 
(see Eq. \eqref{eq:SuccessLG}), where 
${\rm sgn}(\epsilon_{\tau\tau}^{(1)}) = 
{\rm sgn}(\Im(U_{\tau 2}^* U_{\tau 3})\sin 2x)$. 
It is not difficult to check that for the best-fit values of the 
neutrino oscillation parameters $\theta_{12}$, $\theta_{13}$, 
$\theta_{23}$, $\Delta m^2_{21}$ and $\Delta m^2_{31}$ given in Table 
\ref{Tab::BestFit}, we have:\\
i) for $5^\circ < x < 90^\circ$, 
$(1-2p_{1\tau}) > 0$ for $0 \leq \alpha_{23} < 220^\circ$ and 
  $500^\circ < \alpha_{23} \leq 720^\circ$, 
with  $(1-2p_{1\tau}) = 0$ at $\alpha_{23} \cong 220^\circ$ and $500^\circ$,
where the precise values at which $(1-2p_{1\tau}) = 0$ depend somewhat 
on $x$:
those corresponding to $x = 10^\circ$ are given approximately  
by $\alpha_{23} \cong 250^\circ$ and $470^\circ$; 
for  $0 < x \lesssim 5^\circ$ we have 
$(1-2p_{1\tau}) > 0$ for any $\alpha_{23}$  
from the interval $[0,720^\circ]$;
\\
ii)  ${\rm sgn}(\Im(U_{\tau 2}^* U_{\tau 3})\sin 2x) > 0~(< 0)$ for 
 $0 < \alpha_{23} < 360^\circ$ ($360^\circ < \alpha_{23} < 720^\circ$).
\\
We can conclude on the basis if these observations that, 
except for $0 < x \lesssim 5^\circ$,
the plateau values of $\eta_B$ \\
a) are negative approximately 
for  $0 < \alpha_{23} < 220^\circ$ and $360^\circ < \alpha_{23} < 500^\circ$;\\
b) are positive
for  $220^\circ < \alpha_{23} < 360^\circ$ and 
$500^\circ < \alpha_{23} < 720^\circ$.\\
In the case of $x$ in the interval $0 < x \lesssim 5^\circ$, 
the plateau values of $\eta_B$ are negative (positive) 
for  $0 < \alpha_{23} < 360^\circ$ ($360^\circ < \alpha_{23} < 720^\circ$).
However, as our numerical study shows, in this case 
it is possible to reproduce the observed value of $\eta_B$ only  
in a very a narrow interval of values of $x$, namely,
for   $2.5^\circ~(1.6^\circ) \lesssim x \lesssim 5^\circ$ 
with $\alpha_{23}$ in the range $0 < \alpha_{23} < 250^\circ$ 
($450^\circ < \alpha_{23} < 720^\circ$). 
As a consequence, this makes only a relatively small addition to 
regions in the space of parameters of the considered scenario 
for which one can have successful leptogenesis.
Therefore we concentrate further on the case of $5^\circ < x < 90^\circ$.

 It follows from the preceding discussion that
if we denote by $M^{(1,2,3,4)}_{10}$ the values of $M_1$ 
at which $\eta_B = 0$ at the  1-to-2 flavour regime transitions 
taking place when $\alpha_{23}$ lies respectively in the intervals 
$(0,220^\circ)$, $(360^\circ,500^\circ)$, $(220^\circ,360^\circ)$ and 
$(500^\circ,720^\circ)$, we can expect successful leptogenesis 
to occur for certain ranges of values of $M_1 < M^{(1)}_{10}$ 
($M_1 < M^{(2)}_{10}$) if 
  $0 < \alpha_{23} < 220^\circ$ 
($360^\circ < \alpha_{23} < 500^\circ$), and 
of  $M_1 > M^{(3)}_{10}$ ($M_1 > M^{(4)}_{10}$) 
extending into $\eta_B$ plateau region when 
 $220^\circ < \alpha_{23} < 360^\circ$ 
($500^\circ < \alpha_{23} < 720^\circ$).
Moreover, the results for $90^\circ < x < 180^\circ$ can be obtained 
from those derived for $0 < x < 90^\circ$ by  
making the simultaneous change 
$x\rightarrow \pi - x$ and $\alpha_{23}\rightarrow \alpha_{23} \pm 2\pi$ 
and taking into account that the results are invariant with respect 
to the change  $\alpha_{23}\rightarrow \alpha_{23} \pm 4\pi$. 

\begin{figure}[t]
\centering
\includegraphics[width=12cm]{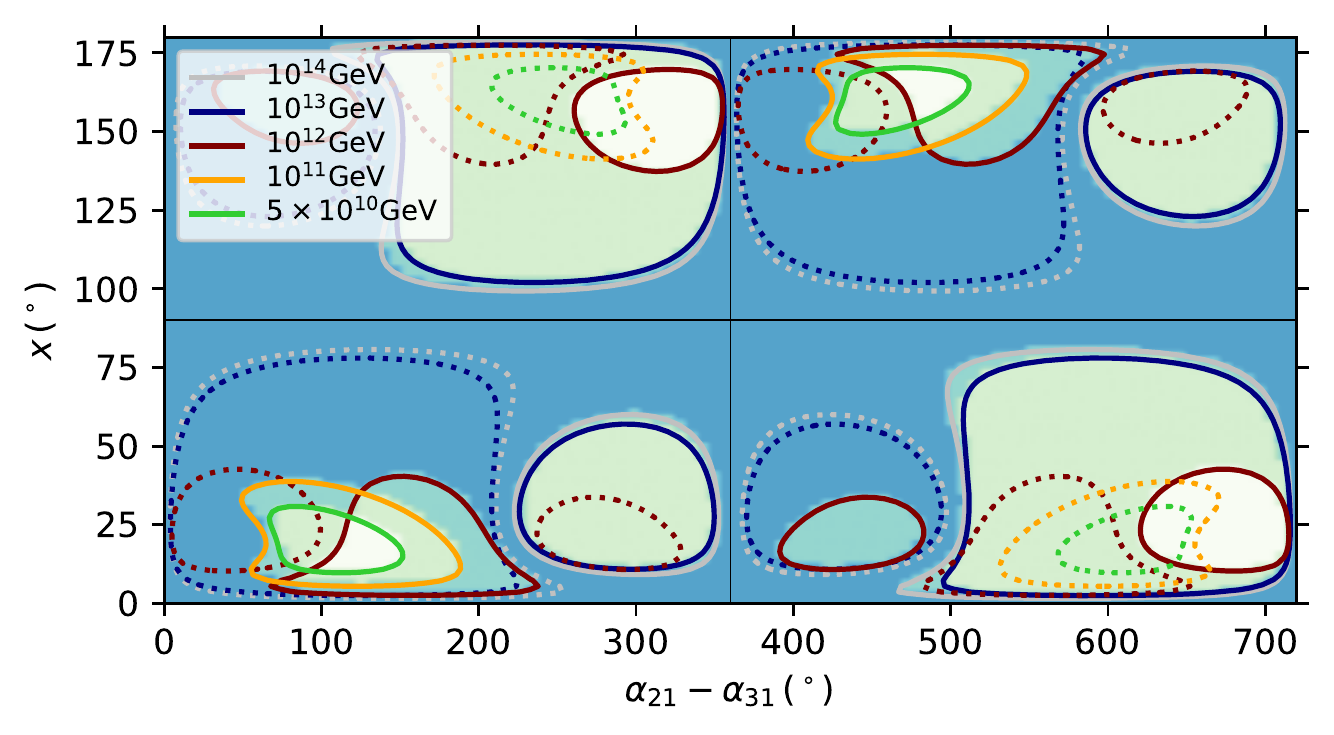}
\caption{Regions of viable leptogenesis 
in the $\alpha_{23} - x$ plane for NH spectrum,  
$y=0$, $\delta = \pi$, 
corresponding to CP violation due to the Majorana phase 
$\alpha_{23}$,
and different $M_1$. The solid contours corresponding to 
fixed values of $M_1 $ surround the regions in which there 
is a combination of values of 
$\alpha_{23}$ and $x$ for which $\eta_B = 6.1\times 10^{-10}$.
The dotted contours surround regions 
where one can have $|\eta_B| = 6.1\times 10^{-10}$ 
but $\eta_B < 0$. The predicted $\eta_B$
outside the contours  is smaller in 
magnitude than the present BAU.
See text for further details.
}
\label{fig:SCNH_scanxa23_d180}
\end{figure}

  These qualitative conclusions are essentially confirmed by 
a thorough numerical analysis, the results of which are shown graphically 
in Fig. \ref{fig:SCNH_scanxa23_d180}. We next summarise briefly these 
results giving the ranges of $\alpha_{23}$ and $M_1$ 
of viable leptogenesis  in the four intervals of values of 
$\alpha_{23}$ identified earlier for $5^\circ < x < 90^\circ$.
We will do it for the representative value of
 $x=20.5^\circ$ at which some of the ranges of interest are maximal,
commenting first the results for the specific values 
of $\alpha_{23}$ at which viable leptogenesis occurs for 
the minimal for the case value of $M_1$.
%

\vspace{0.2cm}
\noindent
{\bf Region I: $0 < \alpha_{23} < 220^\circ$.}\\ 
The minimal  mass scale 
$M_{1\text{min}}$ for successful leptogenesis is 
found at $\alpha_{23} = 102^\circ$:  
$M_{1\text{min}} = 3.3 \times 10^{10}$ GeV.
For $\alpha_{23} = 102^\circ$:\\
i) viable leptogenesis is possible for 
$3.3\times10^{10}~{\rm GeV} \lesssim M_{1} \lesssim 7.7\times 10^{11}$ GeV; 
\\
ii) the maximal value of the baryon asymmetry $\eta_B$  
is reached at $M_1\simeq 2.55\times10^{11}$ GeV and reads 
$\eta_B^\text{max} \simeq 2.2\times10^{-9}$.
\\
The range of $M_1$ and the maximal value of 
$\eta_B$ change when $\alpha_{23}$ increases (decreases) from 
$102^\circ$ to $208^\circ$ ($64.5^\circ$), $208^\circ$ ($64.5^\circ$)
being the maximal (minimal) value of $\alpha_{23}$ at which it 
is possible to have successful leptogenesis (for $x =20.5^\circ$).
As  $\alpha_{23}$  increases to $\alpha_{23} = 170^\circ$,
the range of $M_1$ corresponding to $\alpha_{23}$ 
essentially shifts to larger values, with  $\eta_B^\text{max}$ 
remaining practically unchanged. More specifically,
at  $\alpha_{23} = 120^\circ$, for example,
the range of $M_1$ changes to 
$(3.3\times 10^{10} - 1.1\times 10^{12})$ GeV;
at $170^\circ$ it reads 
$(8.2\times 10^{10} - 1.9\times 10^{12})$ GeV. 
At
$\alpha_{23} = 120^\circ$ ($170^\circ$), we have
$\eta_B^\text{max} \simeq 2.75~(2.32)\times10^{-9}$, which  
occurs at $M_1 \simeq 3.6~(6.3)\times 10^{11}$ GeV.

As $\alpha_{23}$ increases further, $\eta_B^\text{max}$
and the range of $M_1$ begin to decrease.
At  $\alpha_{23} = 195^\circ$ we find
$\eta_B^\text{max} \simeq 1.2\times 10^{-9}$ and 
$1.7\times10^{11} \lesssim M_{1}/\text{GeV} \lesssim 1.7\times 10^{12}$,
with $\eta_B^\text{max}$ taking place at $M_1 \simeq 6.3\times 10^{11}$ GeV. 
Finally, at $\alpha_{23} \simeq 208^\circ$, 
$\eta_B^\text{max}$ coincides with the observed value 
of BAU and the related range of $M_1$ reduced to the point 
$M_1 \simeq  6.3\times10^{11}$. 

 We find a similar pattern when $\alpha_{23}$  decreases from 
$102^\circ$ to $64.5^\circ$. At  $\alpha_{23} = 90^\circ~(75^\circ)$, 
for example, we find for the range of interest of $M_1$:
$3.4\,(4.4)\times 10^{10}\lesssim M_1/\text{GeV}
\lesssim 5.2\,(2.7)\times 10^{11}$.
The maximal asymmetry is  $\eta_B^\text{max} \simeq 1.7~(1.0)\times 10^{-9}$ 
and occurs at $M_1 \simeq 2.1\,(1.4)\times 10^{11}$ GeV.
When $\alpha_{23}$ decreases further, the range of $M_1$ 
and  $\eta_B^\text{max}$ also decrease further and, e.g., at 
$\alpha_{23} = 65^\circ$ we have: 
$7.8\times 10^{10}\lesssim M_1/\text{GeV} \lesssim 1.2\times 10^{11}$,
$\eta_B^\text{max} \simeq 6.3\times 10^{-10}$ which  
occurs at $M_1 \simeq 9.8\times 10^{10}$ GeV.
At $64.5^\circ$ $\eta_B^\text{max}$ coincides with observed value 
of $\eta_B$ and the range of $M_1$ reduces to the point 
$M_1 \simeq 9.0\times10^{10}$ GeV.

 The interval of values of $\alpha_{23}$ and the related 
interval of values of $M_1$, 
where we can have successful leptogenesis, 
depend also on $x$, although this dependence is 
relatively weak. We find that at  $x=7.5^\circ$ 
we get the largest maximal (smallest minimal) value
$\alpha_{23}$ at which we still have successful leptogenesis 
in the considered range of $x$.
These values read: $\alpha^\text{min}_{23} = 37^\circ$, 
$\alpha^\text{max}_{23} = 233^\circ$. The ranges of $M_1$ 
corresponding to  $\alpha^\text{min}_{23}$ 
and  $\alpha^\text{max}_{23}$ are just the points 
$M_1 \simeq 2.98\times 10^{11}$ GeV  
and $M_1 \simeq 6.3\times10^{11}$ GeV, respectively. 
For $\alpha_{23} = 102^\circ$, for example, 
we have at $x=7.5^\circ$: 
i) the range of $M_1$ of successful leptogenesis 
is $6.8\times10^{10} \lesssim M_{1}/\text{GeV}\lesssim 1.4\times10^{12}$,
ii) $\eta_B^\text{max} \simeq 1.69\times10^{-9}$,
takes place at $M_1\simeq 4.33\times10^{11}$ GeV and
is bigger than the observed $\eta_B$ by the factor $C_{M1} = 2.77$.
These values should be compared with those given above for 
$x = 20.5^\circ$.
When $\alpha_{23}$ increases (decreases) to 
$233^\circ$ ($37^\circ$), we find quite similar behaviour 
of the correlation between the values of $\alpha_{23}$ and 
$M_1$ to that described for $x=20.5^\circ$, 
so we are not going to comment on it further.

\vspace{0.2cm}
\noindent
{\bf Region II: $220^\circ < \alpha_{23} < 360^\circ$.}\\
The minimal  mass scale $M_{1\text{min}}$ for successful leptogenesis is 
found at $\alpha_{23} = 301^\circ$:  
$M_{1\text{min}} = 3.1 \times 10^{12}$ GeV.
For $\alpha_{23} = 301^\circ$:\\
i)  the plateau of $\eta_B$ begins at $M_{1P}\simeq 8.9\times10^{13}$ GeV;\\
ii) the asymmetry at the plateau  
 $\eta_B \simeq 3.38\times10^{-9}$.\\
The asymmetry at the plateau is larger than the observed value 
of $\eta_B$ by the factor $C_{M2} \simeq 5.54$.
Knowing this factor allows us to determine the minimal and maximal values 
of $\alpha_{23}$ for having successful leptogenesis. The total range of 
values of $\alpha_{23}$ of interest reads:
 \be 
\label{eq:interval2}
 227^\circ  \lesssim \alpha_{23} \lesssim 352^\circ.
\ee
%
For the corresponding range of $M_1$ we get:
$3.11 \times 10^{12}\lesssim M_1/\text{GeV} \lesssim 10^{14}$. 
with the minimal value obtained for 
$\alpha_{23} \simeq 301^\circ$.

\vspace{0.2cm}
\noindent 
{\bf Region III: $360^\circ < \alpha_{23} < 500^\circ$.}\\
The minimal mass scale of leptogenesis is obtained for 
$\alpha_{23} = 433^\circ$. 
For this choice of $\alpha_{23}$:\\
i) leptogenesis is successful for 
 $1.54\times10^{11} \lesssim M_{1}/\text{GeV} \lesssim 1.59\times 10^{12}$, 
where the maximal $M_1$ corresponds also to $\alpha_{23} = 433^\circ$;\\
ii) the maximal value of $\eta_B$ is reached at 
$M_1\simeq 6.28\times10^{11}$ GeV and reads $\eta_B^\text{max} \simeq 1.37\times10^{-9}$.\\
The asymmetry $\eta_B$ at its maximum is greater than the observed 
value by a factor $C_{M3} \simeq 2.25$.  
The range of 
$\alpha_{23}$ of viable leptogenesis is
\be 
\label{eq:interval3}
382^\circ  \lesssim \alpha_{23} \lesssim 482^\circ\,. 
\ee
%

\vspace{0.2cm}
\noindent
{\bf Region IV: $500^\circ < \alpha_{23} < 720^\circ$.} \\
The minimal $M_1$ for having successful leptogenesis is found  
at $\alpha_{23} = 691^\circ$ and reads $M_{1}\simeq 1.53\times10^{11}$ GeV.
At this value of  $\alpha_{23}$,\\
i) the plateau of $\eta_B$ begins at
$M_{1}\simeq 4.98\times10^{12}$ GeV;\\
ii) the asymmetry at the plateau 
$\eta_B \simeq 4.11\times10^{-9}$.\\
This value is larger than the observed value of $\eta_B$ 
by the factor $C_{M4} \simeq 6.73$. 
Successful leptogenesis is possible for the following  range of 
values of $\alpha_{23}$: 
 \be 
\label{eq:interval4}
 506^\circ  \lesssim \alpha_{23} \lesssim 716^\circ. 
  \ee
%
For the quoted range of $\alpha_{23}$ we have viable 
leptogenesis for $1.54 \times 10^{11}\lesssim M_1/\text{GeV} \lesssim 10^{14}$. 
If $x\lesssim 5^\circ$, the lower bound in the interval \eqref{eq:interval4} 
is somewhat smaller at $\approx 470^\circ$, but successful 
leptogenesis is possible only for $M_1\gtrsim 10^{13}$ GeV.
\begin{figure}[t]
\centering
\includegraphics[width=12cm]{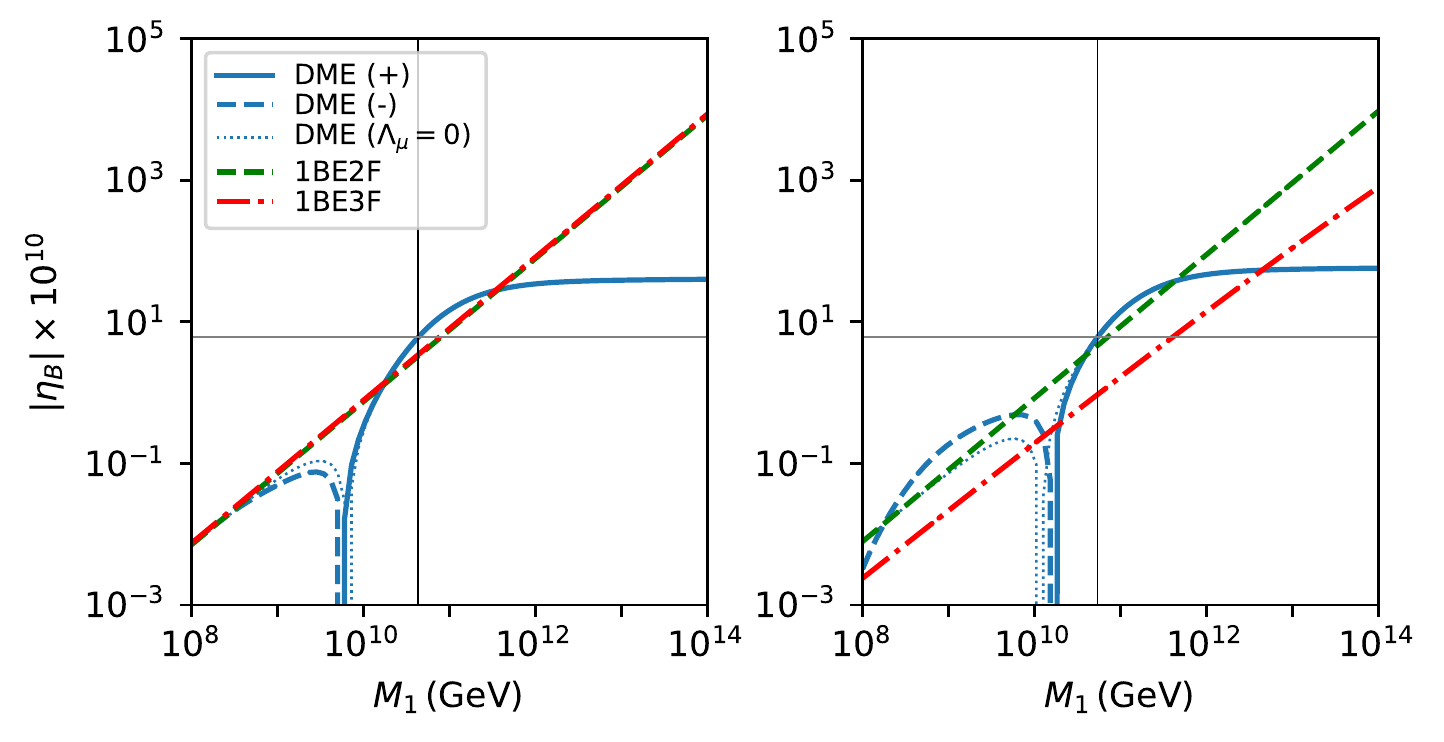} 
\caption{The asymmetry $\eta_B$ versus $M_1$  in the case of IH spectrum
with CP violation due to the Majorana phase $\alpha_{21}$ 
(Majorana and Dirac phases  $\alpha_{21}$ and $\delta$)
of the PMNS matrix.
The results shown in the left (right) panel are obtained for 
$x=k\pi$, $k=0,1,2$, $y\neq 0$ (purely imaginary $R_{11}R_{12}$) and
$\alpha_{21} = 164^\circ$ ($190^\circ$)
$\delta = \pi$ ($3\pi/2$) and
 $y=-\,76^\circ$ ($-\,50^\circ$).  
See text for further details.
}
\label{fig:MajCPVIHy-76_50(Minimal)070621}
\end{figure}
%

In all four cases we have discussed 
there exists a correlation between the value of the CPV phase 
$\alpha_{23}$ and the scale of leptogenesis $M_1$. 
Thus, obtaining constraints on 
$\alpha_{23}$ in low-energy experiments, in principle, can  
constrain the leptogenesis scale of the considered 
scenario, or even rule out this scenario.

\vspace{0.2cm}
\leftline{{\bf IH Spectrum}}

\vspace{0.2cm}
 Viable leptogenesis in the scenario of interest
is possible only for purely imaginary 
product $R_{11}R_{12}$ of elements of the $R$-matrix
($x=k\pi$, $k=0,1,2$, $y\neq 0$).
In Fig. \ref{fig:MajCPVIHy-76_50(Minimal)070621}
we show the modulus of $\eta_B$ versus $M_1$ 
for $\delta = \pi$ ($3\pi/2$)
and values of the  other parameters for which 
successful leptogenesis takes place for the minimal 
for the considered scenario value of $M_1$:
$\alpha_{21} = 164^\circ$ ($190^\circ$), 
$y=-\,76^\circ$ ($-\,50^\circ$),  
for the left (right) panel.
The right panel is obtained for $\delta = 3\pi/2$,
so it illustrates the case of viable leptogenesis with 
CP violation generated by both the Majorana and Dirac 
CPV phases of the PMNS matrix. It also illustrates the 
non-negligible effects the CP violating Dirac phase 
$\delta$ can have in leptogenesis 
when the CP violation is provided by the Majorana 
phase(s).
 
 In both cases illustrated in 
Fig.  \ref{fig:MajCPVIHy-76_50(Minimal)070621},
$\eta_B$ exhibits a ``non-standard'' behaviour 
as a function of $M_1$ going through zero in 
the 1-to-2 flavour transition at 
$M_1\sim 10^{10}~{\rm GeV} \ll 10^{12}$ GeV.
In the left (right) panel the minimal value 
of $M_1$ at which the calculated $\eta_B$ matches the 
observed value of $\eta_B$ is 
$M_1 = 4.3\times 10^{10}~(5.4\times 10^{10})$ GeV.
The plateau value of 
$\eta_B \cong 3.9\times 10^{-9}~(5.6\times 10^{-9})$ and 
is reached at $M_1\cong 1.1\times 10^{12}~(1.8\times 10^{12})$ GeV. 
\begin{figure}[t]
    \centering
\includegraphics[width=0.8\textwidth]{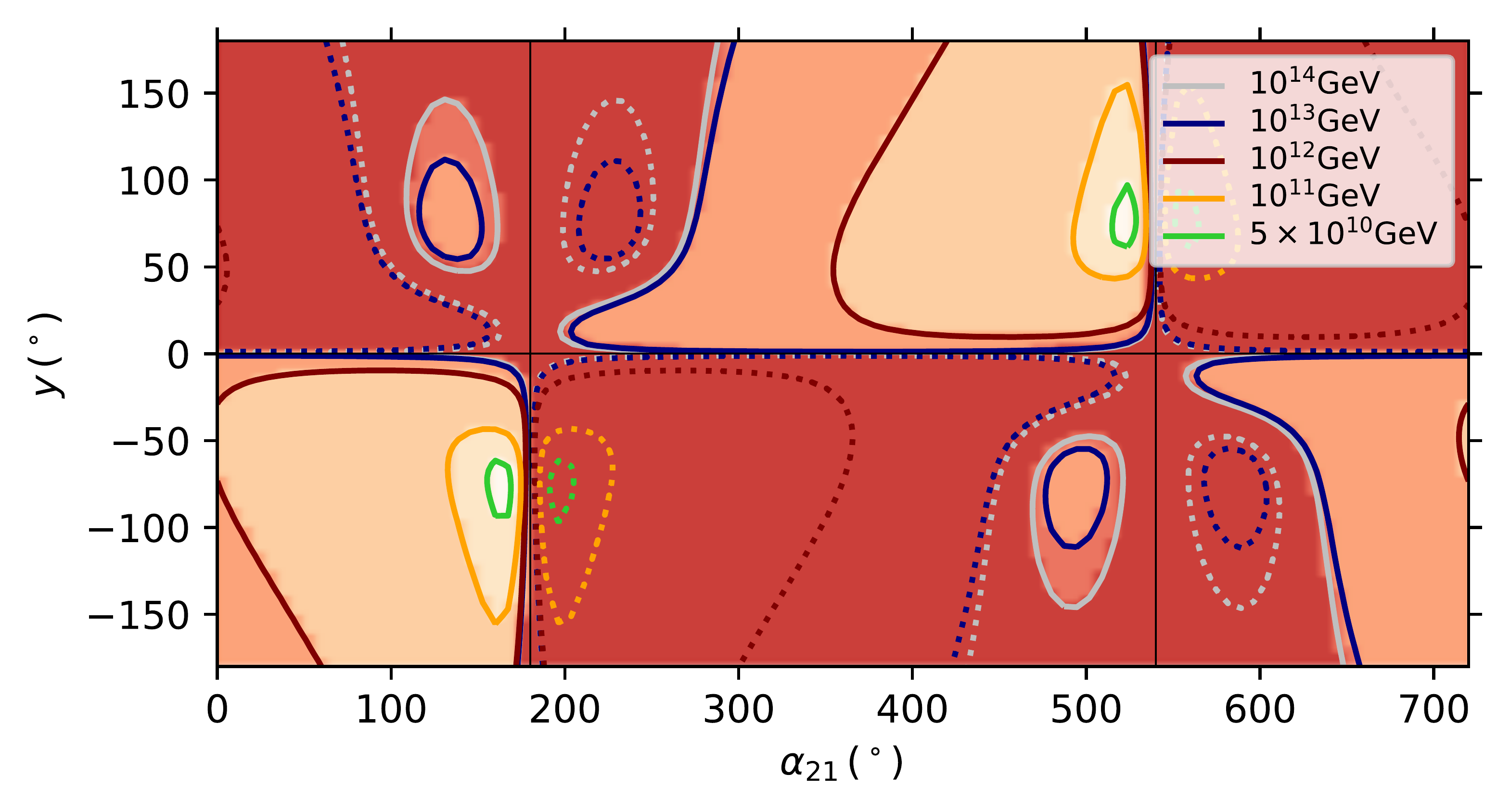}
\caption{The regions of successful leptogenesis 
in the  $\alpha_{21}-y$ plane in the case of IH spectrum and for 
$\delta = \pi$, purely imaginary $R_{11}R_{12}$ 
(i.e. $x = k\pi$, $k=0,\,1,\,2$, $y\neq$) 
and different $M_1$. 
The solid contours corresponding to 
fixed values of $M_1$ surround the regions 
where there exists a combination of values of 
$\alpha_{21}$ and $y$ for which $\eta_B = 6.1\times 10^{-10}$.
The dotted contours surround regions 
where one can have $|\eta_B| = 6.1\times 10^{-10}$ 
but $\eta_B < 0$. The predicted $\eta_B$
outside the contours  is smaller in 
magnitude than the present BAU.
The results for $0\leq y \leq 180^\circ$ 
can be obtained from those derived for 
$-\,180^\circ \leq y \leq 0$ by making 
the simultaneous change $ y\rightarrow -\,y$ 
and $\alpha_{21}\rightarrow \alpha_{21} \pm 2\pi$. 
See text for further details.
}
\label{fig:IHya21d180}
\end{figure}
%
Performing a detailed numerical analysis we have determined 
the regions of viable leptogenesis in the space of parameters 
in the considered scenario with IH spectrum, 
CP violation provided by $\alpha_{21}$, 
$x=k\pi$, $k=0,1,2$, $y\neq 0$
(purely imaginary  $R_{11}R_{12}$) 
and $\delta = \pi$. The values of 
$y$ were varied in the interval 
$[-\,180^\circ,+180^\circ]$. 
For $y=0$ we have, obviously, $\eta_B = 0$.
Due to symmetries of the quantities involved in the 
generation of $\eta_B$,
the results for $0 \leq y \leq 180^\circ$
can formally be obtained from those derived for 
$-\,180^\circ \leq y \leq 0$ by 
making the change $y\rightarrow -\,y$ and 
$\alpha_{21} \rightarrow \alpha_{21} \pm 2\pi$
and using the invariance  with respect 
to $\alpha_{21} \rightarrow \alpha_{21} \pm 4\pi$.
The results of this analysis are shown graphically 
in Fig. \ref{fig:IHya21d180}. A few comments 
are in order.

As we have already pointed out,
the minimal value of $M_1$ for successful leptogenesis 
is found to take place at  $y=-76^\circ$ and
$\alpha_{21} = 164^\circ$.
For $y=-76^\circ$ and $\alpha_{21} = 164^\circ$,
the plateau value of $\eta_B$ given earlier
is larger than the observed value of $\eta_B$ 
by the factor $C_{M5} \simeq 6.39$.
We can determine then the range of $\alpha_{21}$ 
for viable leptogenesis the case of
$y=-76^\circ$ and $\delta = \pi$ from the condition:
\be
\label{eq:conditiona21.1}
\cos\left(\alpha_{21}/2\right)(1-2p_{1\tau}(\alpha_{21}))
\simeq C_{M5}^{-1}\cos(82^\circ)(1-2p_{1\tau}(164^\circ))
\ee
%
For the range of interest we get:
\be
\label{eq:intervala21IH.1}
0 \leq \alpha_{21}\lesssim 177.5^\circ \quad \text{and} \quad 632^\circ\lesssim\alpha_{21}\leq 720^\circ
\ee
%
When $y$ increases from $(-76^\circ)$ to near zero, the lower bound of 
the right interval decreases to $\simeq 560^\circ$.
The corresponding range of values of $M_1$ 
extends from $M_1 \simeq 4.3\times10^{10}$ GeV to 
the beginning of the plateau 
at  $M_1\cong 1.1\times 10^{12}$ and further on the plateau 
at least to $M_1 \simeq 10^{14}$ GeV. 
When $M_1$ decreases 
starting from  $M_1\simeq 1.1\times 10^{12}$ at the plateau to
$M_1 \simeq 4.3\times10^{10}$ GeV, the intervals of values of 
$\alpha_{21}$ of successful leptogenesis also decrease and at 
$M_1 \simeq 4.3\times10^{10}$ GeV shrink to the point 
$\alpha_{21} \simeq 164^\circ$. 

 There is an additional relatively small 
region of values of $\alpha_{21}$  and $M_1$ 
around $\alpha_{21}\sim 500^\circ$, for which it is possible to reproduce 
the observed value of BAU. It is clearly seen in Fig. \ref{fig:IHya21d180}.  
In this case the minimal mass scale 
takes place at $\alpha_{21}\simeq 493^\circ$, $y\simeq -\,76^\circ$ 
and reads $M_1 \simeq 5.9\times 10^{12}$ GeV. 
At the plateau we have $\eta_B \simeq C_{M6}\,6.1\times 10^{-10}$, 
with $C_{M6}\simeq 1.83$. The range of $\alpha_{21}$ of viable 
leptogenesis can be determined using the equation:
\be
\cos\left(\alpha_{21}/2\right)\,(1 - 2p_{1\tau}(\alpha_{21})) 
\simeq C^{-1}_{M6}\, \cos(246.5^\circ)\,(1 - 2p_{1\tau}(493^\circ))\,. 
\ee
%
Solving the preceding equation we get for the range of $\alpha_{21}$:
\be
468^\circ \lesssim \alpha_{21} \lesssim 522^\circ\,.
\ee
%
The corresponding range of the leptogenesis scale is 
$5.9\times 10^{12}\lesssim M_1/\text{GeV} \lesssim 10^{14}$.

 As in the previous scenarios discussed by us, also in 
the scenario considered in the present subsection 
there is a correlation between the low-energy phase -- in this case 
$\alpha_{21}$ -- responsible for the CP violation in leptogenesis and the 
scale of leptogenesis $M_1$. Clearly, obtaining constraints on 
$\alpha_{21}$ in low-energy experiments can rule out the considered scenario 
or constrain the leptogenesis scale of the scenario.

\section{Summary and Conclusions}
\label{sec:concs}
%

We have considered the generation of the baryon asymmetry 
of the Universe $\eta_B$ in the high (GUT) scale leptogenesis based on 
type I seesaw mechanism of neutrino mass generation.
Using the density matrix equations (DME) for high scale leptogenesis 
in which the CP violation is provided by the low-energy Dirac 
or/and Majorana phases of the neutrino mixing (PMNS) matrix, 
we have investigated the 1-to-2 and the 2-to-3 flavour regime 
transitions, where the 1, 2 and 3 flavour regimes are  
described by the Boltzmann equations. Concentrating on the 
1-to-2 flavour transitions in leptogenesis with three heavy Majorana 
neutrinos $N_{1,2,3}$ with hierarchical mass spectrum, 
$M_1 \ll M_2 \ll M_3$, we have determined the general conditions 
under which the baryon asymmetry $\eta_B$ goes through zero and changes 
sign in the transition. 
We have shown, in particular, that the asymmetry $\eta_B$ goes 
through zero changing its sign when leptogenesis proceeds in 
the strong wash-out regime with zero initial abundance 
of the heavy Majorana neutrinos.

 In order to make the discussion of 
all the salient features of the transitions between the different 
flavour regimes of interest as transparent as possible, 
we have investigated further the case of decoupled 
heaviest Majorana neutrinos $N_3$, in which the number 
of parameters in leptogenesis is significantly smaller than 
in the general case of three heavy Majorana neutrinos. 
In particular, the complex orthogonal matrix $R$, 
which makes part of the Casas-Ibarra parametrisation 
of the neutrino Yukawa couplings we have employed 
in the analysis, depends only on one complex angle 
$\theta_{\rm CI} = x + i\,y$, where $x$ and $y$ are 
real parameters.  With only two 
heavy Majorana neutrinos ($N_{1,2}$) active in the seesaw mechanism, 
the light neutrino mass spectrum can only be either normal 
hierarchical (NH) with $m_1\cong 0$, or inverted hierarchical (IH), 
with $m_3 \cong 0$. Furthermore, in the case of interest 
of CP violation in leptogenesis provided only  
by the low-energy CPV phases of the PMNS matrix, 
one can avoid the contributions to the 
CP violation associated with the $R$-matrix 
only if the angle $\theta_{\rm CI}$ is 
such that for NH (IH) spectrum 
$\sin2\theta_{\rm CI}\neq 0$ 
is real (purely imaginary) \cite{Pascoli:2006ci}, 
$\sin2\theta_{\rm CI} = \sin2x$
($\sin2\theta_{\rm CI} = \pm i\,\sinh2y$).

  Analysing in detail the behaviour of $\eta_B$ 
in the transition in the case of two heavy Majorana 
neutrinos $N_{1,2}$ with hierarchical masses, $M_1 \ll M_2$,
allowed us to gain a better understanding of the transitions but also 
to discover new unexpected features of the transitions.  
We have found, in particular, that:\\ 
i) the Boltzmann equations  in many  cases 
fail to describe correctly the generation of the baryon asymmetry 
$\eta_B$ in the 1, 2 and 3 flavour regimes, in particular, 
underestimating $\eta_B$ by a factor $\sim 10$ in certain cases; \\  
ii) depending on the values of the relevant parameters, 
the transitions between the different flavour regimes 
can be ``non-standard'' as, e.g., the 1-to-2 
and the 2-to-3 flavour transitions can take place 
at the same $M_1$, with $\eta_B$ going through a relatively 
shallow minimum at the transition value of $M_1$;\\
iii) the two-flavour regime can persist above  
$5\times 10^{11} - 10^{12}$ GeV (below $\sim 10^9$ GeV), and \\
iv) the flavour effects in leptogenesis persist beyond 
what is usually thought to be the maximum leptogenesis scale for 
these effects of $\sim 10^{12}$ GeV, 
with the requisite CP violation provided by the Dirac or/and 
Majorana phases present in the low-energy PMNS neutrino mixing matrix.

 At $M_1 \sim 10^{12}$ GeV, $|\eta_B|$  
reaches a ``plateau'' where 
it remains practically constant 
as $M_1$ increases and flavour effects are fully operative.
We further have determined the minimal scale $M_{1\text{min}}$ 
at which we can have successful leptogenesis 
when the CP violation is provided only by the Dirac ($\delta$) 
or Majorana ($\alpha_{23}\equiv \alpha_{21} - \alpha_{31}$ or $\alpha_{21}$) phases 
of the PMNS matrix as well as the ranges of the scales 
and the values of the phases for having successful leptogenesis.
In the case of Dirac phase CP violation we found that 
$M_{1\text{min}} \cong 2.5~(1.7)\times 10^{11}$ GeV for 
NH light neutrino mass spectrum 
and $\delta$ lying in the interval $\pi < \delta < 2\pi$ 
($0 < \delta < \pi$) for real (purely imaginary) 
Casas-Ibarra parameter $\sin2\theta_{\rm CI}$  
(Figs. \eqref{fig:NHscanxd_a21_0and360_a31_0} 
and \eqref{fig:SCNH_scanyd_a21_180_a31_0}).   
As $M_{1}$ increases from 
$M_{1\text{min}}$ to $M_{1} \cong 2.7~(2.1)\times 10^{12}$ GeV, 
at which $\eta_B$ reaches the plateau value,
the range of interest of $\delta$  increases from 
the point $\delta = 3\pi/2~(\pi/2)$ to 
$202.4^\circ \lesssim \delta \lesssim 337.6^\circ$ 
($14.6^\circ \lesssim \delta \lesssim 166.4^\circ$) 
and remains practically the same up to $M_1\sim 10^{14}$ GeV. 
We get similar results for  $M_{1\text{min}}$ and the ranges
of $\delta$ and $M_1$ in the case of IH light neutrino mass  
spectrum and purely imaginary Casas-Ibarra 
parameter $\sin2\theta_{\rm CI}$, 
with $M_{1\text{min}} \cong 4.6\times 10^{10}$ GeV for 
$\delta = 3\pi/2$ and  $185^\circ \lesssim \delta \lesssim 355^\circ$ 
for $1.2\times 10^{12}~{\rm GeV}\lesssim  M_1 \lesssim 10^{14}$ GeV 
(Fig. \eqref{fig:SCIH_scanyd_a21_180_a31_0}).
We found also that in the case of NH spectrum 
there is a direct relation 
between the sign of $\sin \delta$ 
and the sign of the baryon asymmetry of the Universe 
in the regions of viable leptogenesis; 
for IH spectrum such a relation holds for 
$M_{1\text{min}} \cong  4.6\times 10^{10} \text{ GeV} \lesssim M_1 \lesssim 10^{13}$ GeV.

  We have investigated also the generation of $\eta_B$ 
when the CP violation is provided solely by the Majorana 
phases  $\alpha_{21}/2$ and $\alpha_{31}/2$
of the PMNS matrix. In the considered scenario with two heavy Majorana 
neutrinos only the phase 
$\alpha_{23} \equiv \alpha_{21} - \alpha_{31}$ ($\alpha_{21}$)
is physically relevant for NH (IH) light neutrino 
mass spectrum. We have performed a thorough analysis 
and have determined the ranges of 
values of the Majorana phase $\alpha_{23}$ ($\alpha_{21}$) 
and the related ranges of the scale $M_1$,   
for which we can successful leptogenesis 
with CP violation provided exclusively by $\alpha_{23}$ ($\alpha_{21}$) 
in the case of NH (IH) spectrum, real (purely imaginary) 
Casas-Ibarra factor $\sin2\theta_{\rm CI} = \sin2x\neq 0$ 
($\sin2\theta_{\rm CI} = \pm i\,\sinh2y \neq 0$) 
and CP conserving value of $\delta = \pi$ 
(the results of these analyses are presented graphically 
in Figs. \eqref{fig:plateau_SCNH_d_180_a21L270_a21R_700},
\eqref{fig:SCNH_scanxa23_d180}, \eqref{fig:MajCPVIHy-76_50(Minimal)070621}
and \eqref{fig:IHya21d180}). 
Our results show, in particular, that there exist relatively 
large regions of the relevant spaces of parameters where it is 
possible to reproduce the observed value of BAU as well as 
that in these regions the values of the respective Majorana phases
providing the requisite leptogenesis CP violation 
are strongly correlated with the value of leptogenesis scale $M_1$.   

 We note that in \cite{Xing:2020erm}, the RGE-corrections to the neutrino 
Yukawa matrices are included in the calculations of the CPV 
lepton asymmetry through a modified Casas-Ibarra parametrisation.
Without these corrections,  within the formalism 
employed in  \cite{Xing:2020erm}, 
which is based on the Boltzmann equations,
all lepton flavour dependence 
(contained in the PMNS matrix) cancels at leptogenesis
scales $M_{\rm LG}\gtrsim 10^{12}$ GeV from the total CPV asymmetries. 
Once the RGE-corrections are included, this cancellation 
no longer occurs and a corrective term approximately proportional to 
the square of the $\tau$-Yukawa coupling is added to the total CPV asymmetry. 
The authors of \cite{Xing:2020erm} then show that this correction is 
sufficient in some regions of the parameter space at 
$M_{\rm LG}\gtrsim 10^{12}$ GeV for successful leptogenesis 
from purely low-scale CP violation due to the phases in the PMNS matrix 
in the absence of the usual flavour effects. 
However, this correction is subdominant to flavour effects discussed 
in this work, typically being by a factor of $\sim 10$ to $\sim 100$ smaller 
in the regions of the parameter space of the scenarios 
we have considered, where leptogenesis successfully generates the 
observed BAU. 
Thus, the mechanism of generation of BAU considered in 
\cite{Xing:2020erm} is subdominant to the mechanism discussed 
in this work.

 It follows from the results obtained in the present article that, 
in particular,  viable leptogenesis based on type I seesaw mechanism with two 
hierarchical in mass heavy Majorana neutrinos 
and CP violation provided by the physical low-energy 
Dirac or/and Majorana phases, present in the PMNS neutrino mixing matrix,  
is possible for rather wide ranges of values of the CP violating 
phases and of the scale of leptogenesis.
The scenarios of leptogenesis investigated by us  
are obviously falsifiable in low-energy experiments on the nature -- 
Dirac or Majorana -- of massive neutrinos. 
As far as the nature of massive neutrinos is not known or if the
massive neutrinos are proven to be Majorana particles, 
the cases of leptogenesis we have considered are still 
testable and falsifiable in low-energy experiments 
on CP violation in neutrino oscillations, on determination of the type 
of spectrum neutrino masses obey and of the absolute neutrino mass scale.
The data from these experiments can severely constrain the corresponding 
leptogenesis parameter spaces 
and even rule out some of, if not all, the cases studied in detail 
by us. We are looking forward to these experimental data 
that can provide crucial tests of the leptogenesis 
scenarios discussed in the present paper.

\section*{Acknowledgements}

This work was supported in part by the INFN program on Theoretical 
Astroparticle Physics (A.G. and S.T.P.) and by the  World Premier 
International Research Center   Initiative (WPI Initiative, MEXT), 
Japan (S.T.P.). K.M. acknowledges the (partial) support from 
the European Research Council under the European Union Seventh 
Framework Programme (FP/2007-2013) / ERC Grant NuMass agreement n. [617143].

\appendix

\section{DMEs from the Three- to the Two-Flavour Basis}
\label{FromDMEs}

We consider here the scenario of two-flavoured leptogenesis within the DMEs. This is equivalent to set $\Gamma_\mu/Hz = 0$ in the DMEs defined in Eqs. \eqref{DME:N}-\eqref{DME:full3}. We describe here in detail how to arrive from the DMEs in the three-flavour basis in Eqs. \eqref{DME:N}-\eqref{DME:full3} to the ones in the two-flavour basis given in Eqs. \eqref{DMEN}-\eqref{DMEtautaup}. We also discuss how the formal solution to those is obtained and how, under the single-flavour approximation, they recover the single-flavoured Boltzmann equations in Eqs. \eqref{1BE1F:N}-\eqref{1BE1F:BL}.

In the calculations that follow, it will prove convenient to use the CP-asymmetry in Eqs. \eqref{eq:eps1ab} also for the $\tau^\perp$-flavour and for the $i^\text{th}$ heavy Majorana neutrino $N_i$, which, in terms of the $C_{i\alpha}$ coefficients defined in Eq. \eqref{eq:Cia}, is given by
\begin{equation}\label{eq:epsiabtaup}
\begin{aligned}
\epsilon^{(i)}_{\alpha\beta}&=\frac{3}{32\pi}
\sum_{j\neq i}\left(Y^{\dagger} Y\right)_{jj}\Bigg\{i\left[C_{i \alpha}C^{*}_{j\beta}(C^{\dagger}C)_{ji}
- C^{*}_{i \beta}C_{j \alpha}(C^{\dagger}C)_{ij}\right] f_1\left(\frac{x_{j}}{x_{i}}\right) \\
&+i\left[C_{i \alpha}C^{*}_{j \beta}(C^{\dagger}C)_{ij}-C^{*}_{i \beta}C_{j \alpha}(C^{\dagger}C)_{ji}\right] f_2\left(\frac{x_{j}}{x_{i}}\right) \Bigg\},
 \end{aligned}
\end{equation}
with $i = 1,\,2,\,3$ and $\alpha,\,\beta = e,\,\mu,\,\tau,\,\tau^\perp$.
Explicitly, the diagonal terms are
\begin{equation}\label{eq:epsiaataup}
\epsilon^{(i)}_{\alpha\alpha}=\frac{3}{16\pi}
\sum_{j\neq i}\left(Y^{\dagger} Y\right)_{jj}\Bigg\{\Im\left[C_{i \alpha}^*C_{j\beta}(C^{\dagger}C)_{ij}\right] f_1\left(\frac{x_{j}}{x_{i}}\right)
+\Im\left[C_{i \alpha}^*C_{j \beta}(C^{\dagger}C)_{ji}\right] f_2\left(\frac{x_{j}}{x_{i}}\right) \Bigg\}.
\end{equation}
We remind also that $\left(Y^\dagger Y\right)_{jj} = \sum_{\gamma} |Y_{\gamma j}|^2$, $\left(C^\dagger C\right)_{ij} = \sum_{\gamma} C_{i \gamma}^*C_{j \gamma}$ and $p_{i\gamma} \equiv |C_{i\gamma}|^2$ with $\gamma = e,\,\mu,\,\tau$, while $p_{i\tau^\perp} \equiv p_{ie} + p_{i\mu}$ and $p_{i\tau} + p_{i\tau^\perp} = p_{i\tau} +  p_{ie} + p_{i\mu} = 1$.
To use Eqs. \eqref{eq:epsiabtaup} and \eqref{eq:epsiaataup} we should define the coefficients $C_{i\tau^\perp}$ and $C_{i\tau^\perp}^*$. However, the relation $\epsilon_{\tau^\perp \tau^\perp}^{(i)} \equiv \epsilon_{ee}^{(i)} + \epsilon_{\mu\mu}^{(i)} $ imposes that
\be\label{eq:CiCj}
C_{i\tau^\perp}C_{j\tau^\perp}^* = C_{ie}C_{je}^* + C_{i\mu}C_{j\mu}^*,
\ee
which for $i=j$ means $|C_{i\tau^\perp}|^2 = |C_{ie}|^2 + |C_{i\mu}|^2 = p_{i\tau^\perp}$. Since the physical quantities (e.g., $N_{B-L}$) depend on $|C_{i\tau^\perp}|^2$ (in our case of interest $i=1$), there is actually no need to define the coefficients $C_{i\tau^\perp}$ and $C_{i\tau^\perp}^*$, apart from imposing the constraint in Eq. \eqref{eq:CiCj}, so we are going to let them free in our calculations.

An important relation that derives from Eqs. \eqref{eq:epsiabtaup} and \eqref{eq:epsiaataup} and that is going to be used further is
\be
2\Re\left[C_{i\beta}C_{i\alpha}^* \epsilon_{\alpha\beta}^{(i)}\right] = p_{i\beta}\epsilon_{\alpha\alpha}^{(i)} + p_{i\alpha}\epsilon_{\beta\beta}^{(i)}\,,
\ee
with $\alpha,\,\beta = e,\,\mu,\,\tau,\,\tau^\perp$.

We now concentrate again on the hierarchical case for which only the decay of the heavy neutrino $N_1$ is relevant for leptogenesis (i.e. $i = 1$).
Firstly, we sum the equations for $N_{ee}$ and $N_{\mu\mu}$, which result from taking $\alpha = \beta = e$, $\mu$ in Eqs. \eqref{DME:full3} respectively, and get an equation for $N_{\tau^\perp\tau^\perp}= N_{ee} + N_{\mu\mu}$:
\be
\begin{split}
\frac{dN_{\tau^\perp\tau^\perp}}{dz} = &\, \epsilon_{\tau^\perp\tau^\perp}^{(1)} D_1 (N_{N_1} - N_{N_1}^{eq}) +\\
&-W_1 \Big\{p_{1e}N_{ee}+p_{1\mu}N_{\mu\mu}+ 2 \Re\left[C_{1e} C_{1\mu}^* N_{\mu e}\right]+\\
& + \Re\left[C_{1e}C_{1\tau}^*N_{\tau e}\right] +\Re\left[C_{1\mu}C_{1\tau}^*N_{\tau \mu}\right]\Big\}.
\end{split}
\ee
The second line of the above equation is actually $p_{1\tau^\perp}N_{\tau^\perp\tau^\perp}$. This can be shown by considering the equations for $p_{1\mu}N_{ee}$, $p_{1e}N_{\mu\mu}$ and $2\Re\left[C_{1e}C_{1\mu}^*N_{ee}\right]$, from which we can write:
\be
\begin{split}
2\Re\left[C_{1e}C_{1\mu}^*\frac{dN_{\mu e}}{dz}\right] =&\; 2\Re\left[C_{1e}C_{1\mu}^*\epsilon_{\mu e}^{(1)}\right] D_1 (N_{N_1} - N_{N_1}^{eq}) \;+ \\
&- W_1 \Re\left[C_{1e}C_{1\mu}^*\left\{P^{0(1)},N\right\}_{\mu e}\right]\\
=&\;(p_{1\mu}\epsilon_{ee}^{(1)} + p_{1e} \epsilon_{\mu\mu}^{(1)}) D_1 (N_{N_1} - N_{N_1}^{eq})\; + \\
&- W_1 \Big\{p_{1e}p_{1\mu}(N_{ee}+N_{\mu\mu})\; +\\ 
&+ (p_{1e} + p_{1\mu})\Re\left[C_{1e}C_{1\mu}^* N_{\mu e}\right]+p_{1\mu}\Re\left[C_{1e}C_{1\tau}^* N_{\tau e}\right]+p_{1e}\Re\left[C_{1\mu}C_{1\tau}^* N_{\tau \mu}\right]\Big\}\;\\
=& \; p_{1\mu}\frac{dN_{ee}}{dz} + p_{1e}\frac{dN_{\mu\mu}}{dz}\,.
\end{split}
\ee
By assuming that at the beginning of leptogenesis ($z_{0}$) all the asymmetries are zero, the following condition must hold at any $z\geq 0$: \footnote{We stress that this is only valid if $\Gamma_\mu/Hz = 0$, as in our case.}
\be
2\Re\left[C_{1e}C_{1\mu}^*N_{\mu e}\right] = p_{1\mu}N_{ee} + p_{1e}N_{\mu\mu}\,,
\ee
which leads to 
\be
p_{1e}N_{ee}+p_{1\mu}N_{\mu\mu}+ 2 \Re\left[C_{1e} C_{1\mu}^* N_{\mu e}\right] = p_{1\tau^\perp}N_{\tau^\perp\tau^\perp}\,.
\ee
We then define
\be\label{eq:NtautaupDEF}
N_{\tau\tau^\perp} \equiv \left(\frac{C_{1e}}{C_{1\tau^\perp}}N_{\tau e} + \frac{C_{1\mu}}{C_{1\tau^\perp}}N_{\tau\mu}\right)
\ee
and $N_{\tau^\perp\tau} = N_{\tau\tau^\perp}^*$, so that the equations for $N_{\tau\tau}$ and $N_{\tau^\perp\tau^\perp}$ can be recast in the forms given in Eqs. \eqref{DMEtau} and \eqref{DMEtaup}.
By using the relation $C_{i\tau^\perp}\epsilon_{\tau\tau^\perp}^{(i)} = C_{ie}\epsilon_{\tau e}^{(i)} + C_{i\mu}\epsilon_{\tau \mu}^{(i)}$ (with $i=1$ in our case), which follows from Eqs. \eqref{eq:epsiabtaup} and \eqref{eq:CiCj}, combined with all the previous relations, we get the equation for $N_{\tau\tau^\perp}$ as in Eq. \eqref{DMEtautaup} and the DMEs in the two-flavour basis are recovered.

The formal expression of $N_{\tau\tau^\perp}$ can be obtained by solving Eq. \eqref{DMEtautaup} with the integrating factor method, which leads to
\be\label{eq:Soltautaup}
\begin{split}
N_{\tau\tau^\perp}(z) = \epsilon_{\tau\tau^\perp}^{(1)}\int_{z_0}^z D_1(z') (N_{N_1}(z')-N_{N_1}^{eq}(z'))e^{-\Lambda_\tau(z-z')}e^{-\frac{1}{2}\int_{z'}^z W_1\,dz''}\,dz' + \\
- \frac{1}{2}C_{1\tau}C_{1\tau^\perp}^*\int_{z_0}^z W_1(z')N_{B-L}(z')e^{-\Lambda_\tau(z-z')}e^{-\frac{1}{2}\int_{z'}^z W_1(z'')\,dz''}\,dz',
\end{split}
\ee
where the initial asymmetry was assumed to be zero, namely $N_{\tau\tau^\perp}(z_0) = 0$.
Notice that the above expression contains a term with $N_{B-L}$ which cannot be ignored in general, if not, e.g., in the limit of $\Gamma_\tau/Hz \to 0$ (see Sec. \ref{sec:sign}).

To get the resulting equation for $N_{B-L} = N_{\tau\tau} + N_{\tau^\perp\tau^\perp}$ we first notice that:
\be
\begin{split}
2\Re\left[C_{1\tau^\perp}C_{1\tau}^*\frac{dN_{\tau\tau^\perp}}{dz}\right] =&\; 2\Re\left[C_{1\tau^\perp}C_{1\tau}^*\epsilon_{\tau\tau^\perp}^{(1)}\right] D_1(N_{N_1}-N_{N_1}^{eq}) \;+\\ &-W_1\left\{\Re\left[C_{1\tau^\perp}C_{1\tau}^*N_{\tau\tau^\perp}\right] + p_{1\tau}p_{1\tau^\perp}N_{B-L}\right\}- 2\Re\left[C_{1\tau^\perp}C_{1\tau}^*N_{\tau\tau^\perp}\frac{\Gamma_\tau}{Hz}\right]\\
=&\;(p_{1\tau^\perp}\epsilon_{\tau\tau}^{(1)} + p_{1\tau}\epsilon_{\tau^\perp\tau^\perp}^{(1)}) D_1(N_{N_1}-N_{N_1}^{eq}) \;+\\ &-W_1\left\{\Re\left[C_{1\tau^\perp}C_{1\tau}^*N_{\tau\tau^\perp}\right] + p_{1\tau}p_{1\tau^\perp}N_{B-L}\right\}- 2\Re\left[C_{1\tau^\perp}C_{1\tau}^*N_{\tau\tau^\perp}\frac{\Gamma_\tau}{Hz}\right]\\
\end{split}
\ee
Then we write the equation for $p_{1\tau^\perp}N_{\tau\tau} + p_{1\tau}N_{\tau^\perp\tau^\perp}$, that is:
\be
\begin{split}
\frac{d}{dz}(p_{1\tau^\perp}N_{\tau\tau} + p_{1\tau}N_{\tau^\perp\tau^\perp}) =&\;
(p_{1\tau^\perp}\epsilon_{\tau\tau}^{(1)} + p_{1\tau}\epsilon_{\tau^\perp\tau^\perp}^{(1)}) D_1(N_{N_1}-N_{N_1}^{eq}) \;+\\
&-W_1\left\{\Re\left[C_{1\tau^\perp}C_{1\tau}^*N_{\tau\tau^\perp}\right] + p_{1\tau}p_{1\tau^\perp}N_{B-L}\right\}\\
=&\; 2\Re\left[C_{1\tau^\perp}C_{1\tau}^*\frac{dN_{\tau\tau^\perp}}{dz}\right] + 2\Re\left[C_{1\tau^\perp}C_{1\tau}^*N_{\tau\tau^\perp}\frac{\Gamma_\tau}{Hz}\right]
\end{split}
\ee
Then, given that
\be
p_{1\tau^\perp}N_{\tau\tau} + p_{1\tau}N_{\tau^\perp\tau^\perp} = N_{B-L} - (p_{1\tau}N_{\tau\tau} + p_{1\tau^\perp}N_{\tau^\perp\tau^\perp})
\ee
we get
\be
p_{1\tau}\frac{dN_{\tau\tau}}{dz} + p_{1\tau^\perp}\frac{dN_{\tau^\perp\tau^\perp}}{dz} + 2\Re\left[C_{1\tau^\perp}C_{1\tau}^*\frac{dN_{\tau\tau^\perp}}{dz}\right] = \frac{dN_{B-L}}{dz} - 2\Re\left[C_{1\tau^\perp}C_{1\tau}^*N_{\tau\tau^\perp}\frac{\Gamma_\tau}{Hz}\right]\,.
\ee
Since all the asymmetries are assumed to be zero at $z_0$, the above relation converts to
\be
\begin{split}
p_{1\tau}N_{\tau\tau}(z)+ p_{1\tau^\perp}N_{\tau^\perp\tau^\perp}(z) + 2\Re\left[C_{1\tau^\perp}C_{1\tau}^*N_{\tau\tau^\perp}(z)\right] = &\; N_{B-L}(z) - \lambda(z)\,,
\end{split}
\ee
with $\lambda(z)$ defined as in Eq. \eqref{lambda}.
We note that $\lambda(z) = 0$ in the single-flavour approximation, namely for $\Gamma_\tau/Hz = 0$.

Finally, by summing Eqs. \eqref{DMEtau} and \eqref{DMEtaup} and using the previous relations we get an equation for $N_{B-L}$ that reads:
\be\label{eq:NB-LFULL}
\frac{dN_{B-L}}{dz} = \epsilon^{(1)} D_1(z) (N_{N_1}(z) - N_{N_1}^{eq}(z)) - W_1(z) N_{B-L}(z) + W_1(z) \lambda(z).
\ee
In the case of $\lambda(z) = 0$, the above equation corresponds to the Boltzmann equation for the $B-L$ asymmetry in the single-flavour approximation given in Eq. \eqref{1BE1F:BL}. Moreover, when $\epsilon^{(1)} = 0$, as in the case of CP violation solely provided by the PMNS phases, Eq. \eqref{eq:NB-LFULL} reduces to Eq. \eqref{dNB-Ldz}.

The formal solution to Eq. \eqref{eq:NB-LFULL} reads:
\be
\begin{aligned}
N_{B-L} (z) =& \int_{z_0}^{z} e^{-\int_{z'}^{z_f}W_1(z'')dz''}\epsilon^{(1)}D_1(z')(N_{N_1}(z')-N_{N_1}^{eq}(z'))\,dz'\\
&+\int_{z_0}^z W_1(z')\lambda(z')e^{-\int_{z'}^z W_1\, dz''}\,dz',
\end{aligned}
\ee
where, as usual, we have assumed vanishing initial asymmetry $N_{B-L}(z_0) = 0$.

\section{Approximated Solutions to the BEs in Various Regimes}\label{Approximations}
In this appendix we illustrate the passages that lead to the analytical approximations to the Boltzmann Equations (BEs) in various regimes. Useful references with similar calculations are \cite{BUCHMULLER2005305, Fong_2012}. 
The BEs are:

\begin{eqnarray}
\label{BEN}
\frac{dN_{N_1}}{dz} &=& - D_1 (N_{N_1} - N_{N_1}^{eq}),\\
\label{BEalpha}
\frac{dN_{\alpha\alpha}}{dz} &=& \epsilon_{\alpha\alpha}^{(1)}D_1(N_{N_1}-N_{N_1}^{eq}) - W_1 p_{1\alpha} N_{\alpha\alpha},
\end{eqnarray}
where $\alpha = \tau,\,\tau^\perp$ or $e,\,\mu,\,\tau$ in the two- or three-flavour basis, respectively. The single-flavour BEs can be recovered by formally substituting $N_{\alpha\alpha}$ with $N_{B-L}$ and setting $p_{1\alpha} = 1$ in \eqref{BEalpha}.
The strength of the decays and inverse decays is quantified by $\kappa_1 p_{1\alpha}$. When $\kappa_1 p_{1\alpha} \gg 1$, the flavour $\alpha$ is said to be in the \textit{strong} wash-out regime. Conversly, if $\kappa_1 p_{1\alpha} \ll 1$, the flavour $\alpha$ is in the \textit{weak} wash-out regime.
The formal solution to the BEs can be found by means of the integrating factor method:

\begin{eqnarray}
\label{BEsolN}
N_{N_1}(z) &=& N_{N_1}(z_0)e^{-\int_{z_0}^z D_1(z')\,dz'} + \int_{z_0}^z D_1(z') N_{N_1}^{eq}(z')e^{-\int_{z'}^z D_1(z'')\,dz''}\,dz',\\
\label{BEsolalpha}
N_{\alpha\alpha}(z) &=&N_{\alpha\alpha}(z_0) 
\begin{aligned}[t]
&e^{-\int_{z_0}^z W_1(z')p_{1\alpha}\,dz'} +\\
&+\; \epsilon_{\alpha\alpha}^{(1)}\int_{z_0}^z D_1(z') \left(N_{N_1}(z')-N_{N_1}^{eq}(z')\right)e^{-\int_{z'}^z W_1(z'')p_{1\alpha}\,dz''}\,dz'.
\end{aligned}
\end{eqnarray}
Assuming zero asymmetry at $z_0$, the first term in Eq. \eqref{BEsolalpha} vanishes.

\subsection{Strong Wash-Out Regime}\label{App:SW}
In the strong wash-out regime for a certain lepton flavour $\alpha$, there is a period $z_{\alpha}^{in}\leq z\leq z_{\alpha}^{out}$ for which $W_1(z) p_{1\alpha} \geq 1$. Assuming that the wash-outs are effective enough, any asymmetry in the flavour $\alpha$ generated before $z_{\alpha}^{in}$ is fully erased by wash-outs. Therefore, there is no dependence on the initial condition in this case. An analytical approximation for the asymmetry for $z_{\alpha}^{in}\leq z\leq z_{\alpha}^{out}$ can then be found by setting the right-hand side of Eq. \eqref{BEalpha} to zero (this corresponds to the so-called \textit{strong wash-out balance approximation} \cite{Fong:2010up, Fong_2012}):
\be
N_{\alpha\alpha}(z) \simeq -\frac{\epsilon_{\alpha\alpha}^{(1)}}{W_1p_{1\alpha}}\frac{dN_{N_1}}{dz}\simeq -\frac{\epsilon_{\alpha\alpha}^{(1)}}{W_1p_{1\alpha}}\frac{dN_{N_1}^{eq}}{dz} = \frac{2N_{\ell}^{eq}}{z\kappa_1p_{1\alpha}}\epsilon_{\alpha\alpha}^{(1)},
\ee
where we have used the approximation $N_{N_1}(z)\simeq N_{N_1}^{eq}(z) = \frac{1}{2} z^2 K_2(z)N_N^{eq}(0)$ valid in this regime and $d(z^2K_2(z))/dz = - z^2K_1(z)$.
After $z_\alpha^{out}$, the asymmetry in the flavour $\alpha$ gets frozen so that:
\be
N_{\alpha\alpha}(\infty)\simeq\frac{2N_{\ell}^{eq}}{\kappa_1}\frac{\epsilon_{\alpha\alpha}^{(1)}}{z_{\alpha}^{out}p_{1\alpha}},
\ee
with $z_\alpha^{out}\simeq 1.25 \ln{(25\kappa_1 p_{1\alpha})}$.

In the two-flavour approximation, since usually $z_d \equiv z_\tau^{out}\simeq z_{\tau^\perp}^{out}$, the final $B-L$ asymmetry is given by
\be
N_{B-L}^\text{1BE2F}(\infty)\simeq\frac{2N_{\ell}^{eq}}{z_d\kappa_1}\frac{\epsilon_{\tau\tau}^{(1)}p_{1\tau^\perp} + \epsilon_{\tau^\perp\tau^\perp}^{(1)}p_{1\tau}}{p_{1\tau}p_{1\tau^\perp}}.
\ee

\subsection{Weak Wash-Out Regime}
In the weak wash-out regime we need to distinguish between two different initial conditions, namely thermal initial abundance (TIA) and vanishing initial abundance (VIA) for which $N_{N_1}(z_0) = N_{N_1}^{eq}(z_0)$ and $N_{N_1}(z_0) = 0$ respectively.

\subsubsection{Vanishing Initial Abundance}\label{App:WWVIA}
The number of heavy neutrinos evolving with $z$ in the VIA case, for which $N_{N_1} (z_0) = 0$, follows from Eq. \eqref{BEsolN}:
\be
\begin{split}
N_{N_1} (z) &\simeq \int_{z_0}^{z} D_1(z') N_{N_1}^{eq}(z')e^{-\int_{z'}^{z}D_1(z'')\,dz''}\,dz'\\
&= 2N^{eq}_{\ell} \int_{z_0}^{z} W_1(z') e^{-\int_{z'}^{z}D_1(z'')\,dz''}\,dz'
\end{split}
\ee
We define $z_{eq}$ as the time at which $N_{N_1} (z_{eq}) = N_{N_1}^{eq} (z_{eq})$, that corresponds to a maximum for $N_{N_1}(z)$. Indeed, from Eq. \eqref{BEN} it follows that, at $z_{eq}$, $dN_{N_1}/dz = 0$ and $d^2N_{N_1}/dz^2 = D_1 dN_{N_1}^{eq}/dz < 0$.
The number of RH neutrinos at $z_{eq}$ can be computed using some analytical approximations such as in \cite{BUCHMULLER2005305}, of which we employ the same result:
\be
N(\kappa_1) \equiv N_{N_1}(z_{eq}) \simeq \frac{9\pi}{16} \kappa_1.
\ee
For $z<z_{eq}$, we can assume $N_{N_1}^{eq}\gg N_{N_1}$. Then, from Eq. \eqref{BEsolalpha} and using Eq. \eqref{eq:W1}, we find that the asymmetry up to $z_{eq}$ reads:
\be
\begin{split}
N_{\alpha\alpha}(z_{eq}) &\simeq \epsilon_{\alpha\alpha}^{(1)}\int_{z_0}^{z_{eq}}D_1(z')N_{N_1}^{eq}(z')e^{-\int_{z'}^{z_{eq}}W_1(z'')p_{1\alpha}\,dz''}\,dz'\\
&= 2N_{\ell}^{eq}\frac{\epsilon_{\alpha\alpha}^{(1)}}{p_{1\alpha}}\left(1-e^{-p_{1\alpha}\frac{N(\kappa_1)}{2N_\ell^{eq}}}\right)\simeq -N(\kappa_1)\epsilon_{\alpha\alpha}^{(1)} + \frac{N(\kappa_1)^2}{4N_\ell^{eq}}\epsilon_{\alpha\alpha}^{(1)}p_{1\alpha}.
\end{split}
\ee
For $z>z_{eq}$, we can instead write the asymmetry as:
\be
\begin{split}
N_{\alpha\alpha}(z) - N_{\alpha\alpha}(z_{eq})
&= -\epsilon_{\alpha\alpha}^{(1)}\int_{z_{eq}}^{z}\frac{dN_{N1}}{dz'}e^{-\int_{z'}^{z}W_1(z'')p_{1\alpha}\,dz''}\,dz'\\
&\simeq -\epsilon_{\alpha\alpha}^{(1)}\int_{z_{eq}}^{z}\frac{dN_{N1}}{dz'}\left(1-p_{1\alpha}\int_{z'}^{z}W_1(z'')\,dz''\right)\,dz'\\
&\simeq \epsilon_{\alpha\alpha}^{(1)}\left(N(\kappa_i) - N_{N_1}(z)\right)-p_{1\alpha}\epsilon_{\alpha\alpha}^{(1)}\int_{z_{eq}}^z dz'\, D_1(z')N_{N_1}(z')\int_{z'}^z dz''\, W_1(z'')\\
&\simeq \epsilon_{\alpha\alpha}^{(1)}\left(N(\kappa_i) - N_{N_1}(z)\right),
\end{split}
\ee
where in the last passage we have neglected the (negative) term proportional to $p_{1\alpha}$. This last approximation may be a bit inaccurate if $10^{-2}<\kappa_1 p_{1\alpha}< 1$ \cite{BUCHMULLER2005305}.

The final asymmetry then reads ($N_{N_1}(\infty) = 0$):
\be
N_{\alpha\alpha}(\infty) \simeq \epsilon_{\alpha\alpha}^{(1)}p_{1\alpha}\frac{N(\kappa_1)^2}{4N_\ell^{eq}} \simeq \frac{81\pi^2}{1024N_\ell^{eq}}\kappa_1^2\epsilon_{\alpha\alpha}^{(1)}p_{1\alpha},
\ee
which, in the two-flavour approximations results in
\be
N_{B-L}^\text{1BE2F} (\infty) \simeq \frac{81\pi^2}{1024N_\ell^{eq}}\kappa_1^2(\epsilon_{\tau\tau}^{(1)}p_{1\tau} + \epsilon_{\tau^\perp\tau^\perp}^{(1)}p_{1\tau^\perp})
\ee

\subsubsection{Thermal Initial Abundance}\label{App:WWTIA}
In the TIA case, $N_{N_1}(z_0) = N_{N_1}^{eq}(z_0)$. We define $z_D$ so that $z_D D_1(z_D) = 2$, i.e. as the time at which decays are in equilibrium against the expanding Universe. In the weak wash-out regime $z_D \gg 1$.

For $z \lesssim 1$ we can consider $N_{N_1}^{eq}(z) \simeq N_{N_1}^{eq} (z_0)$. Hence,
\be
\begin{split}
N_{N_1}(z) &\simeq N_{N_1}^{eq}(z_0) \int_{z_0}^{z}D_1e^{-\int_{z'}^z D_1(z'')\,dz''}\,dz' + N_{N_1}^{eq}(z_0)e^{-\int_{z_0}^z D_1(z')dz'}\\
&= N_{N_1}^{eq}(z_0) \left(1 - e^{-\int_{z_0}^z D_1(z')dz'}\right) + N_{N_1}^{eq}(z_0)e^{-\int_{z_0}^z D_1(z')dz'}=N_{N_1}^{eq}(z_0).
\end{split}
\ee
For $1 < z \leq z_D$, the equilibrium number density is exponentially dropped so that $N_{N_1}^{eq}(z) \ll N_{N_1}(z)$ and we have
\be
N_{N_1}(z) \simeq N_{N_1}^{eq}(z_0)e^{-\int_{1}^z D_1(z')dz'}\simeq N_{N_1}^{eq}(z_0).
\ee
Then the asymmetry up to $z_D$ is roughly zero.
At $z \simeq z_D$ the heavy neutrinos start to decay effectively and their abundance for $z\gtrsim z_D$ is exponentially damped:
\be
N_{N_1}(z) \simeq N_{N_1}^{eq}(z_0)e^{-\int_{z_D}^z D_1(z')dz'}.
\ee
The asymmetry at $z>z_D$ then reads:
\be
\begin{split}
N_{\alpha\alpha}(z) &\simeq -\epsilon_{\alpha\alpha}^{(1)}\int_{z_D}^{z}\frac{dN_{N_1}}{dz'}e^{-\int_{z'}^z W_1(z'')p_{1\alpha}\,dz''}\,dz'\\
&\simeq -\epsilon_{\alpha\alpha}^{(1)}\int_{z_D}^{z}\frac{dN_{N_1}}{dz'}\left(1-\int_{z'}^z W_1(z'')p_{1\alpha}\,dz''\right)\,dz'\\
&\simeq \epsilon_{\alpha\alpha}^{(1)}\left(N_{N_1}^{eq}(z_0) - N_{N_1}(z)\right) -p_{1\alpha}\epsilon_{\alpha\alpha}^{(1)}\int_{z_D}^{z}\,dz'D_1(z')N_{N_1}(z')\int_{z'}^z \,dz'' W_1(z'')\,.
\end{split}
\ee
The final asymmetry $B-L$ is then given by:
\be
N_{\alpha\alpha}(\infty) = \epsilon_{\alpha\alpha}^{(1)} N_{N_1}^{eq}(z_0) - \epsilon_{\alpha\alpha}^{(1)}p_{1\alpha}\int_{z_D}^{z}\,dz'D_1(z')N_{N_1}(z')\int_{z'}^z \,dz'' W_1(z'')\,,
\ee
which in the two-flavour approximation becomes
\be\label{BEsolweakTIA}
N_{B-L}^\text{1BE2F}(\infty) = \epsilon^{(1)} N_{N_1}^{eq}(z_0) - (\epsilon_{\tau\tau}^{(1)}p_{1\tau} + \epsilon_{\tau^\perp\tau^\perp}^{(1)}p_{1\tau^\perp})\mathcal{A}(\kappa_1),
\ee
with 
\be 
\mathcal{A}(\kappa_1) \equiv \int_{z_D}^{\infty}\,dz'D_1(z')N_{N_1}(z')\int_{z'}^z \,dz'' W_1(z'')\,dz'.
\ee
Note that, if not for the second term in \eqref{BEsolweakTIA}, when $\epsilon^{(1)} = 0$ the final asymmetry would vanish.

%
\section{Approximate 1-to-2 Flavour Transitional Mass Scale}
\label{Appendix:Mstar}
%

In this appendix we present an analytical approximation for the mass scale 
of the 1-to-2 flavour transition, $M_{10}$. In the following discussion 
we are not going to consider the effects of the $\mu$-Yukawas interactions, 
thus we set $\Lambda_\mu = \Gamma_\mu/Hz = 0$.
We notice that what controls the scale of the 1-to-2 flavour transition 
is the coefficient that multiplies the integrated wash-out term in 
Eq. \eqref{eq:SDEF}, that is 
\be
\mathcal{T}(M_1, p_{1\tau})\equiv \Lambda_{\tau}p_{1\tau}p_{1\tau^\perp}\,.
\ee 
%
In the ``standard" picture, we find that typically 
$p_{1\tau}p_{1\tau^\perp}\approx 0.25$ and the transition happens at a mass 
scale $M_{10} \approx 10^{12}$ GeV. We then define 
$\mathcal{T}_{12}\equiv \mathcal{T}(M_1 = 10^{12}\text{ GeV}, p_{1\tau} = 0.5)$. 
Our ansatz is that the condition 
\be
\mathcal{T}(M_{10},p_{1\tau}) \approx \mathcal{T}_{12}\,,
\ee
%
gives approximately the mass scale of the 1-to-2 flavour transition. 
In terms of $M_{10}$ the above condition reads:
\be
\label{eq:ansatz}
\frac{M_{10}}{10^{12}\text{ GeV}} \approx  4p_{1\tau}p_{1\tau^\perp}. 
\ee
%
For the values of the parameters used to obtain the bottom-left (right) 
panel of Fig. \ref{fig:SCNH_d_270_a21_0_a31_0}, for example,
we get $4 p_{1\tau}p_{1\tau^\perp}\cong 0.1$ ($6\times 10^{-3}$), and 
correspondingly the mass scale of the transition is found at 
$M_{10} \cong 10^{11}$ GeV ($6\times 10^9$ GeV), in agreement with 
the figure. We find, however, that the above approximation can 
underestimate $M_{10}$ by a factor $\mathcal{O}(1.5-2.5)$ 
when $M_{10}\approx 10^{12}$ GeV. The approximation is more accurate 
when $M_{10}\ll 10^{12}$ GeV, as in the ``non-standard'' scenarios discussed 
in the present paper.

\providecommand{\href}[2]{#2}\begingroup\raggedright\endgroup

\end{document}